\documentclass[pre, aps, twocolumn, longbibliography, notitlepage, nobibnotes]{revtex4-1}
\pdfoutput=1
\usepackage{amsmath, amsfonts}
\usepackage{verbatim}
\usepackage{graphicx}
\usepackage{esint}
\usepackage{float}
\usepackage{bm}
\usepackage{bbm}
\usepackage[subnum]{cases}
\usepackage{aligned-overset}
\usepackage{standalone}

\usepackage{tikz}
\usetikzlibrary{calc, decorations.pathreplacing, matrix, positioning, shapes.misc}
\tikzset{cross/.style={cross out, draw, minimum size=2*(#1-\pgflinewidth), inner sep=0pt, outer sep=0pt}}

\usepackage{xcolor}
\definecolor{darkblue}{rgb}{0,0,0.6}
\definecolor{red}{HTML}{CC78BC}
\definecolor{green}{HTML}{029E73}
\definecolor{blue}{HTML}{0173B2}
\definecolor{lightblue}{HTML}{56B4E9}
\definecolor{brown}{HTML}{CA9161}
\definecolor{grey}{HTML}{949494}
\usepackage[colorlinks, allcolors=darkblue]{hyperref}

\begin{document}

\title{Long-range order in two-dimensional systems with fluctuating active stresses}

\author{Yann-Edwin Keta}
\email{keta@lorentz.leidenuniv.nl}

\author{Silke Henkes}
\email{shenkes@lorentz.leidenuniv.nl}

\affiliation{Instituut-Lorentz for Theoretical Physics, Universiteit Leiden, 2333 CA Leiden, Netherlands}

\begin{abstract}
In two-dimensional tissues, such as developing germ layers, pair-wise forces (or active stresses) arise from the contractile activity of the cytoskeleton, with dissipation provided by the three-dimensional surroundings.
We show analytically how these pair-wise stochastic forces, unlike the particle-wise independent fluctuating forces usually considered in active matter systems, produce conserved centre-of-mass dynamics and so are able to damp large-wavelength displacement fluctuations in elastic systems.
A consequence of this is the stabilisation of long-range translational order in two dimensions, in clear violation of the celebrated Mermin-Wagner theorem, and the emergence of hyperuniformity with a structure factor $S(q) \sim q^2$ in the $q \to 0$ limit.
We then introduce two numerical cell tissue models which feature these pair-wise active forces.
First a vertex model, in which the cell tissue is represented by a tiling of polygons where the edges represent cell junctions and with activity provided by stochastic junctional contractions.
Second an active disk model, derived from active Brownian particles, but with pairs of equal and opposite stochastic forces between particles.
We study the melting transition of these models and find a first-order phase transition between an ordered and a disordered phase in the disk model with active stresses.
We confirm our analytical prediction of long-range order in both numerical models and show that hyperuniformity survives in the disordered phase, thus constituting a hidden order in our model tissue.
Owing to the generality of this mechanism, we expect our results to be testable in living organisms, and to also apply to artificial systems with the same symmetry.
\end{abstract}

\maketitle

\section{Introduction}

Biological tissues are a paradigmatic active or living material, characterised by the competition between crowding effects and active driving. In in-vitro epithelial cell layers, the focus of most models, activity arises primarily from cells crawling over the underlying solid substrate \cite{trepat2009physical,alert2020physical}. This setup has motivated a number of minimal microscopic models of tissues, such as
self-propelled particles \cite{solon2015active,romanczuk2012active,henkes2020dense,martin2021statistical} and self-propelled Voronoi models \cite{bi2016motilitydriven,pasupalak2020hexatic,li2024relaxation}.
In these, the ``self-propelled'' qualifier indicates that a time-persistent and space-independent force acts on each degree of freedom (\textit{e.g.} particle or Voronoi centre). However, in many developmental contexts such as avian \cite{rozbicki2015myosiniimediated,saadaoui2020tensile,sknepnek2023generating} and drosophila
\cite{brauns2024geometric} early development, the driving activity arises from cytoskeletal contractility \cite{mizuno2007nonequilibrium,mackintosh2008nonequilibrium,mackintosh2010active}, and the cells are surrounded by fluid that provides dissipation, but against which the tissue cannot exert active forces.
These models therefore are not suitable here, and instead the active interactions within cell sheets need to be
modelled as stochastic forces which act on pairs of degrees of freedom in an opposite manner, thus respecting action-reaction.
Since the sum of these driving forces cancel, the dynamics they produce conserves the centre of mass, and should be regarded in continuum elastic formulation as deriving from the divergence of an active stress $\underline{\boldsymbol{\sigma}}^{\mathrm{act}}$.
We may then write for these overdamped systems
\begin{equation}
\zeta \dot{\boldsymbol{u}}(\boldsymbol{r}, t) = -\hspace{-1pt}\int \mathrm{d}^2\boldsymbol{r}^{\prime} \, \underline{\boldsymbol{D}}^{\mathrm{el}}(\boldsymbol{r} - \boldsymbol{r}^{\prime}) \boldsymbol{u}(\boldsymbol{r}^{\prime}, t) + \nabla \cdot \underline{\boldsymbol{\sigma}}^{\mathrm{act}}(\boldsymbol{r}, t)
\label{eq:ucont}
\end{equation}
where $\boldsymbol{u}(\boldsymbol{r}, t)$ describes the elastic deformation from position $\boldsymbol{r}$, $\zeta$ is a friction coefficient, and $\underline{\boldsymbol{D}}^{\mathrm{el}}$ is a dynamical matrix \cite{saarloos2024soft} describing elasticity (see Fig.~\ref{fig:models}(a)). More generally, one can envisage a full dissipation matrix $\underline{\boldsymbol{\zeta}}$ that includes additional cell-cell terms, a point we will return to below.

\begin{figure*}
\hspace*{-31pt}
\pgfmathsetmacro{\xcm}{-4.5}
\pgfmathsetmacro{\ycm}{-1.625}
\pgfmathsetmacro{\xcmldiff}{1.5}

\pgfmathsetmacro{\xvm}{{\xcm + 4}}
\pgfmathsetmacro{\yvm}{0}
\pgfmathsetmacro{\xvmldiff}{-2.375}

\pgfmathsetmacro{\xabp}{{\xvm + 4.5}}
\pgfmathsetmacro{\yabp}{-1.125}
\pgfmathsetmacro{\xabpldiff}{-1.25}

\pgfmathsetmacro{\columns}{4}
\pgfmathsetmacro{\rows}{4}
\pgfmathsetmacro{\angle}{40}
\pgfmathsetmacro{\lx}{1.625}
\pgfmathsetmacro{\ly}{{0.75*\lx}}
\pgfmathsetmacro{\depth}{{0.2*\lx}}
\pgfmathsetmacro{\depthbis}{{2*\depth}}
\pgfmathsetmacro{\zoomrec}{0.2}
\pgfmathsetmacro{\zoomarrow}{0.1}
\pgfmathsetmacro{\ylabel}{-2.5}

\begin{tikzpicture}


\draw ({\xcm - \columns*\lx + \xcmldiff}, \ylabel) node[anchor=west]{\strut{(a) 2D continuous medium}};

\foreach \col in {1, ..., \columns} {
    \foreach \row in {1, ..., \rows} {
        \coordinate (G\col\row) at
            ({\xcm - \col*\lx + \row*cos(\angle)*\ly}, {\ycm + \row*sin(\angle)*\ly});
    }
}

\foreach \col in {1, ..., \columns} {
    \coordinate (G\col0) at ($(G\col1)+(0,-\depth)$);
    \draw (G\col1) -- (G\col\rows);
    \draw (G\col1) -- (G\col0);
}
\foreach \row in {1, ..., \rows} {
    \coordinate (G0\row) at ($(G1\row)+(0,-\depth)$);
    \draw (G1\row) -- (G\columns\row);
    \draw (G1\row) -- (G0\row);
}
\draw (G10) -- (G\columns0);
\draw (G01) -- (G0\rows);
\draw ($(G\columns1)!0.5!(G\columns\rows)$) node[anchor=center, rotate=\angle, yshift=5]{2D active tissue};

\draw[opacity=0.1, fill=red] (G\columns0) --  ($(G\columns0) + (0, {-\depthbis})$) -- ($(G10) + (0, {-\depthbis})$) -- (G01) -- cycle;
\draw[opacity=0.1, fill=red]  ($(G0\rows) + (0, {-\depthbis})$) -- (G0\rows) -- (G01) -- ($(G01) + (0, {-\depthbis})$) -- cycle;
\draw[red] ($(G11)!0.5!(G1\rows)$) node[anchor=center, rotate=\angle, yshift=-13.75, xshift=-10]{dissipative support};

\fill[opacity=0.20, grey] (G32) -- (G42) -- (G43) -- (G33) -- cycle;
\pgfmathsetmacro{\bis}{"b"}
\coordinate (G32\bis) at ($(G32)!\zoomrec!(G43)$);
\coordinate (G42\bis) at ($(G42)!\zoomrec!(G33)$);
\coordinate (G43\bis) at ($(G43)!\zoomrec!(G32)$);
\coordinate (G33\bis) at ($(G33)!\zoomrec!(G42)$);
\draw[green, thick, dashed] (G32) -- (G42) -- (G43) -- (G33) -- cycle;
\draw[green, ultra thick] (G32\bis) -- (G42\bis) -- (G43\bis) -- (G33\bis) -- cycle;
\draw[green] ($(G32)!0.5!(G43)$) node{$\underline{\boldsymbol{D}}^{\mathrm{el}}$};
\foreach \p in {G32, G42, G43, G33}
    \draw[green, ultra thick, ->] (\p\bis) -- ($(\p)!0.125!(\p\bis)$);

\fill[opacity=0.20, grey] (G11) -- (G21) -- (G22) -- (G12) -- cycle;
\fill[opacity=0.15, grey] (G11) -- (G10) -- (G20) -- (G21) -- cycle;
\fill[opacity=0.35, grey] (G11) -- (G01) -- (G02) -- (G12) -- cycle;
\draw[blue, ultra thick, <->] ($(G12)!\zoomarrow!(G11)$) -- ($(G11)!\zoomarrow!(G12)$);
\draw[blue, ultra thick, <->] ($(G22)!\zoomarrow!(G12)$) -- ($(G12)!\zoomarrow!(G22)$);
\draw[blue, ultra thick, <->] ($(G22)!\zoomarrow!(G21)$) -- ($(G21)!\zoomarrow!(G22)$);
\draw[blue, ultra thick, <->] ($(G21)!\zoomarrow!(G11)$) -- ($(G11)!\zoomarrow!(G21)$);
\draw[blue] ($(G11)!0.5!(G22)$) node{$\underline{\boldsymbol{\sigma}}^{\mathrm{act}}$};

\draw[dashed] (G20) -- ++(0, {-\depthbis});
\coordinate (U) at (G20);
\draw[->, ultra thick, red] (U) -- ++({+0.7*\lx}, 0) node[midway, below]{$-\zeta \dot{\boldsymbol{u}}$};
\draw[fill, brown] (U) circle(0.1);


\draw ({\xvm + \xvmldiff}, \ylabel) node[anchor=west]{\strut{(b) junctional tension VM (jtVM)}};

\coordinate (A) at (\xvm, \yvm);
\draw (A) circle (0.1) node[left]{$i$};
\coordinate (B) at ($(A) + ({1.5*cos(30 + 0.2*60) + 0.0375}, {1.125*sin(30 + 0*60) - 0.025})$);
\coordinate (C) at ($(A) + ({1.0875*cos(30 + 1*60)}, {1.1625*sin(30 + 1*60)})$);
\draw[fill=black] (C) circle (0.1) node[above left, xshift=0pt, yshift=-1pt]{$\nu$};
\coordinate (D) at ($(A) + ({1.125*cos(30 + 2.1*60)}, {1.125*sin(30 + 1.5*60)})$);
\draw[fill=black] (D) circle (0.1);
\coordinate (E) at ($(A) + ({1.3125*cos(30 + 3*60)}, {1.5*sin(30 + 3*60)})$);
\draw[fill=black] (E) circle (0.1);
\coordinate (F) at ($(A) + ({0.375 + 1.125*cos(30 + 4*60)}, {1.125*sin(30 + 4*60)})$);
\draw[fill=black] (F) circle (0.1);
\coordinate (G) at ($(A) + ({1.125*cos(30 + 5*60)}, {1.125*sin(30 + 5*60)})$);
\draw[fill=black] (G) circle (0.1) node[below, xshift=2pt, yshift=-2pt]{$\nu^{\prime}$};
\coordinate (H) at ($(B) + (1.25, 0.5)$);
\draw[fill=black] (H) circle (0.1) node[right, xshift=2pt]{$\nu^{\prime\prime}$};
\fill[opacity=0.20, grey] (B) -- (C) -- (D) -- (E) -- (F) -- (G) -- (B);
\draw[thin] (B) -- (C) -- (D) -- (E) -- (F) -- (G) -- (B);
\draw[thin] (B) -- (H);
\draw[thin] (C) -- ++(0, 0.5);
\draw[thin] (D) -- ++(-0.5, 0.1);
\draw[thin] (E) -- ++(-0.45, -0.9);
\draw[thin] (F) -- ++(0.15, -0.75);
\draw[thin] (G) -- ++(0.75, -0.3);
\draw[brown, fill=brown] (B) circle (0.1) node[below left, xshift=0pt, yshift=-1pt]{$\mu$};
\draw[blue, ultra thick, <->] (B) -- (C) node[midway, above right, xshift=-6pt]{$T_{\mu\nu}$};
\draw[blue, ultra thick, <->] (B) -- (G) node[midway, right, xshift=0pt, yshift=-3pt]{$T_{\mu\nu^{\prime}}$};
\draw[blue, ultra thick, <->] (B) -- (H) node[midway, below right, yshift=3pt]{$T_{\mu\nu^{\prime\prime}}$};


\draw ({\xabp + \xabpldiff}, \ylabel) node[anchor=west]{\strut{(c) pair ABP (pABP)}};

\coordinate (a) at (\xabp, \yabp);
\draw[opacity=0.20, fill=grey] (a) circle (1);
\draw (a) circle (1) node[opacity=1, right, brown]{$\mu$};
\draw (a) node[cross=2.5pt, brown]{};
\coordinate (b) at ({\xabp + 0.5}, {\yabp + 2});
\draw[opacity=0.25, fill=grey] (b) circle (1);
\draw (b) circle (1) node[left, brown]{$\nu$};
\draw (b) node[cross=2.5pt, brown]{};
\draw[lightblue, ->] (a) -- ($(a) + (-0.75, 1)$) node[midway, right]{$f \boldsymbol{u}(\theta_{\mu})$};
\draw[lightblue, ->] (b) -- ($(b) + (0.75, 1)$) node[midway, left]{$f \boldsymbol{u}(\theta_{\nu})$};
\draw[lightblue, dashed, ->] ($(a) + (-0.75, 1)$) -- ++(-0.75, -1) node[midway, left]{$-f \boldsymbol{u}(\theta_{\nu})$};
\draw[blue, ultra thick, ->] (a) -- ++(-1.5, 0) node[midway, below, xshift=11pt]{$\boldsymbol{\lambda}^{\mathrm{act}}_{\mu \leftarrow \nu}$};
\draw[lightblue, dashed, ->] ($(b) + (0.75, 1)$) -- ++(0.75, -1) node[midway, right]{$-f \boldsymbol{u}(\theta_{\mu})$};
\draw[blue, ultra thick, ->] (b) -- ++(1.5, 0) node[midway, below, xshift=-11pt]{$\boldsymbol{\lambda}^{\mathrm{act}}_{\nu \leftarrow \mu}$};

\end{tikzpicture}
\caption{(a) Sketch of a two-dimensional continuous system with fluctuating active stress described by \eqref{eq:ucont}.
(b) Vertex model with stochastic junction tension (\hyperref[sec:vm]{jtVM}). (c) Active Brownian particles with pair-wise propulsion forces (\hyperref[sec:abp]{pABP}).}
\label{fig:models}
\end{figure*}

Recent studies within a field theoretical \cite{ikeda2023correlated} and a granular context \cite{maire2024enhancing} have linked stochastic active stresses and the emergence of crystalline phases with long-range translational order and hyperuniform density fluctuations -- fluctuations which vanish on large length scales.
In the active matter context, long-range translational order and hyperuniformity have also been uncovered in systems with anti-alignment \cite{boltz2024hyperuniformity}, chiral or non-reciprocal interactions \cite{lei2019nonequilibrium,kuroda2025longrange,chen2024emergent}, oscillatory driving forces \cite{ikeda2024continuous}, pulsating contractions \cite{li2024fluidization}, in systems of driven-dissipative hard spheres \cite{lei2019hydrodynamics}, as well as in active turbulence \cite{backofen2024nonequilibrium} -- see a recent review \cite{lei2025nonequilibrium}.
However, a broader understanding together with a minimal model of these processes is lacking, and the link to the biological context has only very recently been made \cite{li2024fluidization}.

It should be noted that within a continuum active fluid paradigm, Eq.~\eqref{eq:ucont} with pair forces would belong to the well-understood class of dry or mixed dry-wet active nematics or active gels which have long been proposed as tissue models \cite{prost2015active,doostmohammadi2018active}. The difference here is the active \emph{solid} starting point with purely fluctuating activity, very unlike the typical active nematic state with moving defects. Moreover, we take microscopic order into account, allowing us to access translational and hexatic order parameters -- see however \cite{armengol-collado2023epithelia,armengol-collado2024hydrodynamics} for an intrinsically hexatic active fluid approach.

In this paper we first derive the active elastic theory of long-range correlations at linear order. We then show in two models of dense biological tissues how pair-wise fluctuating active forces stabilise long-range translational order.
In addition, we show numerically that hyperuniformity is displayed both in the low-noise ordered solid and the large-noise disordered liquid.
We contrast these results with simulations of analogous models with particle-wise fluctuating active forces where translational order is at most quasi-long-range with algebraically decreasing correlations \cite{shi2023extreme} and diverging structural fluctuations in the infinite size limit.
The paper is organised as follows.
In Sec.~\ref{sec:analysis} we derive analytically the structural fluctuations of models described by \eqref{eq:ucont}.
In Sec.~\ref{sec:simulations} we introduce our models and perform the numerical study of their structural characteristics, including their melting transition.
In Sec.~\ref{sec:conclusion} we gather our findings and discuss their future applications.

\section{Long-range translational order and hyperuniformity}
\label{sec:analysis}

At thermal equilibrium, the Mermin-Wagner theorem indicates that there can be no spontaneous breaking of a continuous symmetry in dimensions $d = 1,2$ at finite temperature \cite{halperin2019hohenberg}.
A consequence of this is that thermally induced long-range displacement fluctuations grow with system size in theses low dimensions, and long-range translational order is unachievable in the thermodynamic limit \cite{kosterlitz2016commentary,illing2017mermin}. Moving flocks however (almost certainly \cite{benvegnen2023metastability}) display true long-range polar \emph{orientational} order \cite{toner1998flocks}, the first indication that activity allows for exceptions of this law.

Structural fluctuations can be quantified through the counting of the number $N(R)$ of particles in a volume $R^d$ as a function of $R$.
For typical disordered systems such as liquids, the variance $\mathrm{Var}(N ; R) = \langle N(R)^2 \rangle -\langle N(R) \rangle^2$ grows as $R^d$, while $\mathrm{Var}(N ; R) \sim R^{d - 1}$ in periodic crystals \cite{torquato2018hyperuniform}.
Many active systems, in particular those which display flocking behaviours, were reported to display \emph{giant number fluctuations}, in the sense that $\mathrm{Var}(N ; R)$ grows faster than $R^d$ \cite{toner2005hydrodynamics,chate2006simple,narayan2007longlived,deseigne2010collective,zhang2010collective,henkes2011active}.
More recently, long-range translational order in two-dimensional active systems has been a subject of interest.
Galliano \textit{et al.} have shown how such order arises in a nonequilibrium particle model driven by pair-wise random displacements \cite{galliano2023twodimensionala}.
Maire and Plati have derived this property in underdamped granular systems driven by a momentum-conserving noise \cite{maire2024enhancing}, using an approach similar to the one detailed below.
In both cases, long-range translational order is associated with \textit{hyperuniformity}. This
designates the anomalous vanishing of density fluctuations on large length scales \cite{torquato2003local}, associated with a variance $\mathrm{Var}(N ; R)$ which decreases slower than $R^d$, and is known to occur in systems with absorbing state transitions \cite{hexner2017noise,deluca2024hyperuniformity,fernandez-nieves2024hyperuniform}.

In this Section we derive, for a two-dimensional continuous elastic medium, the fluctuations of the displacement field at linear order.
In a second part we show how these relate, in a discrete system, to the translational order correlation function and the scaling of the structure factor over large wavelengths.

\subsection{Two-dimensional continuous medium}

We introduce the Fourier transform in space and time of the displacement field
\begin{equation}
\begin{aligned}
\boldsymbol{u}(\boldsymbol{r}, t) &= \frac{1}{(2\pi)^2} \int \mathrm{d}^2\boldsymbol{q} \, e^{\mathrm{i} \boldsymbol{q} \cdot \boldsymbol{r}} \, \tilde{\boldsymbol{u}}(\boldsymbol{q}, t)\\
&= \frac{1}{(2\pi)^3} \int \mathrm{d}^2\boldsymbol{q} \, \int \mathrm{d}\omega \, e^{\mathrm{i} \boldsymbol{q} \cdot \boldsymbol{r} + \mathrm{i} \omega t} \, \tilde{\boldsymbol{U}}(\boldsymbol{q}, \omega),
\end{aligned}
\label{eq:ftdef}
\end{equation}
\mbox{}\\
and write \eqref{eq:ucont} in Fourier space as
\begin{equation}
\mathrm{i} \omega \zeta \tilde{\boldsymbol{U}}(\boldsymbol{q}, \omega) = -\tilde{\underline{\boldsymbol{D}}}^{\mathrm{el}}(\boldsymbol{q}) \tilde{\boldsymbol{U}}(\boldsymbol{q}, \omega) + \tilde{\boldsymbol{\Lambda}}(\boldsymbol{q}, \omega),
\label{eq:ucontft}
\end{equation}
where $\tilde{\boldsymbol{\Lambda}}(\boldsymbol{q}, \omega)$ is the Fourier transform of the noise term
\begin{equation}
\boldsymbol{\lambda}(\boldsymbol{r}, t) = \nabla \cdot \underline{\boldsymbol{\sigma}}^{\mathrm{act}}(\boldsymbol{r}, t).
\label{eq:stress}
\end{equation}

We assume (i) that the Fourier transform of the dynamical matrix $\tilde{\underline{\boldsymbol{D}}}^{\mathrm{el}}(\boldsymbol{q})$ can be decomposed as follows \cite{henkes2020dense}
\begin{equation}
\begin{aligned}
&\tilde{\underline{\boldsymbol{D}}}^{\mathrm{el}}(\boldsymbol{q}) \tilde{\boldsymbol{U}}(\boldsymbol{q}, \omega) =\\
&\qquad |\boldsymbol{q}|^2 \left[(B + \mu) \tilde{\boldsymbol{U}}_{\parallel}(\boldsymbol{q}, \omega) + \mu \tilde{\boldsymbol{U}}_{\perp}(\boldsymbol{q}, \omega)\right],
\end{aligned}
\label{eq:duk}
\end{equation}
where \mbox{$\hat{\boldsymbol{q}} = \boldsymbol{q}/|\boldsymbol{q}|$}, \mbox{$\tilde{\boldsymbol{U}}_{\parallel}(\boldsymbol{q}, \omega) = (\tilde{\boldsymbol{U}}(\boldsymbol{q}, \omega) \cdot \hat{\boldsymbol{q}}) \, \hat{\boldsymbol{q}}$} and \mbox{$\tilde{\boldsymbol{U}}_{\perp}(\boldsymbol{q}, \omega) = \tilde{\boldsymbol{U}}(\boldsymbol{q}, t) - \tilde{\boldsymbol{U}}_{\parallel}(\boldsymbol{q}, \omega)$} are respectively the longitudinal and transverse displacements in Fourier space, and $B$ and $\mu$ are respectively the bulk and shear moduli.

We assume (ii) that the fluctuations of the active stress \eqref{eq:stress} follow \cite{ikeda2023correlated,maire2024enhancing,cavagna2024noise}
\begin{equation}
\left<\boldsymbol{\lambda}(\boldsymbol{r}, t) \cdot \boldsymbol{\lambda}(\boldsymbol{r}^{\prime}, t^{\prime})\right> = -\sigma^2 a^2 \, e^{-|t - t^{\prime}|/\tau} \, \nabla^2 \delta(\boldsymbol{r} - \boldsymbol{r}^{\prime}),
\label{eq:stresscor}
\end{equation}
where $\sigma$ is an energy scale, $a$ a coarse-graining length scale, and $\tau$ a persistence time.
This assumption stems from the idea that the stress is uncorrelated in space but autocorrelated in time.
We highlight here that, from the point of view of numerical particle-based models, the divergence of the stress in \eqref{eq:stress} should be interpreted as the discrete divergence of a noise vector field defined over the ensemble of pairs of particles \cite{cavagna2024noise}.
Assuming that these noise vectors are uncorrelated between pairs leads to \eqref{eq:stresscor} with $\nabla^2$ interpreted as a discrete Laplacian.
We provide an exact derivation for a triangular lattice in the ESI (Sec.~\ref{app:discreteuk2}).
We write \eqref{eq:stresscor} in Fourier space as
\begin{widetext}
\begin{equation}
\left<\tilde{\boldsymbol{\Lambda}}(\boldsymbol{q}, \omega) \cdot \tilde{\boldsymbol{\Lambda}}(\boldsymbol{q}^{\prime}, \omega^{\prime})^*\right> = (2\pi)^3 \frac{2 \sigma^2 \tau |\boldsymbol{q}|^2}{1 + \omega^2 \tau^2} \, \delta(\omega - \omega^{\prime}) \, a^2\delta(\boldsymbol{q} - \boldsymbol{q}^{\prime}).
\label{eq:stresscorft}
\end{equation}

\begin{subequations}
We use \eqref{eq:ucontft}, \eqref{eq:duk}, and \eqref{eq:stresscorft}, to compute the equal-time displacement fluctuations in Fourier space (see ESI, Sec.~\ref{app:fulluk2}, for a full derivation) and obtain
\begin{align}
\left<\tilde{\boldsymbol{u}}(\boldsymbol{q}, t) \cdot \tilde{\boldsymbol{u}}(\boldsymbol{q}^{\prime}, t)^*\right> = \frac{\sigma^2 \tau^2}{2 \zeta^2} \left[\frac{\xi_{\parallel}^{-2}}{1 + |\boldsymbol{q}|^2 \xi_{\parallel}^2} + \frac{\xi_{\perp}^{-2}}{1 + |\boldsymbol{q}|^2 \xi_{\perp}^2}\right] (2 \pi a)^2 \delta(\boldsymbol{q} - \boldsymbol{q}^{\prime}),
\label{eq:uk2stress}
\end{align}
where we have used the longitudinal and transversal correlation length scales $\xi_{\parallel} = \sqrt{(B + \mu) \tau/\zeta}$ and $\xi_{\perp} = \sqrt{\mu \tau/\zeta}$.
We compare this result to the case of a space-uncorrelated fluctuating force field with correlations $\left<\boldsymbol{\lambda}(\boldsymbol{r}, t) \cdot \boldsymbol{\lambda}(\boldsymbol{r}^{\prime}, t^{\prime})\right> = f^2 a^2 \, e^{-|t - t^{\prime}|/\tau} \, \delta(\boldsymbol{r} - \boldsymbol{r}^{\prime})$, for which the spectrum reads \cite{henkes2020dense}
\begin{align}
\left<\tilde{\boldsymbol{u}}(\boldsymbol{q}, t) \cdot \tilde{\boldsymbol{u}}(\boldsymbol{q}^{\prime}, t)^*\right> =  \frac{f^2 \tau^2}{2 \zeta^2} \frac{1}{|\boldsymbol{q}|^2} \left[\frac{\xi_{\parallel}^{-2}}{1 + |\boldsymbol{q}|^2 \xi_{\parallel}^2} + \frac{\xi_{\perp}^{-2}}{1 + |\boldsymbol{q}|^2 \xi_{\perp}^2}\right] (2 \pi a)^2 \delta(\boldsymbol{q} - \boldsymbol{q}^{\prime}).
\label{eq:uk2force}
\end{align}
\label{eq:uk2}%
\end{subequations}
\end{widetext}
Spectrum \eqref{eq:uk2stress} importantly differs from \eqref{eq:uk2force} by the absence of the factor $1/|\boldsymbol{q}|^2$ which diverges on large length scales $|\boldsymbol{q}| \to 0$.
This difference arises solely from the spectrum of stochastic forces \eqref{eq:stresscorft} whose fluctuations vanish in the limit $|\boldsymbol{q}| \to 0$.

We define the variance of displacement for a square two-dimensional system of linear size $L$ as
\begin{equation}
\begin{aligned}
&\left<u^2\right>\\
&= \lim_{t \to \infty} \frac{1}{(2\pi)^4} \iint_{2\pi/L}^{2\pi/a} \mathrm{d}^2\boldsymbol{q} \, \mathrm{d}^2\boldsymbol{q}^{\prime} \left<\tilde{\boldsymbol{u}}(\boldsymbol{q}, t)\cdot\tilde{\boldsymbol{u}}(\boldsymbol{q}^{\prime}, t)^*\right>,
\end{aligned}
\end{equation}
where we introduced a lower-wave-vector cuttof $q_{\mathrm{min}} = 2\pi/L$ due to finite system sizes $L$ and upper-wave-vector cutoff $q_{\mathrm{max}} = 2\pi/a$ due to coarse-graining over length scale $a$.
The difference in $\boldsymbol{q}$-scaling we have stressed between spectra \eqref{eq:uk2} leads to the following respective variances scaling in the thermodynamic limit
\begin{subequations}
\label{eq:u2}
\begin{align}
\label{eq:u22}
\left<u^2\right>^\sigma \underset{L \to \infty}&{=} \left<u^2\right>^\sigma_{\infty} = \mathrm{cst},\\
\left<u^2\right>^f \underset{L \to \infty}&{=} C^f \log(L/a),
\end{align}
\end{subequations}
where the $\sigma$ and $f$ subscript refer to the fluctuating stress and force cases respectively.
In the latter case of a fluctuating force field, this variance diverges for infinite systems~\cite{peierls1935quelques,imry1979longrange,maire2024enhancing}.

\subsection{Two-dimensional regular lattice}

We just showed how two-dimensional systems with fluctuating active stresses have finite displacement fluctuations in the thermodynamic limit $L \to \infty$.
This motivates us to investigate structural fluctuations in discrete ordered systems.
We then consider a system of $N$ particles with displacements $\boldsymbol{u}_i = \boldsymbol{r}_i - \boldsymbol{r}_i^0$ from their equilibrium lattice points $\boldsymbol{r}_i^0$.
We highlight that the displacement spectra \eqref{eq:uk2} follow the transformation from continuous to discrete Fourier space \mbox{$\left<\tilde{\boldsymbol{u}}(\boldsymbol{q}, t)\cdot\tilde{\boldsymbol{u}}(\boldsymbol{q}^{\prime}, t)^*\right> \to a^4 \left<\tilde{\boldsymbol{u}}_{\boldsymbol{q}}(t) \cdot \tilde{\boldsymbol{u}}_{\boldsymbol{q}^{\prime}}(t)^*\right>$}, with the coarse-graining length $a = L/\sqrt{N}$, through the substitution rule $\delta(\boldsymbol{q} - \boldsymbol{q}^{\prime}) \to (\Delta q)^{-2} \delta_{\boldsymbol{q},\boldsymbol{q}^{\prime}}$ with $\Delta q = 2\pi/L$ the wave-vector spacing \cite{henkes2020dense}.

We introduce the local hexatic orientational order parameter $\psi_{6,i}$ and the local translational order parameter $\psi_{\boldsymbol{q}_0,i}$ in order to quantify structural order \cite{nelson1979dislocationmediated,kardar2007statistical,engel2013harddisk,shi2023extreme},
\begin{eqnarray}
\label{eq:psi6}
\psi_{6,i} = \frac{1}{z_i} \sum_{j \in \mathcal{N}_i} e^{6\mathrm{i}\theta_{ij}},~\theta_{ij} = \mathrm{arg}(\boldsymbol{r}_j - \boldsymbol{r}_i),\\
\label{eq:psiq0}
\psi_{\boldsymbol{q}_0,i} = e^{\mathrm{i} \boldsymbol{q}_0 \cdot (\boldsymbol{r}_i - \boldsymbol{r}_0)},
\end{eqnarray}
where $\mathcal{N}_i$ is the ensemble of nearest neighbours of particle $i$ and $z_i$ the coordination number of particle $i$ \textit{i.e.} the number of particles in $\mathcal{N}_i$, $\boldsymbol{r}_0$ is the position of one of the particles, and $\boldsymbol{q}_0$ is a reciprocal vector of the lattice.
We infer the latter from the position of the maximum of the structure factor \eqref{eq:S} $S(\boldsymbol{q}_0) = \max_{\boldsymbol{q}} S(\boldsymbol{q})$ \cite{li2019accurate}.

The translation by $-\boldsymbol{r}_0$ in \eqref{eq:psiq0} is chosen so that $\psi_{\boldsymbol{q}_0,i}=1$ everywhere in a perfect crystal lattice (see Fig.~\ref{fig:snap}). This works since for any vector $\boldsymbol{v}$ linking two lattice points, $\boldsymbol{q}_0 \cdot \boldsymbol{v}=0$ modulo $2\pi$. As this simply corresponds to choosing a particular origin, it does not affect the correlations \eqref{eq:cr}, only the argument of \eqref{eq:psiq0}.

We quantify the global order parameter with the following ensemble averages ($\mathrm{x}=6,\boldsymbol{q}_0$)
\begin{equation}
\Psi_{\mathrm{x}} = \left<\left|\frac{1}{N}\sum_{i=1}^N \psi_{\mathrm{x},i}\right|\right>,
\label{eq:Psi}
\end{equation}
and the following spatial correlation functions
\begin{equation}
C_{\mathrm{x}}(r) = \left<\frac{\sum_{i \neq j}\psi_{\mathrm{x},i} \, \psi_{\mathrm{x},j}^* \, \delta(r - |\boldsymbol{r}_j - \boldsymbol{r}_i|)}{\sum_{i \neq j} \delta(r - |\boldsymbol{r}_j - \boldsymbol{r}_i|)}\right>.
\label{eq:cr}
\end{equation}
The latter function, considering translational order, is related to the scalings \eqref{eq:u2} (see ESI, Sec.~\ref{app:ctr}) as follows
\begin{subequations}
\label{eq:cq0}
\begin{align}
\label{eq:cq02}
C^\sigma_{\boldsymbol{q}_0}(r) &\underset{\substack{L\to\infty\\r\to\infty}}{=} e^{-\frac{1}{2}|\boldsymbol{q}_0|^2\left<u^2\right>_{\infty}^\sigma},\\
\label{eq:cq01}
C^f_{\boldsymbol{q}_0}(r) &\underset{\substack{L\to\infty\\r\to\infty}}{\sim} r^{-\frac{1}{2}|\boldsymbol{q}_0|^2 C^f}.
\end{align}
\end{subequations}
These scalings show that, in the case a stochastic stress field, \emph{true} long-range translational order is possible, as opposed to the quasi-long-range order in the case of a stochastic force field.

Finally, we define the structure factor \cite{binder2005glassy}
\begin{equation}
S(\boldsymbol{q}) = \frac{1}{N} \sum_{i,j=1}^N \left<e^{\mathrm{i} \boldsymbol{q}\cdot(\boldsymbol{r}_j - \boldsymbol{r}_i)}\right>
\label{eq:S}
\end{equation}
which characterises density fluctuations.
Under the assumptions of normally distributed displacements (\emph{e.g.} satisfied for normally distributed stochastic forces and harmonic interactions) and isotropy, the structure factor can be reduced in the large system size and large wavelength limit to (see ESI, Sec.~\ref{app:S}, for a derivation)
\begin{equation}
S(\boldsymbol{q}) \underset{\substack{L\to\infty\\|\boldsymbol{q}|\sim2\pi/L}}{=} \frac{1}{2N} |\boldsymbol{q}|^2 \left<|\tilde{\boldsymbol{u}}_{\boldsymbol{q}}|^2\right>.
\label{eq:Staylor}
\end{equation}
This result relies on a Taylor expansion of the exponentials in \eqref{eq:S}, in the large-system-size and small-wavevector limits, and is consistent with previous derivations\cite{kim2018effect}.
Therefore, given the displacement spectra \eqref{eq:uk2}, we infer the following scalings for the structure factor \eqref{eq:S}
\begin{subequations}
\begin{align}
\label{eq:Sk2}
S^{\sigma}(\boldsymbol{q}) \underset{\substack{L\to\infty\\|\boldsymbol{q}|\sim2\pi/L}}&{\sim} |\boldsymbol{q}|^2,\\
S^{f}(\boldsymbol{q}) \underset{\substack{L\to\infty\\|\boldsymbol{q}|\sim2\pi/L}}&{\sim} 1,
\end{align}
\label{eq:Sk}%
\end{subequations}
indicating that, in the case of a stochastic stress field, the system is ``maximally hyperuniform'' similarly to other systems with conserved centre-of-mass dynamics \cite{torquato2003local,ikeda2023correlated,deluca2024hyperuniformity,hexner2017noise}.

In summary, we have presented three emerging properties for two-dimensional systems in the thermodynamic limit: finite displacement fluctuations with respect to the lattice positions \eqref{eq:u22}, long-range translational order \eqref{eq:cq02}, and hyperuniformity \eqref{eq:Sk2}.
We stress that these properties derive from considering a space-uncorrelated stochastic stress field rather than a force field: they rely on the driving fluctuations (here the active stress field) decaying on large wavelengths while competing against a dissipation mechanism acting independently on all individuals in the system.
It is thus noteworthy that these properties would disappear if we included a white noise term in \eqref{eq:ucont} (see ESI, Sec.~\ref{sec:thermallimituq2}) -- such a term would be expected to counterbalance the white drag if the system were to follow the second fluctuation-dissipation theorem \cite{kubo1966fluctuationdissipation,doi1988theory,tothova2022overdamped}.
We finally highlight that these properties are not affected if we include an additional viscosity-like cell-cell dissipation term (see ESI, Sec.~\ref{sec:thermallimituq2}), i.e. a full dissipation matrix $\underline{\boldsymbol{\zeta}}$. This adds realism to cell models as it counterbalances the fluctuating stresses, and we can show that this modification as expected only affects the small-wavelength fluctuations.

\section{Tissue models with pair-wise stochastic forces}
\label{sec:simulations}

Our predictions were derived for harmonic systems with particles arranged in regular lattices, and so are potentially susceptible to renormalisation group flow to different exponents from nonlinear terms \cite{ikeda2023correlated,fernandez-nieves2024hyperuniform}.
In order to test the robustness of these predictions outside of the linear and ordered regime, as well as the applicability of these results, we introduce two numerical models suited for the description of dense cell tissues (see Fig.~\ref{fig:models}) and study their structural fluctuations.

\subsection{Model definitions}

In Sec.~\ref{sec:vm}, we introduce a model based on a polygonal tiling of the two-dimensional space, where each tile represents a single cell.
This kind of model has been widely used to represent confluent cell sheets \cite{farhadifar2007influence,fletcher2014vertex,alert2020physical} and allows to represent complex interactions, between cells and between cells and their substrate, as well as constraints on cell shapes.
We highlight here two general classes of these models: Voronoi models \cite{bi2016motilitydriven,barton2017active,pasupalak2020hexatic,li2024relaxation} where the tiling of cells is determined through a Voronoi tesselation from cell centres, and vertex models (VM) \cite{spencer2017vertex,tong2022linear,lin2023structure} where cell corners (or \emph{vertices}) are the degrees of freedom.
The model we first introduce belongs to this second class.

In Sec.~\ref{sec:abp}, we introduce a model based on interacting disks.
Despite such models lacking the possibility to represent accurate cell shapes and deformations, they have been applied with great success to understand mesoscale correlations in active tissues \cite{henkes2020dense}.
Moreover, the simplicity of these models allow for the numerical simulation of the large systems needed to characterise long wavelength fluctuations.

For both models studied here, we introduce the following overdamped equation of motion of their degrees of freedom (respectively cell vertices and disk centres), which is the discrete analogue of \eqref{eq:ucont},
\begin{equation}
\zeta \dot{\boldsymbol{r}}_{\mu} = - \partial_{\boldsymbol{r}_{\mu}} U^{\mathrm{int}} + \boldsymbol{\lambda}^{\mathrm{act}}_{\mu},
\label{eq:dumodel}
\end{equation}
where the degrees of freedom are indexed by $\mu$, $\zeta$ is a friction coefficient, $U^{\mathrm{int}}$ is an interaction potential, and the $\boldsymbol{\lambda}^{\mathrm{act}}_{\mu}$ are the active driving forces.

\subsubsection{Junctional tension vertex model (jtVM)}
\label{sec:vm}

Vertex models are defined as meshes of vertices $\mu$, at positions $\boldsymbol{r}_{\mu}$, linked by edges which enclose faces which describe the cells $i$ (see Fig.~\ref{fig:models}(b)).
By extension we will denote $i$ the centroid of cell $i$, whose position is
\begin{equation}
\boldsymbol{r}_i = \frac{1}{6 A_i} \sum_{\mu~\in~\mathrm{cell}~i} (\boldsymbol{r}_{\mu} + \boldsymbol{r}_{\mu+1})(x_{\mu}y_{\mu+1} - x_{\mu+1}y_{\mu}),
\end{equation}
where $\boldsymbol{r}_{\mu} = (x_{\mu}, y_{\mu})$, vertices $\mu$ are ordered in an anticlockwise direction relative to the cell centre $i$, and where $A_i$ is the area of cell $i$; we denote its perimeter by $P_i$.
We stress that the centroid corresponds to the centre of mass of the cell, \textit{i.e.} the arithmetic mean of all the points on its surface, and should be distinguished from the mean of its corners' positions.

We introduce the following vertex model interaction potential \cite{barton2017active}
\begin{equation}
U^{\mathrm{int}} = \sum_{\mathrm{cells}~i} \left[\frac{1}{2} \frac{k}{A_0} \, (A_i - A_0)^2 + \frac{1}{2} k \, (P_i - P_0)^2\right],
\end{equation}
with $k$ a spring constant, $A_0$ the target area and $P_0$ the target perimeter.

We consider active forces which are pair-wise or particle-wise, respectively (see Fig.~\ref{fig:models}(b) for a visual representation)
\begin{subequations}
\begin{align}
\boldsymbol{\lambda}^{\mathrm{act}}_{\mu} &= \sum_{\nu \wedge \mu} T_{\mu\nu} \hat{\boldsymbol{\ell}}_{\mu\nu},\\
\boldsymbol{\lambda}^{\mathrm{act}}_{\mu} &= f \, \boldsymbol{u}(\theta_{\mu}),
\end{align}
\end{subequations}
where $f$ is the stochastic force amplitude, $\boldsymbol{u}(\theta_{\mu}) = (\cos\theta_{\mu}, \sin\theta_{\mu})$, $\nu \wedge \mu$ designates the vertices $\nu$ linked by an edge to vertex $\mu$, $T_{\mu\nu}$ is the tension of edge $\mu\nu$, $\boldsymbol{\ell}_{\mu\nu} = \boldsymbol{r}_{\nu} - \boldsymbol{r}_{\mu}$ and $\hat{\boldsymbol{\ell}}_{\mu\nu} = \boldsymbol{\ell}_{\mu\nu}/|\boldsymbol{\ell}_{\mu\nu}|$.
These forces evolve stochastically:
tensions $T_{\mu\nu}$ follow an Ornstein-Uhlenbeck process \cite{curran2017myosin,krajnc2020solid,duclut2021nonlinear,yamamoto2022nonmonotonic} and angles $\theta_{\mu}$ diffuse similarly to self-propelled Voronoi models \cite{pasupalak2020hexatic,li2024relaxation},
\begin{subequations}
\begin{align}
\tau \dot{T}_{\mu\nu} &= -T_{\mu\nu} + \sqrt{2 f^2 \tau} \, \eta_{\mu\nu},\\
\dot{\theta}_{\mu} &= \sqrt{1/\tau} \, \eta_{\mu},
\end{align}
\end{subequations}
where $\tau$ is the persistence time, and $\eta_{\mu\nu}$ and $\eta_{\mu}$ are Gaussian white noises with zero means and variances
$\left<\eta_{\mu\nu}(0)\eta_{\alpha\beta}(t)\right> = \delta_{\mu\alpha}\delta_{\nu\beta}\delta(t)$
and $\left<\eta_{\mu}(0)\eta_{\nu}(t)\right> = \delta_{\mu\nu} \delta(t)$.

We perform simulations of $N$ cells with periodic boundary conditions.
We define the unit length $\sqrt{A_0}=1$, energy $k A_0 = 1$, and time $\zeta/k = 1$.
Simulations are started from a regular hexagonal packing which satisfies cell areas $A_i = A_0$.
This implies that the area of the system is $L^2 = N A_0$.
We enforce the shape index $s_0 = P_0/\sqrt{A_0} = 3.72$ such that in absence of forcing the regular hexagonal packing of cells is both rigid and force-free.
We use $\tau = 5$ for all data presented.

This model is integrated using the \textsc{MIT}-licensed library \texttt{cells} \cite{cells}.

\subsubsection{Pair active Brownian particles (pABP)}
\label{sec:abp}

We consider $N$ disk particles with diameters $D$ and positions $\boldsymbol{r}_{\mu}$ which interact via a harmonic pair-wise potential
\begin{equation}
U^{\mathrm{int}} = \sum_{\substack{\mu,\nu=1 \\ \mu \neq \nu}}^N \frac{1}{2} \, k \, (D - |\boldsymbol{r}_{\nu} - \boldsymbol{r}_{\mu}|)^2 \, \Theta(D - |\boldsymbol{r}_{\nu} - \boldsymbol{r}_{\mu}|),
\end{equation}
where $\Theta$ is the Heaviside function.
For this model the cell centres are the disk centres with positions $\boldsymbol{r}_i = \boldsymbol{r}_{\mu}$.

We consider active forces which are pair-wise or particle-wise, respectively (see Fig.~\ref{fig:models}(c) for a visual representation)
\begin{subequations}
\begin{align}
\boldsymbol{\lambda}^{\mathrm{act}}_{\mu} &= f \sum_{\substack{\nu=1 \\ \mu \neq \nu}}^N (\boldsymbol{u}(\theta_{\mu}) - \boldsymbol{u}(\theta_{\nu})) \, \beta(|\boldsymbol{r}_{\nu} - \boldsymbol{r}_{\mu}|),\\
\boldsymbol{\lambda}^{\mathrm{act}}_{\mu} &= f \, \boldsymbol{u}(\theta_{\mu}),
\end{align}
\end{subequations}
where $f$ is the stochastic force amplitude and with the kernel function $\beta(r) = (1 - r/D^{\prime}) \Theta(D^{\prime} - r)$ where $D^{\prime} = 1.5 \, D > D$ to avoid singularities at the point of contact.
These forces evolve stochastically:
angles $\theta_{\mu}$ diffuse as in active Brownian particles \cite{fily2012athermal,solon2015active,romanczuk2012active}
\begin{equation}
\dot{\theta}_{\mu} = \sqrt{1/\tau} \, \eta_{\mu},
\end{equation}
where $\tau$ is the persistence time, and $\eta_{\mu}$ is a Gaussian white noise with zero mean and variance $\left<\eta_{\mu}(0)\eta_{\nu}(t)\right> = \delta_{\mu\nu} \delta(t)$.
We stress that the particle-wise force model corresponds exactly to active Brownian particles (ABPs).

We perform simulations of $N$ cells with periodic boundary conditions.
We define the unit length $D/2=1$, energy $k (D/2)^2 = 1$, and time $\zeta/k = 1$.
Except where noted, simulations are started from a regular hexagonal packing with packing fraction $\phi = N \pi (D/2)^2 /L^2 = 1$, corresponding to a thoroughly solid state.
We use $\tau = 25$ for all data presented.

This model is integrated using the \textsc{GPL}-licensed library \texttt{SAMoS} \cite{samos}.

\subsection{Location and properties of the melting transition}

We expect a disordered \emph{liquid} phase at large stochastic force amplitude $f$, and an ordered \emph{solid} phase at small $f$.
While in equilibrium, it is expected that structural and dynamical characteristics coincide and so determine the phase of the system, we cannot exclude that out of equilibrium, structure and dynamics may indicate different phases (see e.g. the flowing crystals of \cite{briand2016crystallization}).
We thus consider two structural characteristics, the global hexatic and translational order parameters $\Psi_{\mathrm{6}}$ and $\Psi_{\boldsymbol{q}_0}$ \eqref{eq:Psi}, and a dynamical characteristics, the mean square displacement (MSD) \cite{kob1995testing}
\begin{equation}
\begin{aligned}
&\mathrm{MSD}(t)\\
&= \frac{1}{N} \sum_{i=1}^N \left<|[\boldsymbol{r}_i(t) - \overline{\boldsymbol{r}}(t)] - [\boldsymbol{r}_i(0) - \overline{\boldsymbol{r}}(0)]|^2\right>,
\end{aligned}
\label{eq:msd}
\end{equation}
with $\overline{\boldsymbol{r}}(t) = (1/N) \sum_{i=1}^N \boldsymbol{r}_i(t)$ the position of the centre of mass at time $t$, to determine the phase of our models as a function of the force amplitudes.

\begin{figure}[!t]
\centering
\begin{tikzpicture}
\matrix (p)[row sep=0mm, column sep=0mm, inner sep=0mm,  matrix of nodes] at (0,0) {
\includegraphics[width=0.238\textwidth]{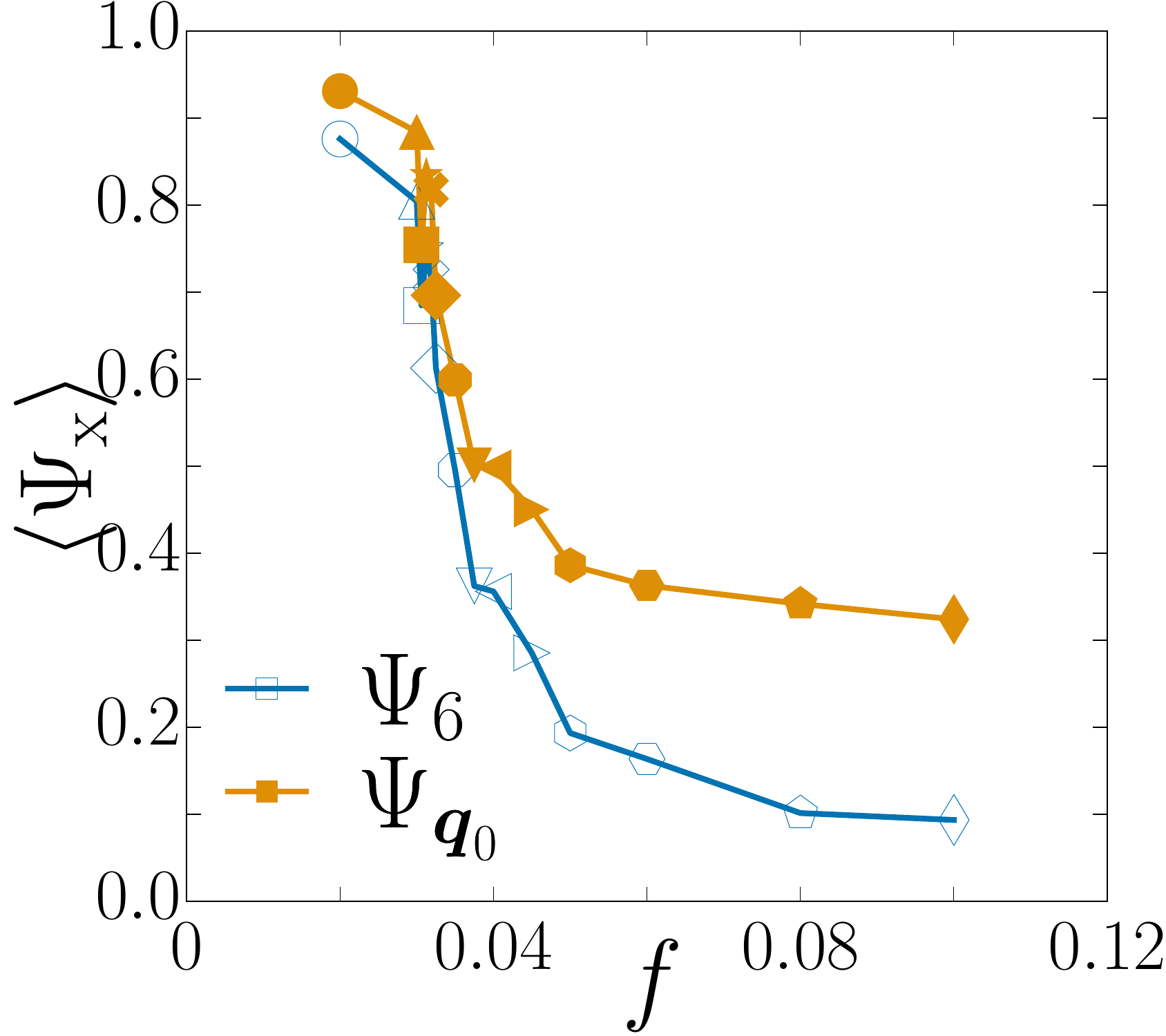}&
\includegraphics[width=0.238\textwidth]{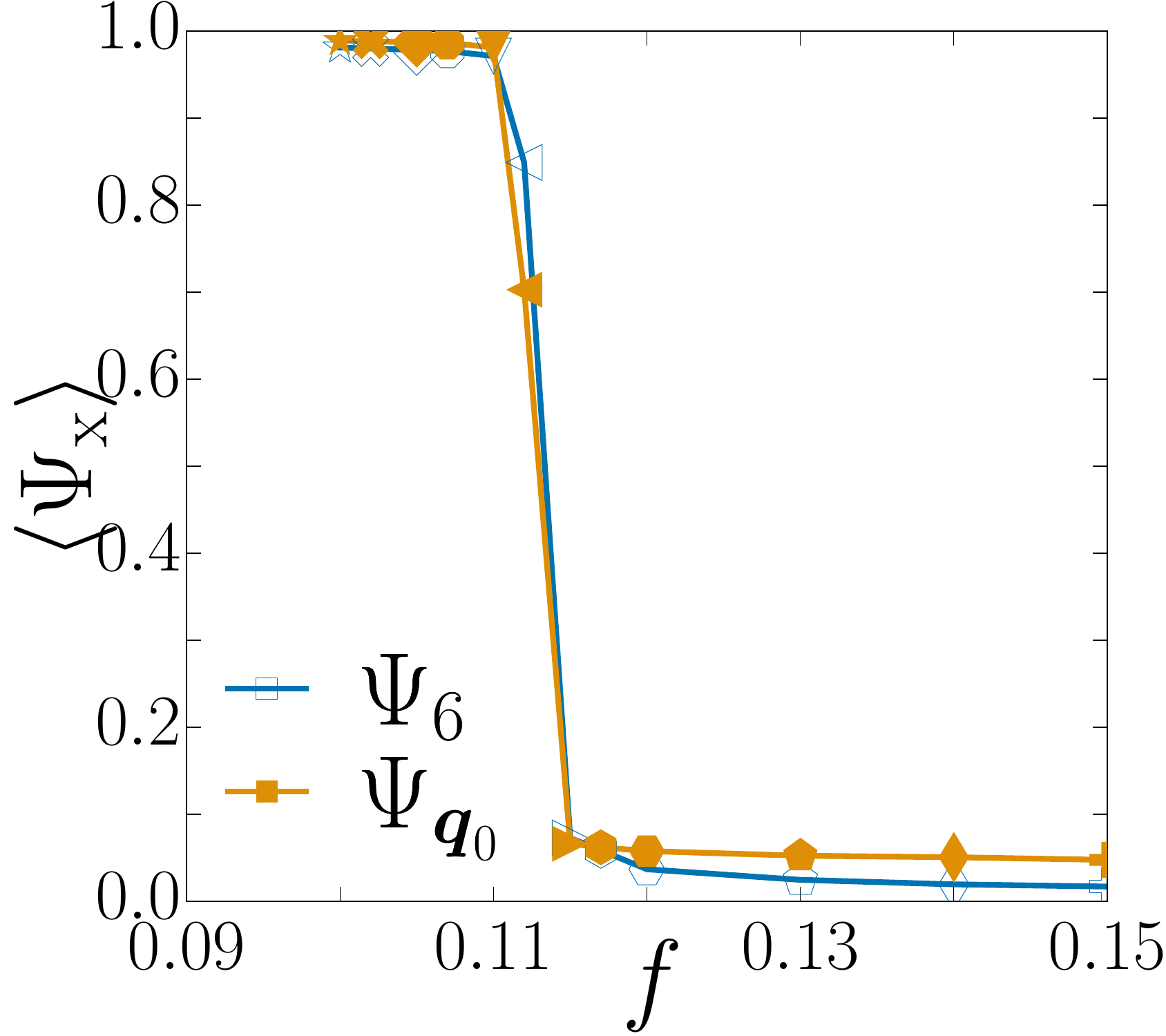}\\
\includegraphics[width=0.238\textwidth]{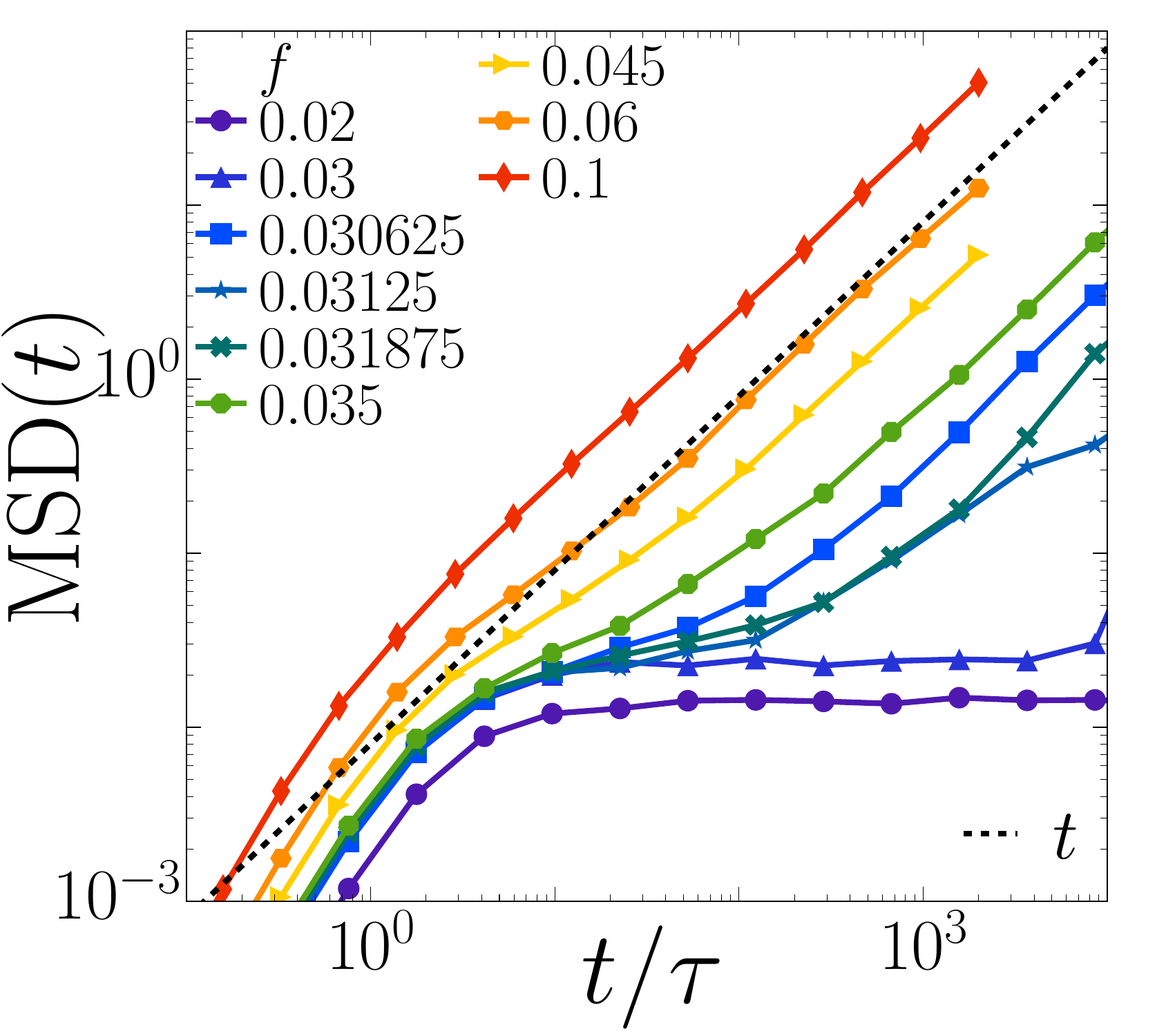}&
\includegraphics[width=0.238\textwidth]{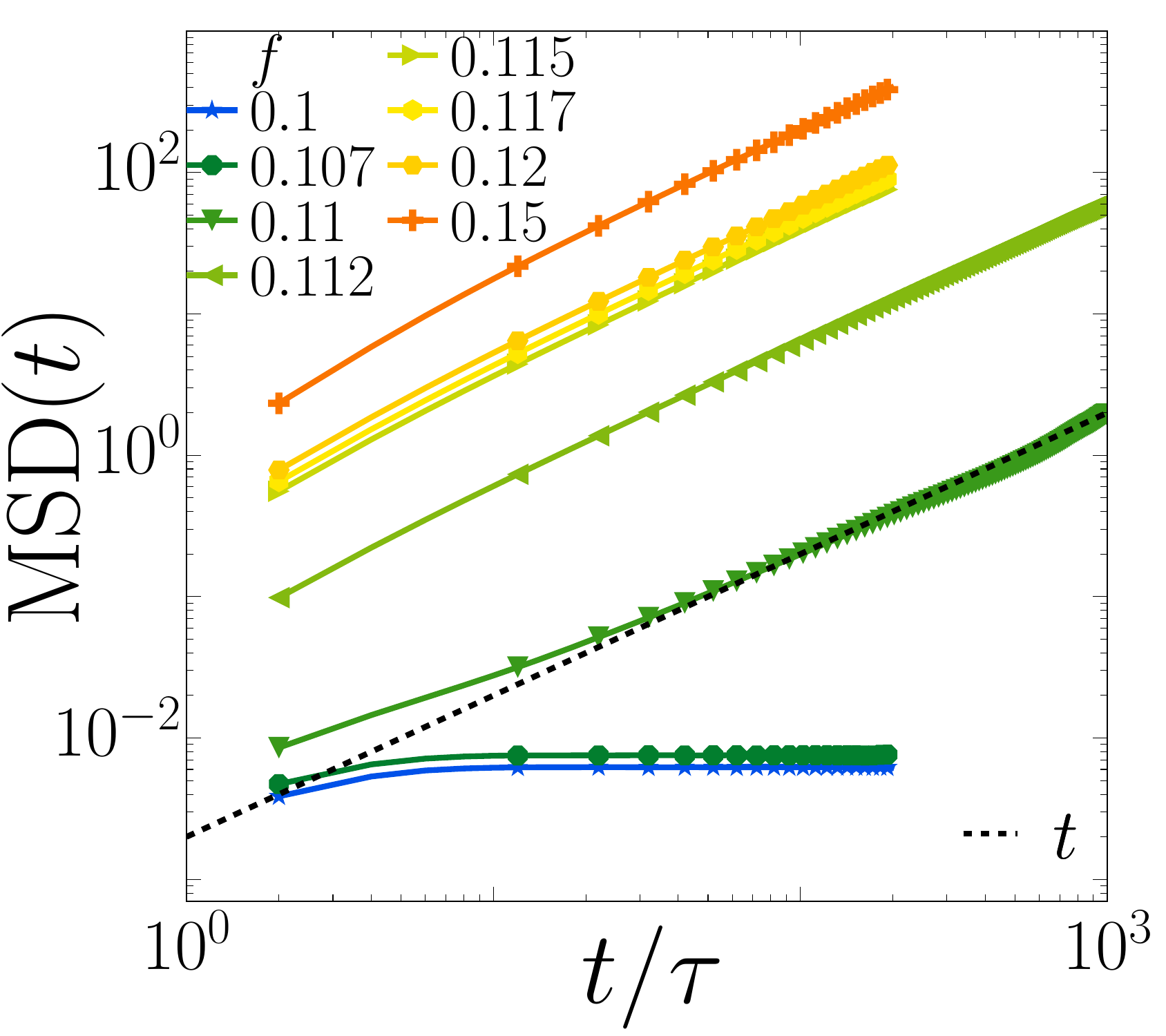}\\
};
\draw (p-1-1.north) node[above, xshift=5pt, yshift=-5pt]{\strut{\bf jtVM}};
\draw (p-1-2.north) node[above, xshift=5pt, yshift=-5pt]{\strut{\bf pABP}};
\draw (p-1-1.south west) node[fill=white, inner sep=0.5pt, xshift=30pt, yshift=47pt]{(a)};
\draw (p-1-2.south west) node[fill=white, inner sep=0.5pt, xshift=30pt, yshift=47pt]{(b)};
\draw (p-2-1.south west) node[fill=white, inner sep=0.5pt, xshift=30pt, yshift=41pt]{(c)};
\draw (p-2-2.south west) node[fill=white, inner sep=0.5pt, xshift=30pt, yshift=41pt]{(d)};
\end{tikzpicture}
\caption{(a, b) Global structural order parameters \eqref{eq:Psi} -- hexatic order parameter $\Psi_6$ (blue open symbols) and translational order parameter $\Psi_{\boldsymbol{q}_0}$ (orange full symbols) -- in steady state as functions of the stochastic force amplitude $f$.
(c, d) Mean squared displacements as functions of time $\mathrm{MSD}(t)$ \eqref{eq:msd} in steady state for different stochastic force amplitudes $f$.
Dashed lines are linear functions of time as guides to the eye.
(a, c) Junctional tension vertex model (\hyperref[sec:vm]{jtVM}) with $N = 108$ cells and persistence time $\tau = 5$.
(b, d) Pair active Brownian particles (\hyperref[sec:abp]{pABP}) with $N = 16384$ particles and persistence time $\tau = 25$.
Identical markers between (a) and (c) and between (b) and (d) correspond to identical data sets.}
\label{fig:Psi}
\end{figure}

\begin{figure}[!t]
\centering
\begin{tikzpicture}
\matrix (snap)[row sep=0mm, column sep=0mm, inner sep=0mm,  matrix of nodes] at (0,0) {
\includegraphics[width=0.238\textwidth]{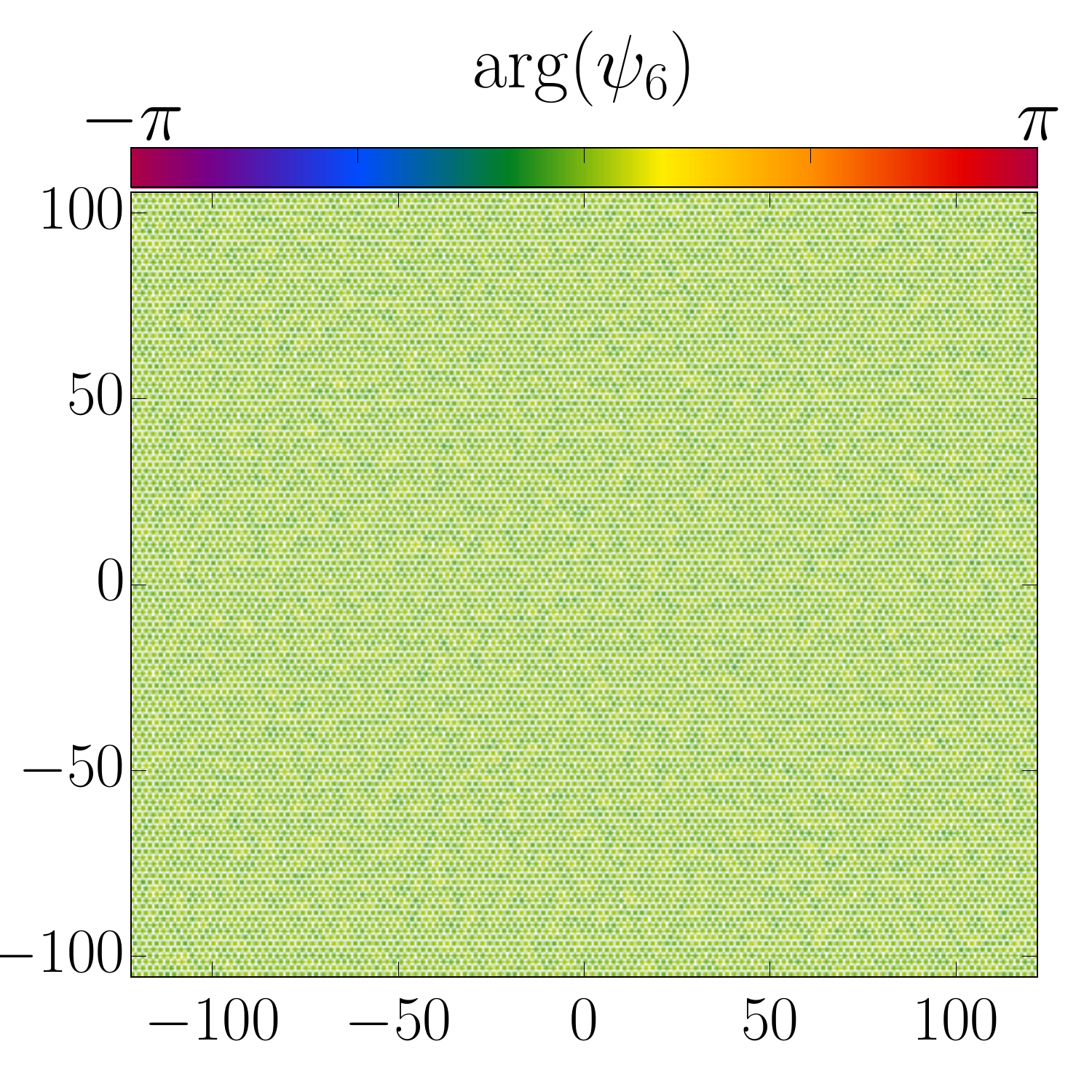}&
\includegraphics[width=0.238\textwidth]{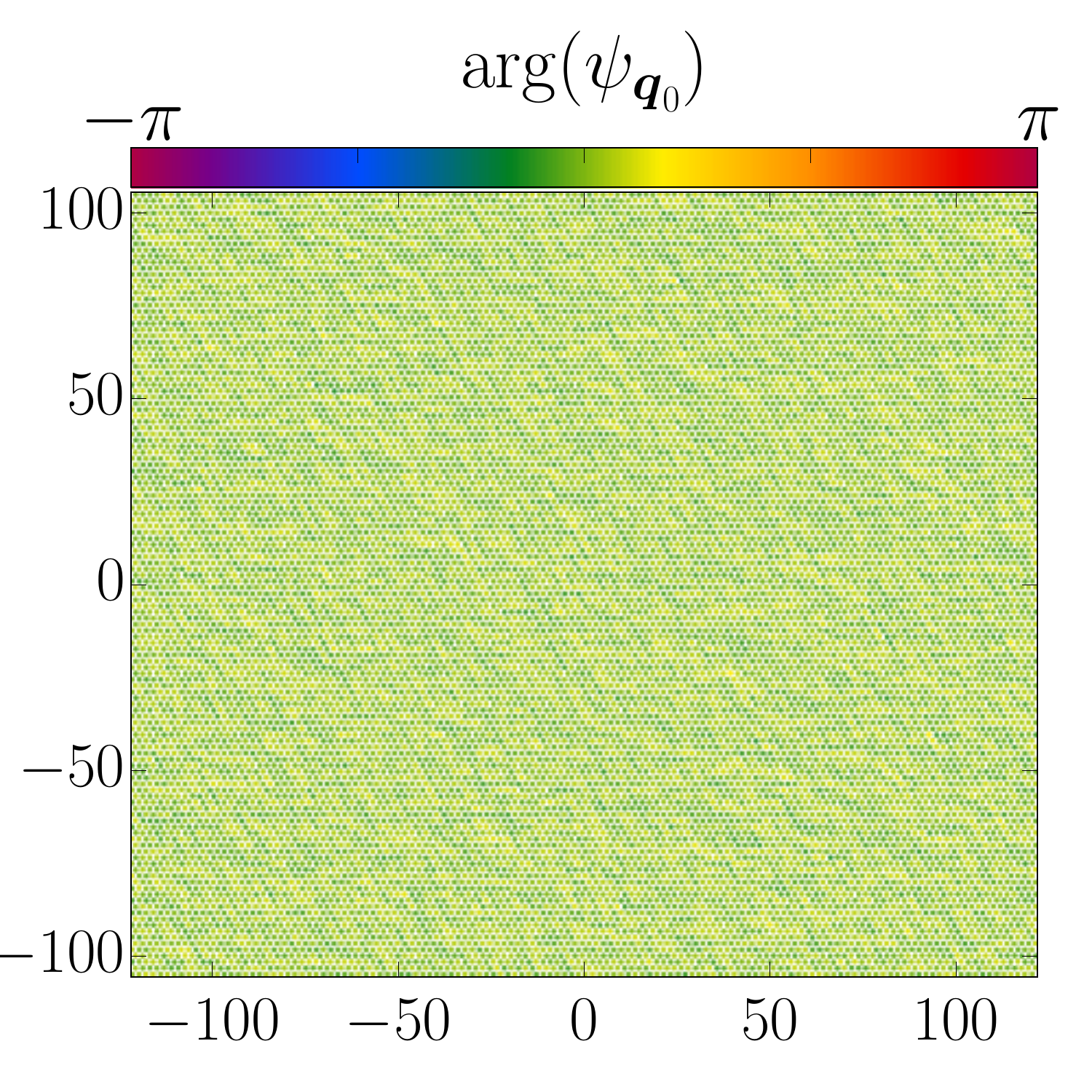}\\
\includegraphics[width=0.238\textwidth]{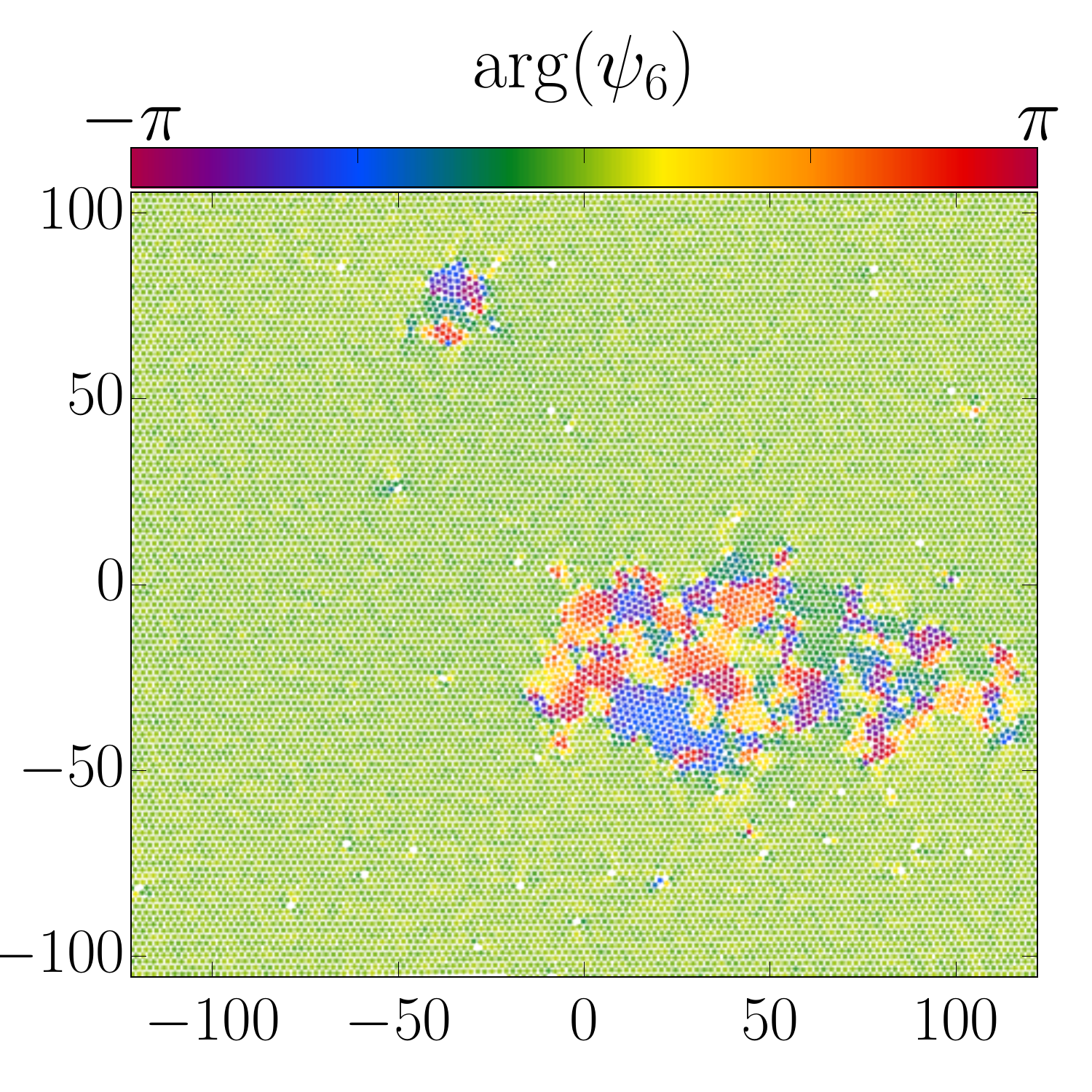}&
\includegraphics[width=0.238\textwidth]{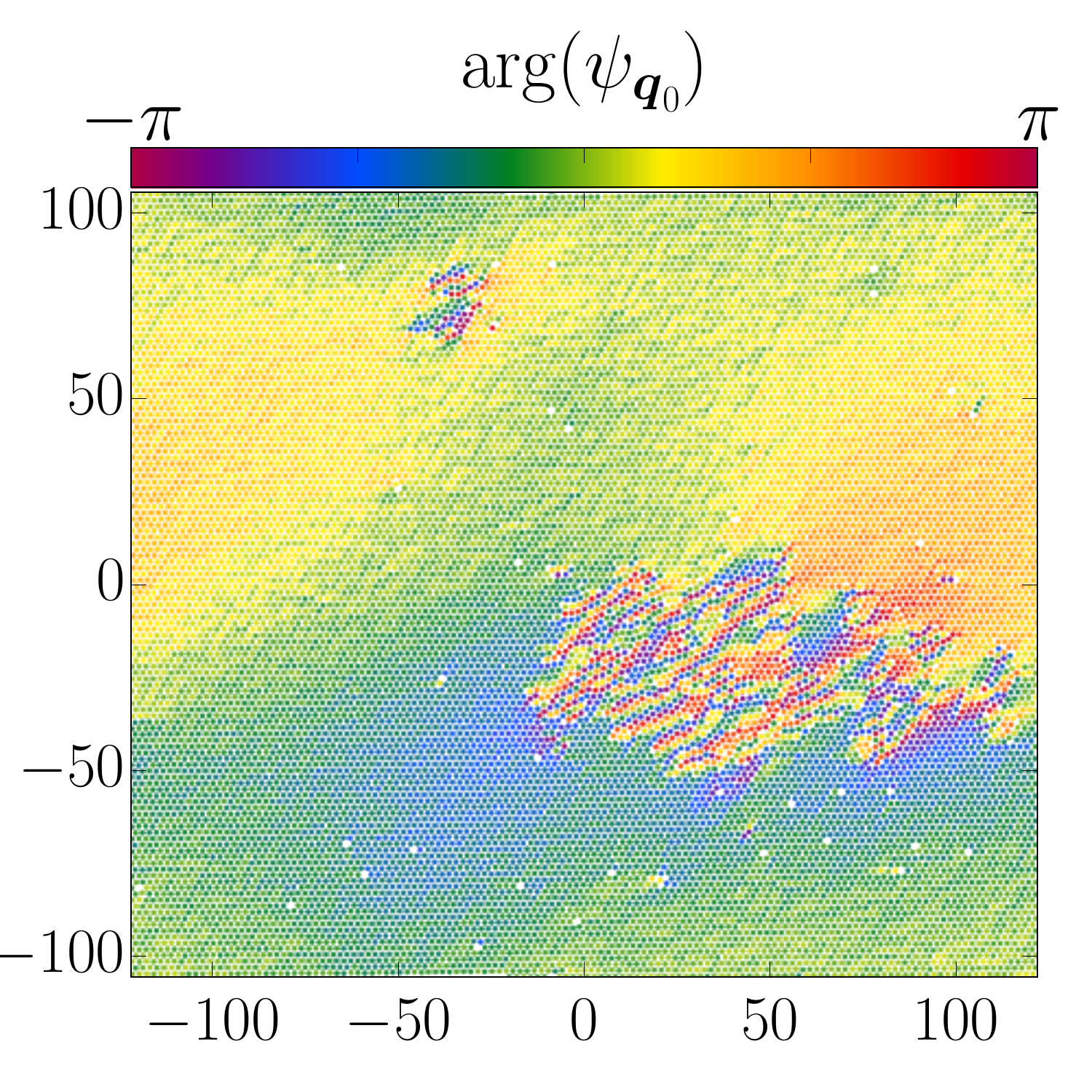}\\
\includegraphics[width=0.238\textwidth]{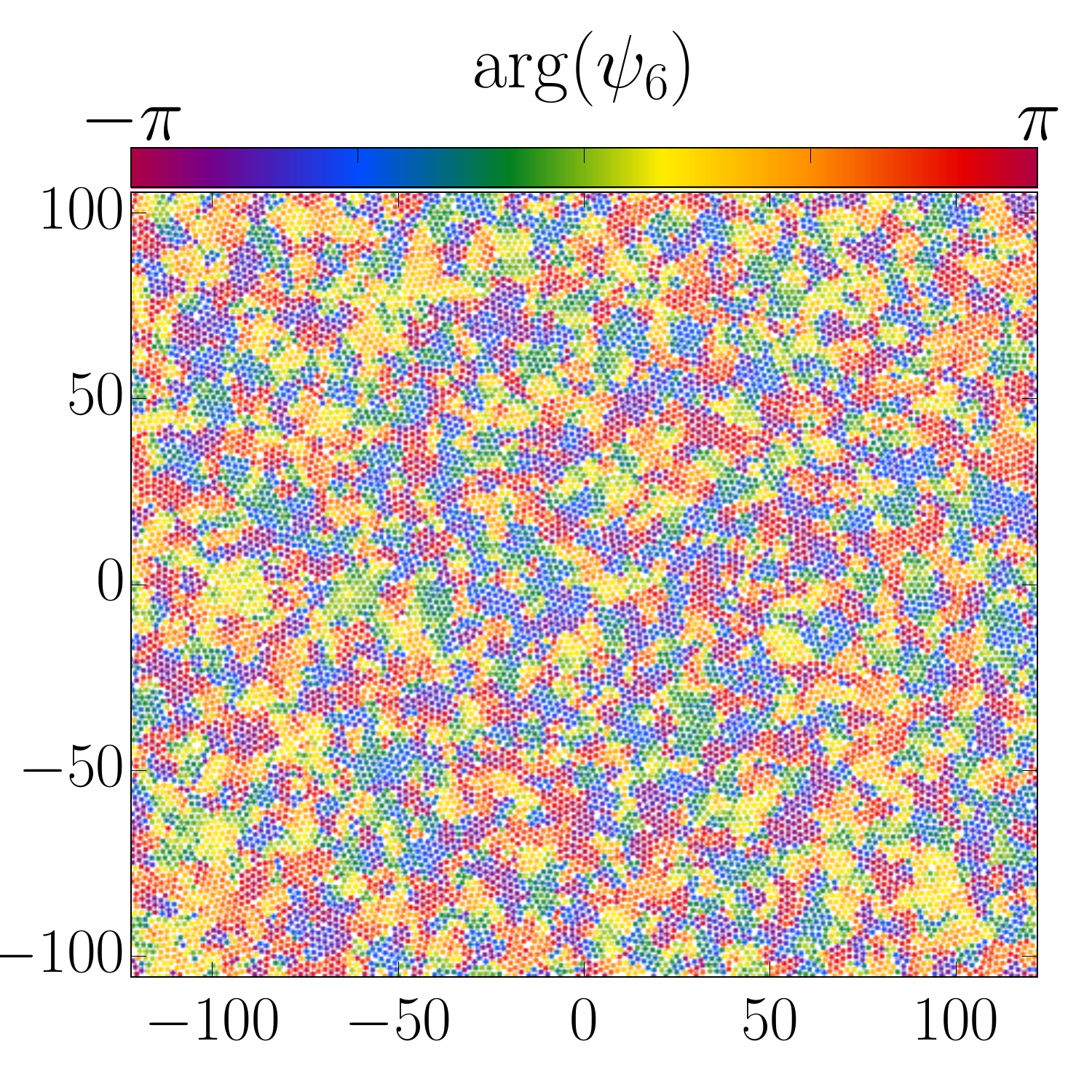}&
\includegraphics[width=0.238\textwidth]{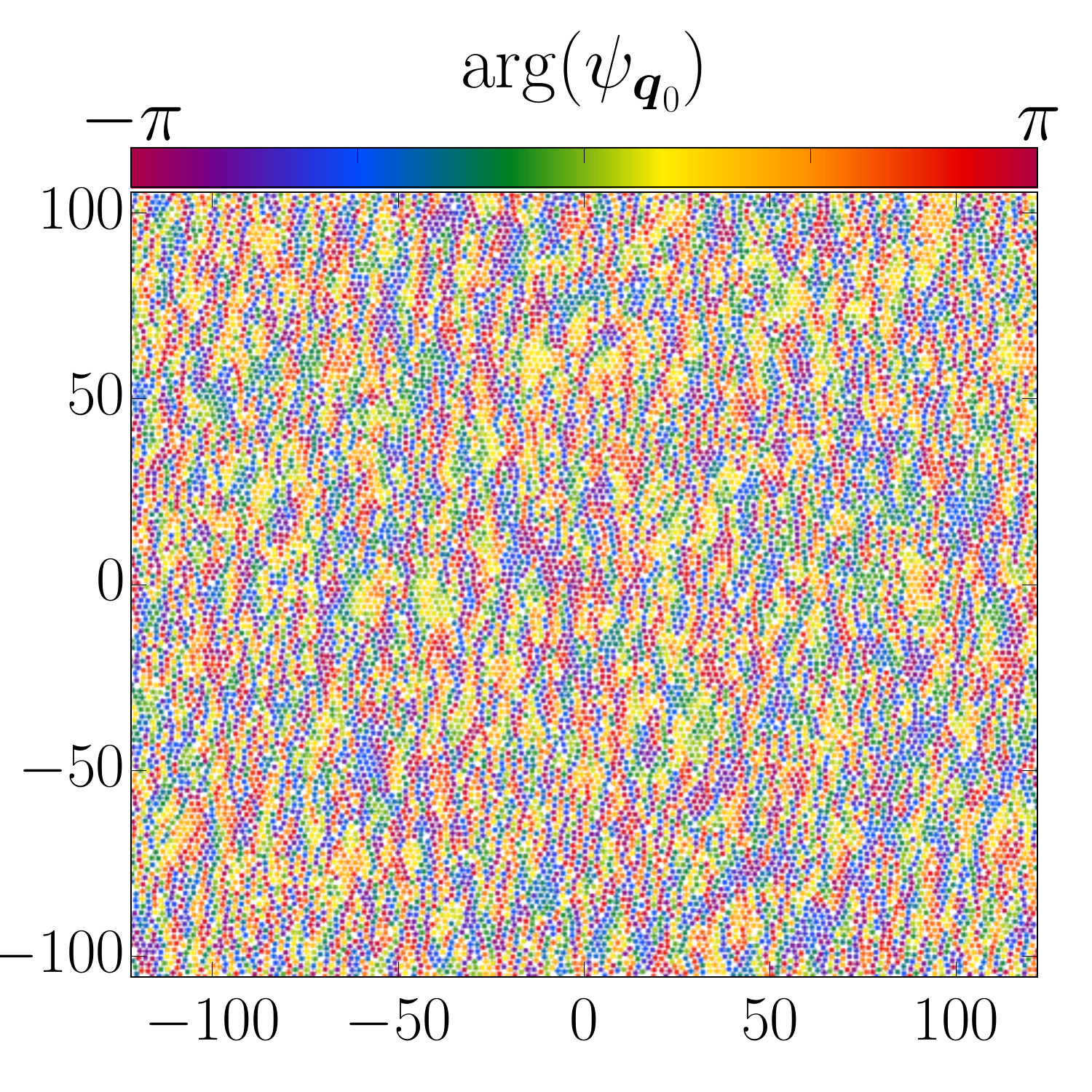}\\
};
\draw (snap-1-1.west) node[rotate=90, xshift=-5pt, yshift=0pt]{\strut{$f = 0.1$}};
\draw (snap-2-1.west) node[rotate=90, xshift=-5pt, yshift=0pt]{\strut{$f = 0.112$}};
\draw (snap-3-1.west) node[rotate=90, xshift=-5pt, yshift=0pt]{\strut{$f = 0.15$}};
\draw (snap-1-1.north east) node[fill=white, inner sep=0.5pt, xshift=-15pt, yshift=-29pt]{(a)};
\draw (snap-1-2.north east) node[fill=white, inner sep=0.5pt, xshift=-15pt, yshift=-29pt]{(b)};
\draw (snap-2-1.north east) node[fill=white, inner sep=0.5pt, xshift=-15pt, yshift=-29pt]{(c)};
\draw (snap-2-2.north east) node[fill=white, inner sep=0.5pt, xshift=-15pt, yshift=-29pt]{(d)};
\draw (snap-3-1.north east) node[fill=white, inner sep=0.5pt, xshift=-15pt, yshift=-29pt]{(e)};
\draw (snap-3-2.north east) node[fill=white, inner sep=0.5pt, xshift=-15pt, yshift=-29pt]{(f)};
\end{tikzpicture}
\begin{tikzpicture}
\matrix (psi6)[row sep=0mm, column sep=0mm, inner sep=0mm,  matrix of nodes] at (0,0) {
\includegraphics[width=0.9075\columnwidth]{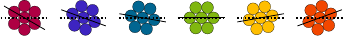}\\
};
\draw (psi6-1-1.west) node[xshift=-12.5pt]{(g)};
\end{tikzpicture}
\caption{Order and disorder for pair active Brownian particles (\hyperref[sec:abp]{pABP}). Visualisation of the argument of the local hexatic order parameter $\mathrm{arg}(\psi_{6,i})$ \eqref{eq:psi6} (left column, (a, c, e)), and the argument of the local translational order parameter $\mathrm{arg}(\psi_{\boldsymbol{q}_{0,i}})$ \eqref{eq:psiq0} (right column, (b, d, f)).
(g) shows different hexagonal packing orientations with the colour corresponding to the argument of the hexatic order parameter.
Colours associated to the argument of the translational order parameter illustrate deviations from the regular lattice corresponding to $\psi_{\boldsymbol{q}_0}$: an ordered lattice should display $\mathrm{arg}(\psi_{\boldsymbol{q}_{0,i}})=0$ everywhere.
We used $N=16384$ particles and persistence time $\tau = 25$.
Stochastic force amplitude is (a, b) $f=0.1$, (c, d) $f=0.112$, (e, f) $f=0.15$.}
\label{fig:snap}
\end{figure}

For both jtVM and pABP models there is, as the force amplitude is increased, a sharp simultaneous decrease in both translational and hexatic order (Fig.~\ref{fig:Psi}(a, b)).
We thus observe a structurally ordered phase, in the jtVM model for $f \lesssim 0.03$, and in the pABP model for $f < 0.11$. For the pABP model, dynamics is slow near the transition, and we have confirmed that configurations at $f=0.115$ and $f=0.117$ are disordered in steady-state by using long simulations of both ordered and disordered initial configurations.

In Fig.~\ref{fig:snap}, we plot configurations of the pABP model corresponding to the solid (a, b), the liquid (e, f), and a configuration at an intermediate value of $f$ (c, d). The intermediate force amplitudes $f = 0.112$ and to a small extent $f=0.11$ show phase coexistence at steady state after liquid bubbles slowly nucleate in the ordered solid (see ESI, Video 1-2).

This structural ordering partially coincides with a dynamical transition from a diffusing liquid state at large $f$, where correlations decay over times $t \gtrsim \tau$, to an arrested solid state at small $f$ as evidenced by the plateau in MSD (Fig.~\ref{fig:Psi}(c, d)).
For the pABP model, we observe a narrow range $f=0.11-0.112$ of very slow liquid dynamics while the system stays long range ordered while bubbles of disordered phase nucleate in the system. In the ESI (Sec.~\ref{app:abpdyn}), we also show how the self-intermediate scattering function $F_s(t)$ relaxes, and finally, we also provide the dynamical characteristics of the ABP model, for comparison.

The abrupt structural relaxation together with the equally abrupt transition in the MSD and the coexisting phase is indicative of a first-order melting transition in the pABP system.
We now turn to our observations for the structural fluctuations.

\subsection{Structural fluctuations}
\label{sec:strucfluc}

\begin{figure}[!t]
\centering
\begin{tikzpicture}
\matrix (u2)[row sep=0mm, column sep=0mm, inner sep=0mm,  matrix of nodes] at (0,0) {
\includegraphics[width=0.238\textwidth]{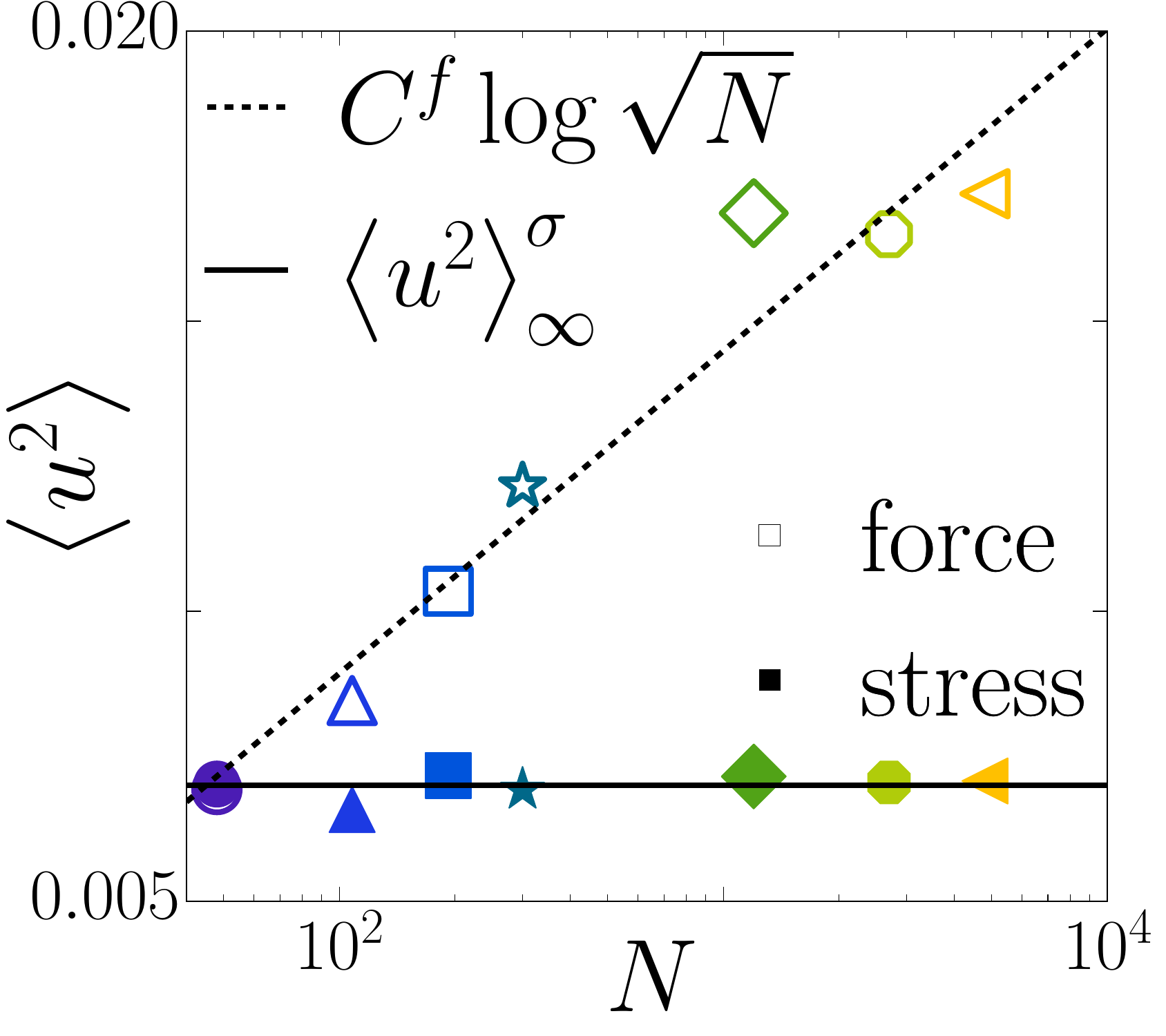}&
\includegraphics[width=0.238\textwidth]{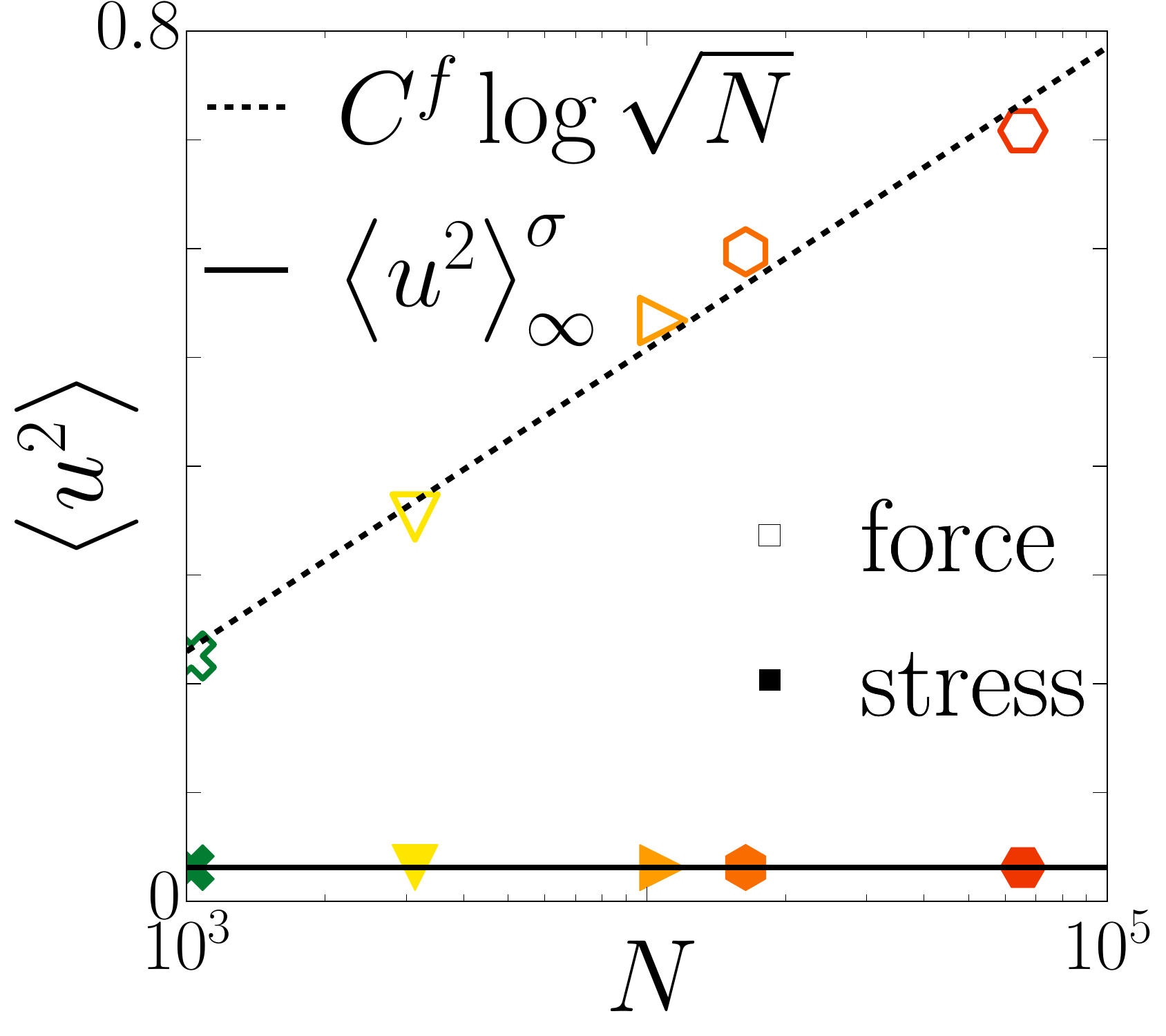}\\
\includegraphics[width=0.238\textwidth]{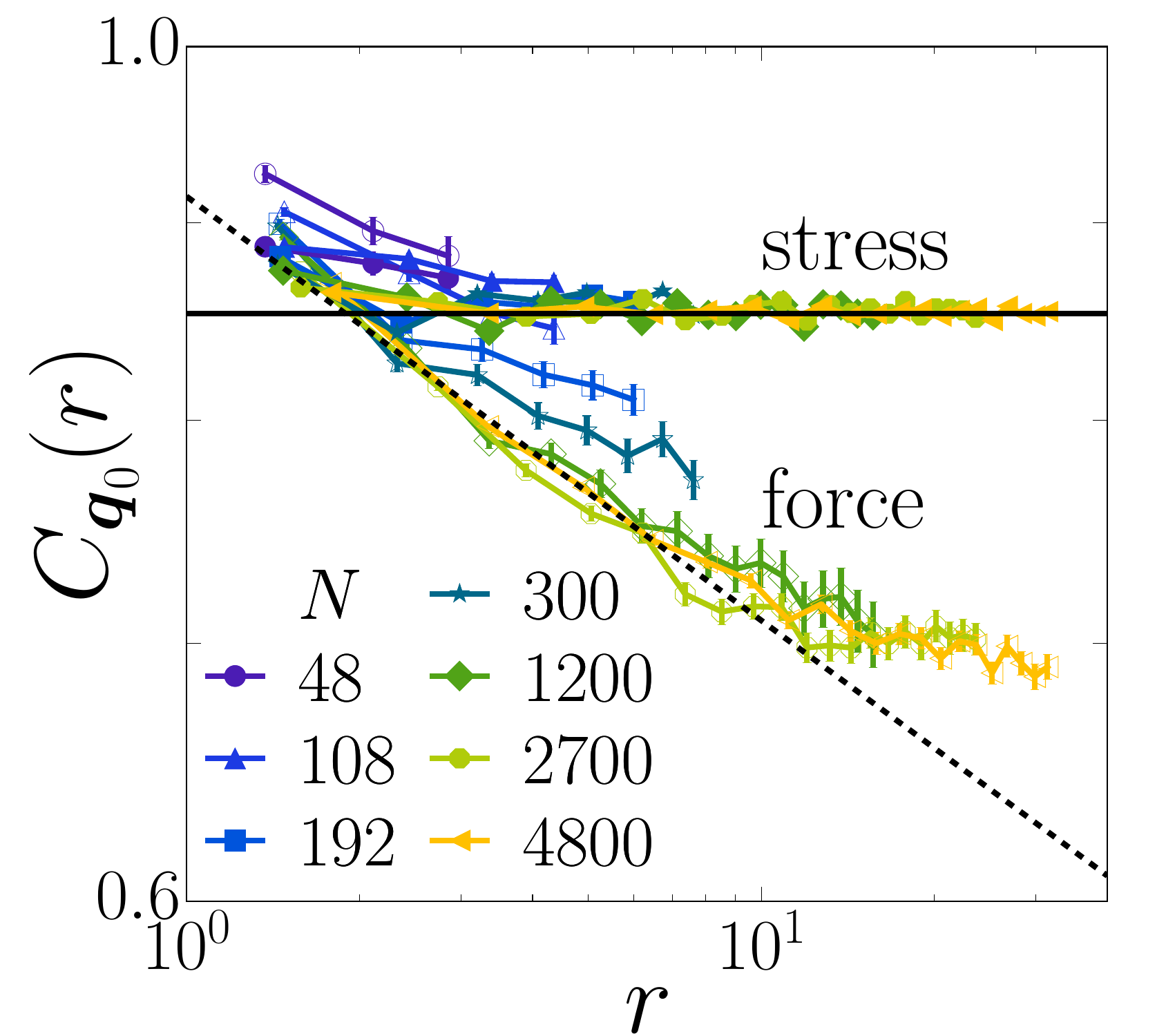}&
\includegraphics[width=0.238\textwidth]{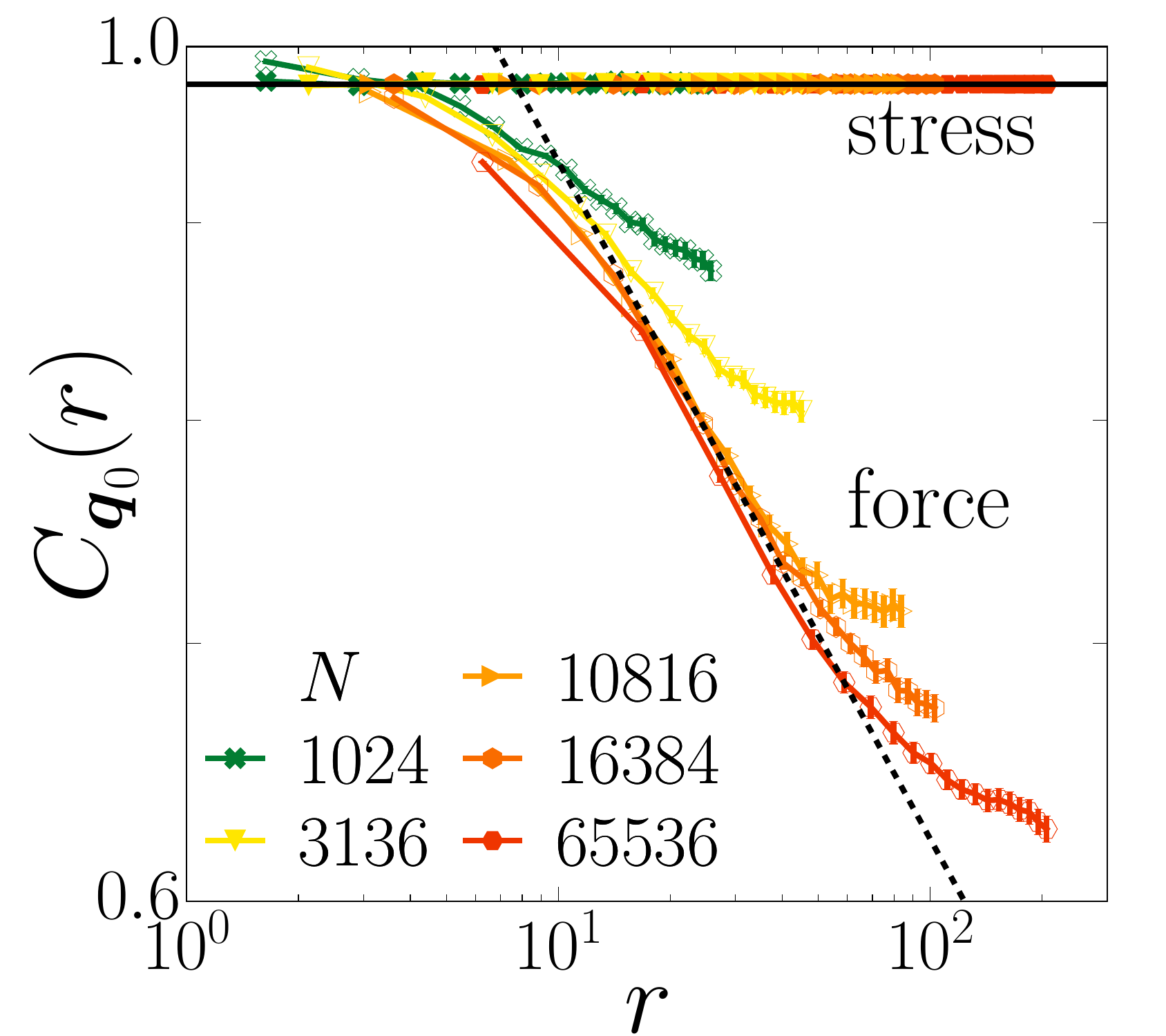}\\
};
\draw (u2-1-1.north) node[above, xshift=5pt, yshift=-5pt]{\strut{\bf (jt)VM}};
\draw (u2-1-2.north) node[above, xshift=5pt, yshift=-5pt]{\strut{\bf (p)ABP}};
\draw (u2-1-1.north east) node[fill=white, inner sep=0.5pt, xshift=-15pt, yshift=-45pt]{(a)};
\draw (u2-1-2.north east) node[fill=white, inner sep=0.5pt, xshift=-15pt, yshift=-45pt]{(b)};
\draw (u2-2-1.north east) node[fill=white, inner sep=0.5pt, xshift=-15pt, yshift=-45pt]{(c)};
\draw (u2-2-2.north east) node[fill=white, inner sep=0.5pt, xshift=-15pt, yshift=-45pt]{(d)};
\end{tikzpicture}
\caption{(a, b) Displacement fluctuations $\left<u^2\right>$ \eqref{eq:u2} computed in steady state and plotted for (a) the vertex model with particle-based self-propulsion and junctional tension (\hyperref[sec:vm]{jtVM}), and (b) particle and pair active Brownian particles (\hyperref[sec:abp]{pABP}) as a function of the number of particles $N$. Full symbols represent models with pair-wise stochastic forces, which we fit to a constant $\left<u^2\right>_{\infty}^\sigma$ (solid lines). In contrast, for models with particle-wise stochastic forces (plotted in open symbols), we fit to $C^f \log\sqrt{N}$ (dashed lines).
These plots are logarithmic on the x-axis and linear on the y-axis.
(c, d) Translational order correlations $C_{\boldsymbol{q}_0}$ \eqref{eq:cr}
for the same data as in (a, b) with pair-wise models (full symbols) and particle-wise models (open symbols), together with our predictions
for the large-distance scalings of the correlations \eqref{eq:cq0}: a constant (cst., solid lines) for pair-wise and an algebraic decay (alg., dashed lines) for particle-wise stochastic forces.
These plots are logarithmic on both axes.
We used persistence time $\tau = 5$ and force amplitude $f = 0.02$ for both vertex models (a, c)
and persistence time $\tau = 25$ and force amplitudes $f = 0.1$ and $f = 0.05$ for the pair-wise and particle-wise particle models respectively (b, d).
Identical colours and markers between plots correspond to identical data sets.}
\label{fig:u2}
\end{figure}

Deep in the solid phase, we expect our derivations for harmonic lattices to hold.
We first compute the displacement variance $\left<u^2\right>$ for both models in the solid phase, with particle-wise and pair-wise stochastic forces, as a function of system size (Fig.~\ref{fig:u2}(a, b)).
These confirm our predictions \eqref{eq:u2}. For systems with pair-wise stochastic forces, $\left<u^2\right>$ converges to a finite value at large $N$. In contrast, for systems with particle-wise stochastic forces, $\left<u^2\right>$ diverges as $\log L \sim \log \sqrt{N}$. We fit $\left<u^2\right>(N)$ to a constant $\left<u^2\right>_{\infty}^\sigma$ for models with pair-wise stochastic forces, and to $C^f \log\sqrt{N}$ for models with particles-wise stochastic forces.
We extract $C^f \approx 4.83 \times 10^{-3}$ and $\left<u^2\right>_{\infty}^\sigma \approx 7.00 \times 10^{-3}$ in the vertex model with self-propelled vertices and junction tensions respectively (Fig.~\ref{fig:u2}(a)), and $C^f \approx 2.41 \times 10^{-2}$ and $\left<u^2\right>_{\infty}^\sigma \approx 3.10 \times 10^{-3}$ for active and pair active Brownian particles respectively (Fig.~\ref{fig:u2}(b)).
It is noteworthy that $\left<u^2\right>$ can be derived from the displacement spectra \eqref{eq:uk2} (see ESI, Sec.~\ref{app:ctr}).

\begin{figure}[!t]
\centering
\begin{tikzpicture}
\matrix (abphexatic)[row sep=0mm, column sep=0mm, inner sep=0mm,  matrix of nodes] at (0,0) {
\includegraphics[width=0.245\textwidth]{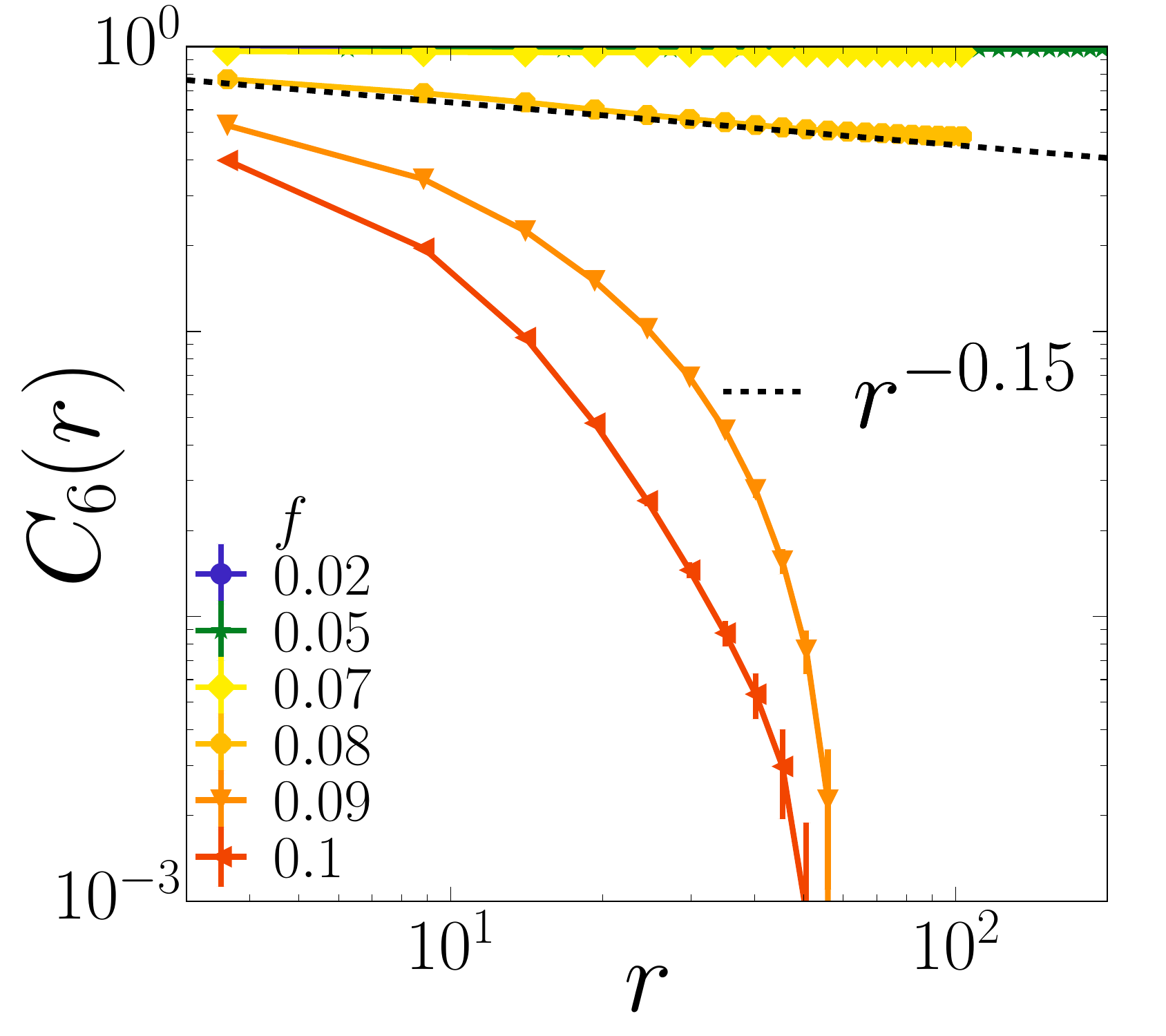}&
\includegraphics[width=0.245\textwidth]{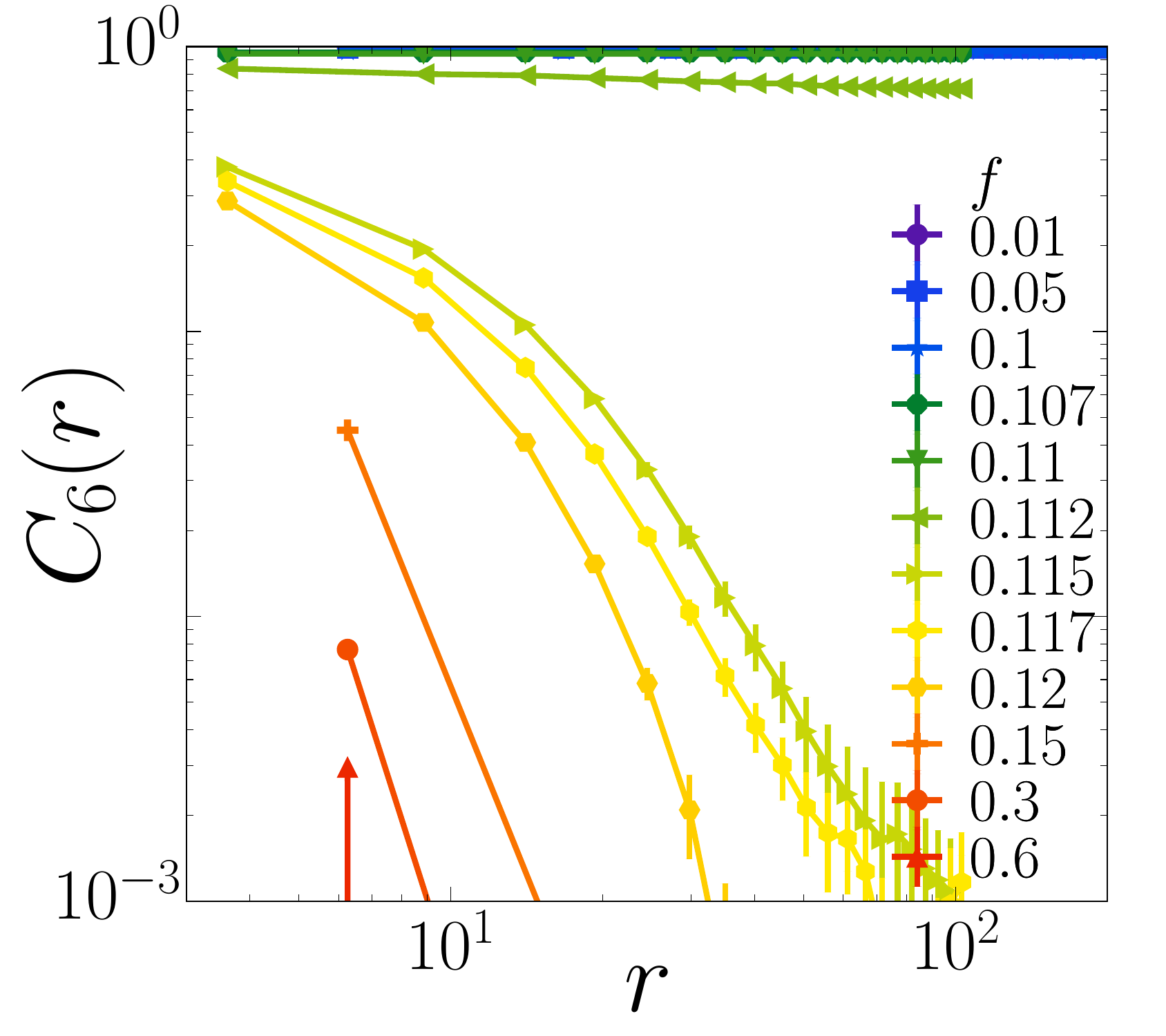}\\
\includegraphics[width=0.245\textwidth]{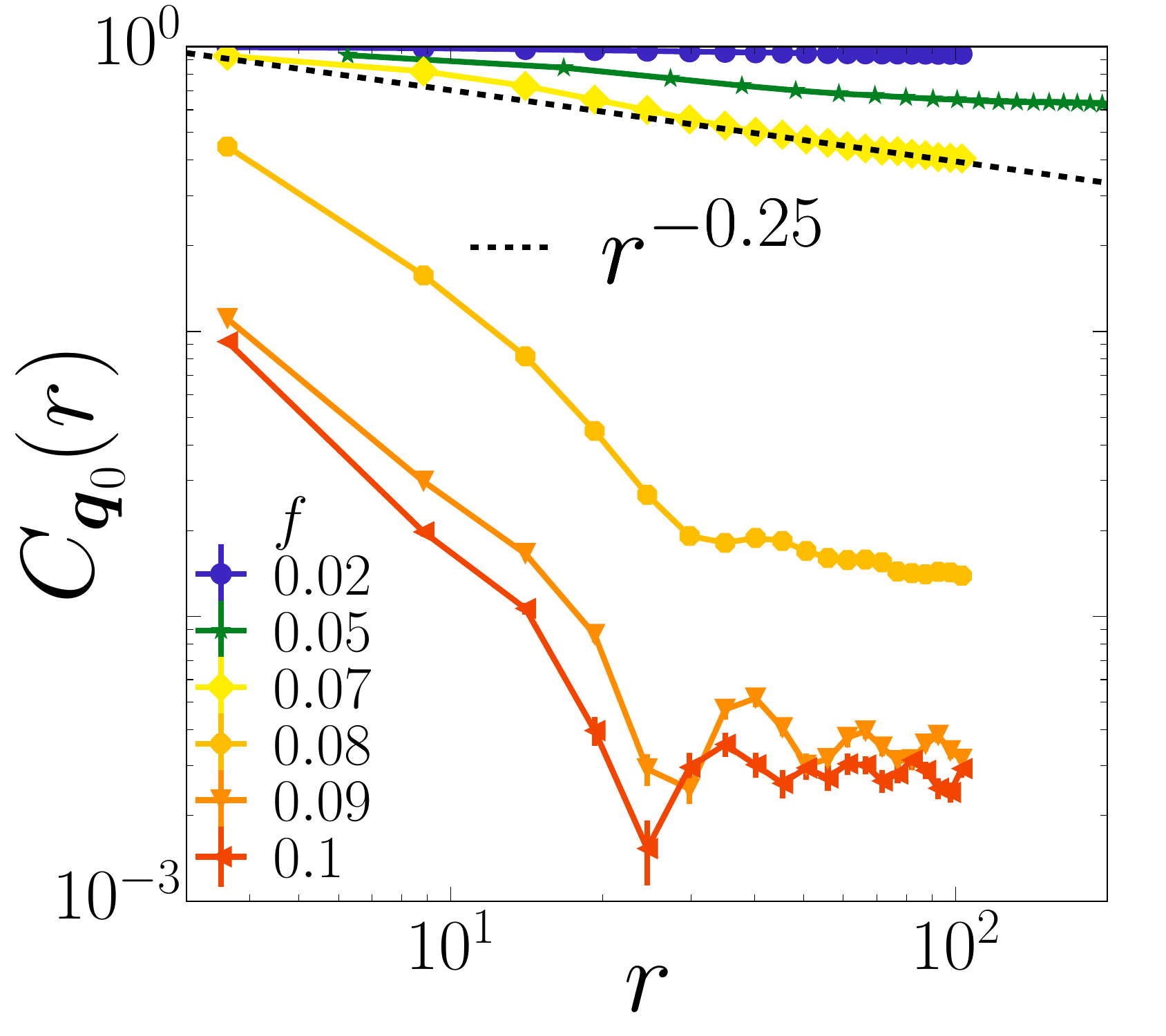}&
\includegraphics[width=0.245\textwidth]{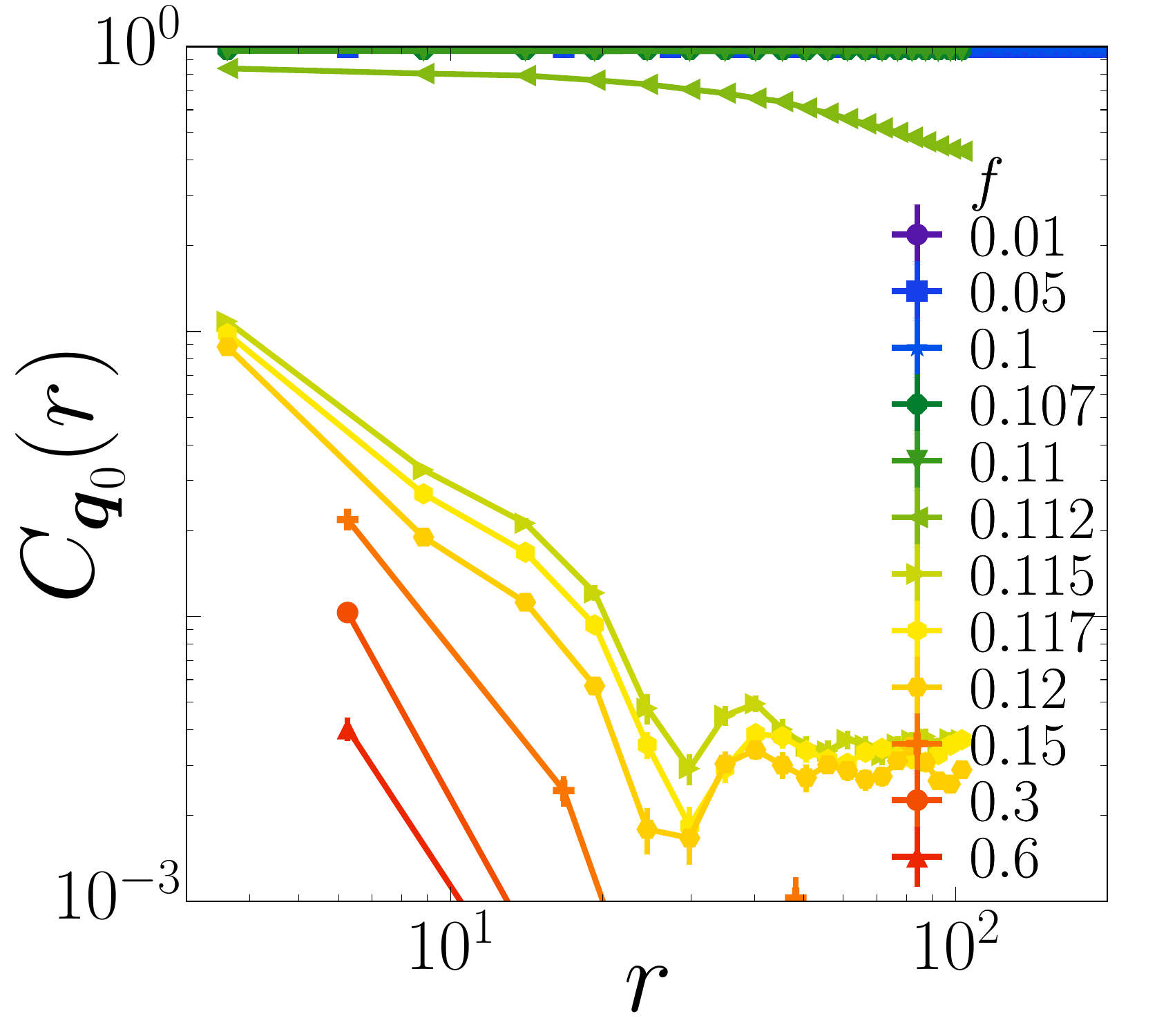}\\
};
\draw (abphexatic-1-1.north) node[above, xshift=5pt, yshift=-5pt]{\strut{\bf ABP}};
\draw (abphexatic-1-2.north) node[above, xshift=5pt, yshift=-5pt]{\strut{\bf pABP}};
\draw (abphexatic-1-1.south west) node[fill=white, inner sep=0.5pt, xshift=29pt, yshift=73pt]{(a)};
\draw (abphexatic-1-2.south west) node[fill=white, inner sep=0.5pt, xshift=29pt, yshift=73pt]{(b)};
\draw (abphexatic-2-1.south west) node[fill=white, inner sep=0.5pt, xshift=29pt, yshift=87pt]{(c)};
\draw (abphexatic-2-2.south west) node[fill=white, inner sep=0.5pt, xshift=29pt, yshift=87pt]{(d)};
\end{tikzpicture}
\caption{(a, b) Orientational order correlation functions $C_6(r)$,and (c, d) translational order correlations $C_{\boldsymbol{q}_0}(r)$ as functions of space for different stochastic force amplitudes $f$ for particle-wise (ABP, left) and pair active forcing (pABP, right).
We used $N = 16384$ or $65536$ where available, and persistence time $\tau = 25$.
Dashed lines are algebraic functions of distance as guides to the eye.
}
\label{fig:psicor}
\end{figure}

Next, we compute the translational order correlation function $C_{\boldsymbol{q}_0}$ in the solid phase (Fig.~\ref{fig:u2}(c, d)). We confirm that the solid phase of the models with pair-wise stochastic forces exhibits long-range translational order, with a large-distance correlation value deriving from the (finite) displacement variance.
We observe, in the case of particle-wise stochastic forces, that correlations decay algebraically, consistently with quasi-long-range translational order \cite{shi2023extreme}.
Our linear response predictions \eqref{eq:cq0} fit this data if we use the values of $C^f$ and $\left<u^2\right>_{\infty}^\sigma$ determined from the system-size dependence of displacement fluctuations (Fig.~\ref{fig:u2}(a, b)), where we also use the measured reciprocal lattice vector squared norms $|\boldsymbol{q}_0|^2 \approx 45.6$ for the vertex model and $|\boldsymbol{q}_0|^2 \approx 14.6$ for the particle model. Therefore, we confirm that the exponent of this decay is directly related to the factor with which the displacement variance diverges \eqref{eq:cq01}.
We report the following exponents for the algebraic decay of the correlations $C_{\boldsymbol{q}_0} \sim r^{-\eta}$ in the case of systems with stochastic forces: $\eta^{\mathrm{ABP}} = \frac{1}{2} |\boldsymbol{q}_0|^2 C_f \approx 0.176$ for disk ABPs and $\eta^{\mathrm{VM}} \approx 0.110$ for the self-propelled vertex model.
Both exponents are below their expected upper limit of $\frac{1}{3}$ within the KTHNY theory, consistently with previous studies on self-propelled particles \cite{klamser2018thermodynamic} and unlike active particles with small amounts of polar alignment \cite{shi2023extreme}.

In Fig.~\ref{fig:psicor} we compute the order correlation functions through the melting transition, using the largest system size $N$ we have available. As shown in Fig.~\ref{fig:psicor}(a, b), the orientational order correlation function $C_6(r)$ in both ABP and pABP models shows long-range correlations in the solid phase and short-ranged correlations in the liquid phase. We observe indications of a power-law decay near the melting transition, and notably we find a significant drop between $f=0.11$ and $f=0.112$ for the pABP model, both liquids in the coexistence region. As shown in Fig.~\ref{fig:psicor}(c, d), the translational order correlation $C_{\boldsymbol{q}_0}$ has the same transition points, though in the solid phase the pABP has long range order while the ABP is power law (note the extended y range compared to Fig. \ref{fig:u2}(d)).
We stress that regions with short-range structural order are rich in defects, as evidenced by coordination numbers $z_i \neq 6$, while regions with long-range structural order are exempt of these same defects (see ESI, Sec.~\ref{app:abpdefects}), in accordance with the established understanding of the melting of two-dimensional solids \cite{digregorio2022unified}.
For the pABP model, we confirm thus that the diffusing liquid state corresponds to a molten state with short-range structural correlations, while the arrested solid state displays long-range structural order.
At our current numerical resolution, in the pABP model, we do not see evidence of a \emph{hexatic} phase \cite{klamser2018thermodynamic,pasupalak2020hexatic,durand2019thermally} with finite hexatic order and vanishing translational order, with the potential candidates $f=0.11-0.112$ showing solid-liquid coexistence instead.

\begin{figure}[!t]
\centering
\begin{tikzpicture}
\matrix (S)[row sep=0mm, column sep=0mm, inner sep=0mm,  matrix of nodes] at (0,0) {
\includegraphics[width=0.238\textwidth]{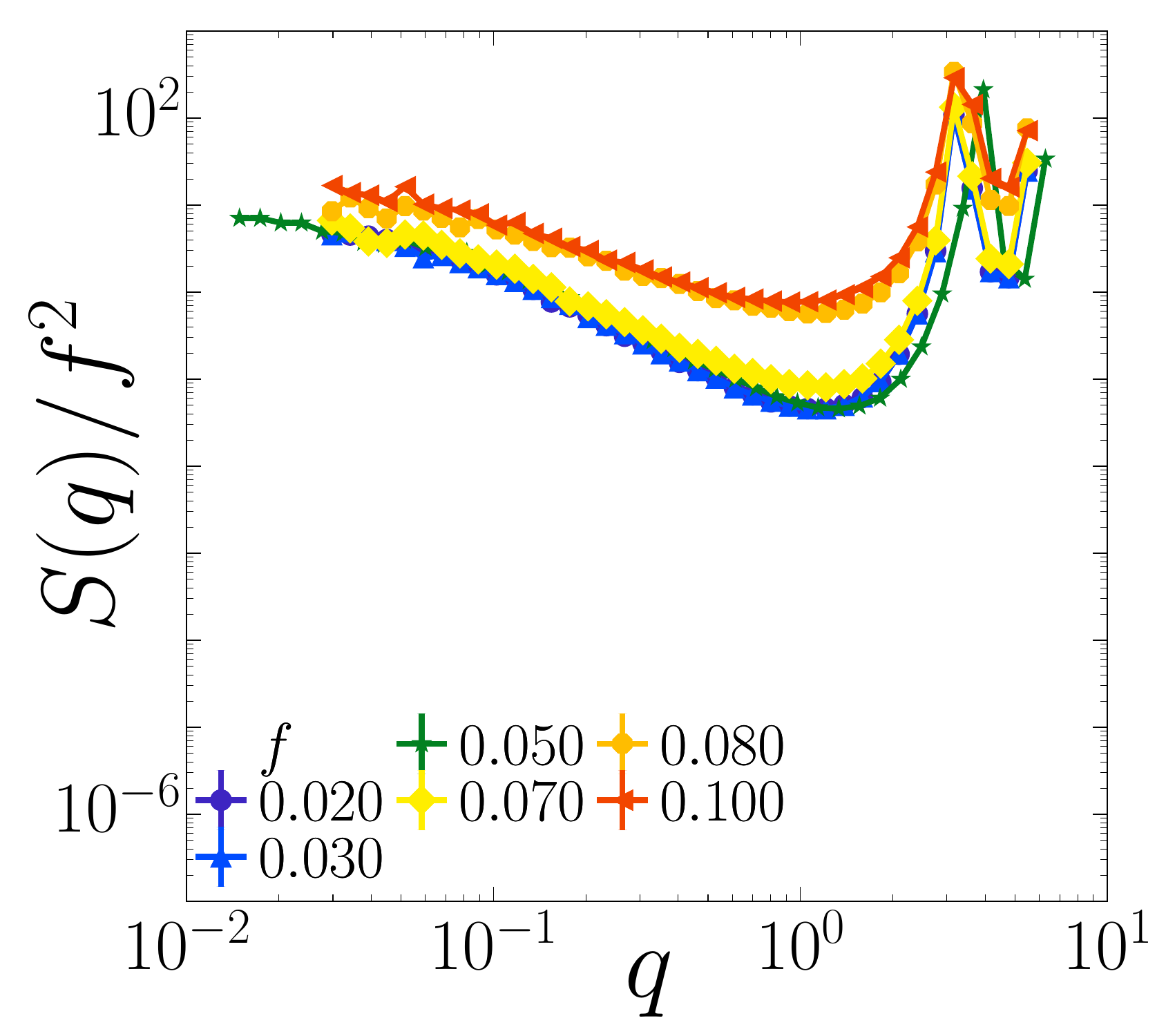}&
\includegraphics[width=0.238\textwidth]{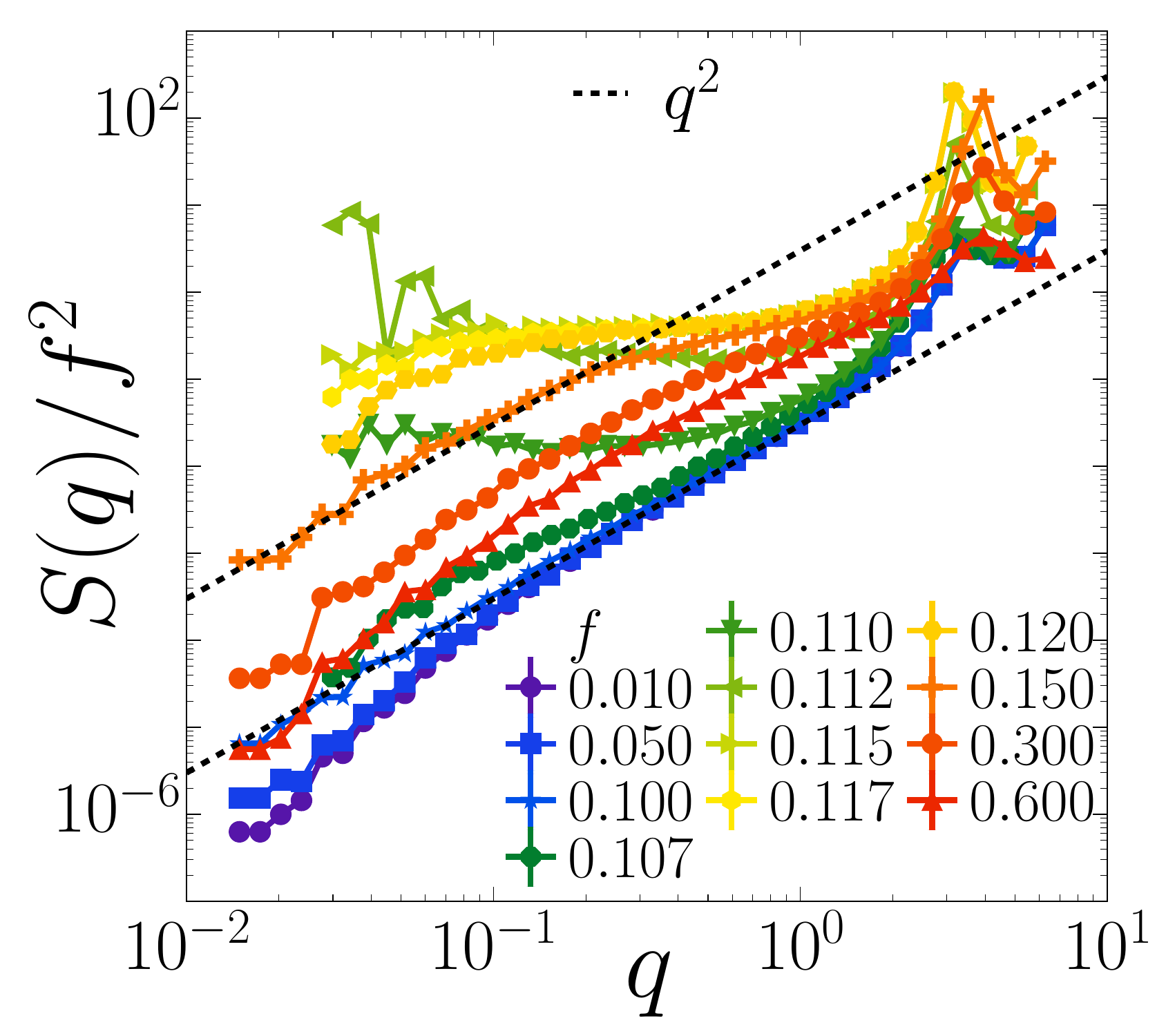}\\
};
\draw (S-1-1.north) node[above, xshift=5pt, yshift=-5pt]{\strut{\bf ABP}};
\draw (S-1-2.north) node[above, xshift=5pt, yshift=-5pt]{\strut{\bf pABP}};
\draw (S-1-1.north east) node[fill=white, inner sep=0.5pt, xshift=-15pt, yshift=-50pt]{(a)};
\draw (S-1-2.north east) node[fill=white, inner sep=0.5pt, xshift=-15pt, yshift=-50pt]{(b)};
\end{tikzpicture}
\caption{Scaled structure factor $S(q)/f^2$ \eqref{eq:S} defined as cylindrical average of $S(\boldsymbol{q})$ over wave-vectors $\boldsymbol{q} = (2\pi m/L_x, 2\pi n/L_y)$ which satisfy $|\boldsymbol{q}| \in [q -\delta q/2, q + \delta q/2]$ with $\delta q=10^{-2}$ and divided by the square of the stochastic force amplitude $f^2$.
We plot the structure factor for different stochastic force amplitudes $f$ in (a) for the particle model with particle-wise stochastic force (ABP), and in (b) for the particle model with pair-wise stochastic forces (\hyperref[sec:abp]{pABP}).
We used $N=16384$ or $65536$ particles, and persistence $\tau = 25$.}
\label{fig:S}
\end{figure}

Finally, we compute the radially-averaged structure factor $S(q)$ to test our predictions of hyperuniformity \eqref{eq:S}.
We present results exclusively for the (p)ABP models because the large systems sizes needed to observe the large-wavelength decay of structural fluctuations are not yet attainable with our current vertex model algorithm.
In Fig.~\ref{fig:S}, we plot the structure factor for particle-wise (a) and pair-wise (b) stochastic forces.
In the former case, we observe that $S(q)$ converges to a finite constant as $q \to 0$, which is the expected behaviour for self-propelled particles systems at moderate activities away from spontaneous phase separation \cite{demacedobiniossek2018static,dulaney2021isothermal}.
In the latter case, we do observe the expected scaling $S(q) \sim q^2$ as $q \to 0$ away from the order-to-disorder transition, which confirms that these systems are indeed hyperuniform.
Most interestingly, we observe that this scaling applies not only in the solid phase at $f\leq 0.1$, where our derivation is expected to hold, but also in the liquid phase at $f\geq 0.15$. Therefore, despite this phase showing orientational and translational disorder on small length scales only (Fig.~\ref{fig:snap}(e, f)), this measurement confirms that there is a ``hidden order'' on large length scales \cite{torquato2018hyperuniform}, which has also recently been observed in phase separating droplets \cite{wilken2023spatial}.
We also see that the scaling $S(q) \sim f^2$ of the continuum prediction works well in the deeply solid region.
In both regions, we observe an apparent $\sim q^\alpha$ scaling with $\alpha>2$ for some values of $f$. However, as we showed, long wavelength relaxations are heavily suppressed for fluctuating active stresses, and relaxation to the true steady state of $S(q)$ may occur only on an inordinately long time scale.
Meanwhile, near the transition region where order and disorder coexist, on both sides, the system displays more complex, but not hyperuniform scaling of $S(q)$.

\section{Discussion and conclusion}
\label{sec:conclusion}

In this paper we first demonstrated how,
within linear elasticity theory, two-dimensional active solids featuring fluctuating active stresses
display finite displacement fluctuations $\left<u^2\right>$ in the thermodynamic limit $N \to \infty$ \eqref{eq:u2}, long-range translational order \eqref{eq:cq0}, and vanishing large-wavelength density fluctuations, \textit{i.e.} hyperuniformity \eqref{eq:Sk}.

These properties derive from, on the one hand, the absence of particle-wise white noise counterbalancing the instantaneous particle-wise friction, thus violating the second fluctuation-dissipation theorem \cite{kubo1966fluctuationdissipation,doi1988theory,tothova2022overdamped}, and on the other hand, the fluctuations of the driving force vanishing for large wavelengths \eqref{eq:stresscorft}.

Microscopically, these fluctuating active stresses take the form of pair-wise stochastic forces, forces which respect action-reaction rather than the more usual particle-wise active driving.
We introduced two numerical models which incorporate these pair-wise stochastic forces: a vertex model with fluctuating junction tension (jtVM) and a particle-based model with pair-wise additive stochastic forces respecting action-reaction (pABP).
We showed that all of our predictions held in both models.
These models are built from widespread active matter models used to describe the physics of dense cell tissues, therefore we expect these results to shed light on both out-of-equilibrium ordering transitions as well as on structural fluctuations of living systems.

Since an ordered symmetry-broken phase is possible within our model, it is questionable if the two-step melting scenario of two-dimensional solid from the KTHNY theory applies here.
This has been the subject of recent interest in the active matter community with several experimental systems being designed to address these questions \cite{briand2016crystallization,massana-cid2024multiple,das2024flockinga}.
To the best of our current numerical efforts, we have not been able to identify a hexatic phase which would feature short-range translational order and quasi-long-range orientational order.
Moreover we observe phase-separated configurations close to the transition (Fig.~\ref{fig:snap}(c, d)) which indicate a first-order transition between a disordered liquid and an ordered solid as in three-dimensional equilibrium systems.

Our numerical results also show that disordered liquid states, outside the scope of linear elasticity theory, display hyperuniformity as well \cite{lei2019nonequilibrium,lei2019hydrodynamics,li2024fluidization}.
Previously it was shown that force-balanced configurations (ground states of the potential energy) of Voronoi models of biological tissues have suppressed density fluctuations and are thus hyperuniform \cite{li2018biological,zheng2020hyperuniformity,tang2024tunable}.
In contrast, our pathway is more general as it is based on the symmetry of the active fluctuations and does not depend on the specific interaction potential.
This outlines a larger class of active systems which encompasses non-motile active matter models, \textit{e.g.} featuring pulsating contractions \cite{li2024fluidization} or state-dependent interaction potentials \cite{alston2022intermittent}.
Therefore we expect our results to be widely applicable.

We note that there is evidence of hyperuniform organisation in biological systems, \textit{e.g.} cell nuclei in brain tumours \cite{jiao2011spatial} or suspensions of algae \cite{huang2021circular}.
Further studies may explore if the mechanism we have described is exploited in cell tissues in order to suppress density fluctuations.

\section*{Conflicts of interest}

There are no conflicts to declare.

\section*{Data availability}

All relevant data supporting the findings of this study are included in the manuscript and its ESI and were generated using the free software libraries \texttt{cells} \cite{cells} and \texttt{SAMoS} \cite{samos}.

\section*{Acknowledgements}

We wholeheartedly thank Henry Alston, Konstantinos Andreadis, Calvin Bakker, Arthur Hernandez, Sander Kammeraat, Juliane Klamser, Qun-Li Lei, Marko Popović, Jan Rozman, and Rastko Sknepnek for insightful discussions.
We acknowledge fruitful discussions at the workshop ``Computational Advances in Active Matter'' held in December 2023 at the Lorentz Center in Leiden which we thank. SH would like to acknowledge the 2024 Active Solids KITP program, and this research was supported in part by grant NSF PHY-2309135 to the Kavli Institute for Theoretical Physics (KITP).
This work was performed using the computational resources from Instituut-Lorentz and from the Academic Leiden Interdisciplinary Cluster Environment (ALICE) provided by Universiteit Leiden.

\bibliography{ref}

\clearpage
\onecolumngrid
\appendix
\makeatletter
\def\thesection{\Alph{section}}
\def\thesubsection{\Alph{section}.\arabic{subsection}}
\def\p@subsection{}
\makeatother
\counterwithin{figure}{section}
\renewcommand\theequation{\thesection\arabic{equation}}
\renewcommand\thefigure{\thesection\arabic{figure}}

\section{Displacement fluctuations in a two-dimensional continuous medium}
\label{app:fulluk2}

We consider the following equation of motion
\begin{equation}
\zeta \dot{\boldsymbol{u}}(\boldsymbol{r}, t) = - \int_{\mathbb{R}^2} \mathrm{d}^2 \boldsymbol{r}^{\prime} \underline{\boldsymbol{D}}^{\mathrm{el}}(\boldsymbol{r} - \boldsymbol{r}^{\prime}) \boldsymbol{u}(\boldsymbol{r}^{\prime}, t) + \boldsymbol{\lambda}(\boldsymbol{r}, t)
\label{eq:ucontA}
\end{equation}
where $\boldsymbol{u}(\boldsymbol{r}, t)$ describes the elastic deformation from position $\boldsymbol{r}$, $\zeta$ is a friction coefficient, $\underline{\boldsymbol{D}}^{\mathrm{el}}$ is a dynamical matrix \cite{saarloos2024soft} describing elasticity, and $\boldsymbol{\lambda}$ a stochastic term.

Using the Fourier transform in space and time of the displacement field \eqref{eq:ftdef}
\begin{equation}
\boldsymbol{u}(\boldsymbol{r}, t) = \frac{1}{(2\pi)^2} \int_{\mathbb{R}^2} \mathrm{d}^2\boldsymbol{q} \, e^{\mathrm{i} \boldsymbol{q} \cdot \boldsymbol{r}} \, \tilde{\boldsymbol{u}}(\boldsymbol{q}, t) = \frac{1}{(2\pi)^3} \int_{\mathbb{R}^2} \mathrm{d}^2\boldsymbol{q} \, \int_{\mathbb{R}} \mathrm{d}\omega \, e^{\mathrm{i} \boldsymbol{q} \cdot \boldsymbol{r} + \mathrm{i} \omega t} \, \tilde{\boldsymbol{U}}(\boldsymbol{q}, \omega),
\end{equation}
\mbox{}\\
we write \eqref{eq:ucontA} in Fourier space
\begin{equation}
\mathrm{i} \omega \zeta \tilde{\boldsymbol{U}}(\boldsymbol{q}, \omega) = -\tilde{\underline{\boldsymbol{D}}}^{\mathrm{el}}(\boldsymbol{q}) \tilde{\boldsymbol{U}}(\boldsymbol{q}, \omega) + \tilde{\boldsymbol{\Lambda}}(\boldsymbol{q}, \omega).
\label{eq:ucontftA}
\end{equation}
We assume that the Fourier transform of the dynamical matrix $\tilde{\underline{\boldsymbol{D}}}^{\mathrm{el}}(\boldsymbol{q})$ can be decomposed as follows \cite{henkes2020dense}
\begin{equation}
\tilde{\underline{\boldsymbol{D}}}^{\mathrm{el}}(\boldsymbol{q}) \tilde{\boldsymbol{U}}(\boldsymbol{q}, \omega) = |\boldsymbol{q}|^2 \left[(B + \mu) \tilde{\boldsymbol{U}}_{\parallel}(\boldsymbol{q}, \omega) + \mu \tilde{\boldsymbol{U}}_{\perp}(\boldsymbol{q}, \omega)\right],
\label{eq:dukA}
\end{equation}
where \mbox{$\hat{\boldsymbol{q}} = \boldsymbol{q}/|\boldsymbol{q}|$}, \mbox{$\tilde{\boldsymbol{U}}_{\parallel}(\boldsymbol{q}, \omega) = (\tilde{\boldsymbol{U}}(\boldsymbol{q}, \omega) \cdot \hat{\boldsymbol{q}}) \, \hat{\boldsymbol{q}}$} and \mbox{$\tilde{\boldsymbol{U}}_{\perp}(\boldsymbol{q}, \omega) = \tilde{\boldsymbol{U}}(\boldsymbol{q}, t) - \tilde{\boldsymbol{U}}_{\parallel}(\boldsymbol{q}, \omega)$} are respectively the longitudinal and transverse displacements in Fourier space, and $B$ and $\mu$ are respectively the bulk and shear moduli.
We can thus now write
\begin{equation}
\left<\tilde{\boldsymbol{U}}(\boldsymbol{q}, \omega)\cdot\tilde{\boldsymbol{U}}(\boldsymbol{q}^{\prime}, \omega^{\prime})^*\right> = \frac{\left<\tilde{\boldsymbol{\Lambda}}_{\parallel}(\boldsymbol{q}, \omega) \cdot \tilde{\boldsymbol{\Lambda}}_{\parallel}(\boldsymbol{q}^{\prime}, \omega^{\prime})^*\right>}{[\mathrm{i} \omega \zeta + |\boldsymbol{q}|^2 (B + \mu)][-\mathrm{i}\omega^{\prime}\zeta + |\boldsymbol{q}^{\prime}|^2 (B + \mu)]} + \frac{\left<\tilde{\boldsymbol{\Lambda}}_{\perp}(\boldsymbol{q}, \omega) \cdot \tilde{\boldsymbol{\Lambda}}_{\perp}(\boldsymbol{q}^{\prime}, \omega^{\prime})^*\right>}{[\mathrm{i} \omega \zeta + |\boldsymbol{q}|^2 \mu][- \mathrm{i}\omega^{\prime}\zeta + |\boldsymbol{q}^{\prime}|^2 \mu]},
\end{equation}
where $\tilde{\boldsymbol{\Lambda}}_{\parallel} = (\tilde{\boldsymbol{\Lambda}} \cdot \hat{\boldsymbol{q}})\hat{\boldsymbol{q}}$ and $\tilde{\boldsymbol{\Lambda}}_{\perp} = \tilde{\boldsymbol{\Lambda}} - \tilde{\boldsymbol{\Lambda}}_{\parallel}$ are respectively the longitudinal and transverse stochastic terms in Fourier space.
We assume that the stochastic term is either the divergence of a tensor field uncorrelated in space, or is a vector field uncorrelated in space, with correlations respectively
\begin{subequations}
\begin{align}
\left<\boldsymbol{\lambda}(\boldsymbol{r}, t) \cdot \boldsymbol{\lambda}(\boldsymbol{r}^{\prime}, t^{\prime})\right> &= -\sigma^2 a^2 \, e^{-|t - t^{\prime}|/\tau} \, \nabla^2 \delta(\boldsymbol{r} - \boldsymbol{r}^{\prime}),\\
\left<\boldsymbol{\lambda}(\boldsymbol{r}, t) \cdot \boldsymbol{\lambda}(\boldsymbol{r}^{\prime}, t^{\prime})\right> &= f^2 a^2 \, e^{-|t - t^{\prime}|/\tau} \, \delta(\boldsymbol{r} - \boldsymbol{r}^{\prime}),
\end{align}
\end{subequations}
where $\sigma$ is an energy scale, $f$ a force scale, $a$ a coarse-graining length scale, and $\tau$ a persistence time.
Using the following identities
\begin{subequations}
\begin{gather}
\int_{\mathbb{R}} \mathrm{d}t \, e^{-\mathrm{i} (\omega - \omega^{\prime})t} = 2\pi \, \delta(\omega - \omega^{\prime})\\
\int_{\mathbb{R}} \mathrm{d}t \int_{\mathbb{R}} \mathrm{d}t^{\prime} \, e^{-\mathrm{i}(\omega t - \omega^{\prime} t^{\prime})} e^{-|t - t^{\prime}|/\tau} \underset{t^{\prime}=t+s}{=} \int_{\mathbb{R}} \mathrm{d}t \, e^{-\mathrm{i}(\omega - \omega^{\prime}) t} \int_{\mathbb{R}} \mathrm{d}s \, e^{\mathrm{i}\omega^{\prime}s} e^{-|s|/\tau} = 2\pi \frac{2\tau}{1 + \omega^2 \tau^2} \, \delta(\omega - \omega^{\prime}),
\end{gather}
\end{subequations}
we write these correlations in Fourier space
\begin{subequations}
\begin{align}
\left<\tilde{\boldsymbol{\Lambda}}(\boldsymbol{q}, \omega)\cdot\tilde{\boldsymbol{\Lambda}}(\boldsymbol{q}^{\prime}, \omega^{\prime})^*\right> = (2\pi)^3 \frac{2 \sigma^2 \tau |\boldsymbol{q}|^2}{1 + \omega^2 \tau^2} \, \delta(\omega - \omega^{\prime}) \, a^2 \delta(\boldsymbol{q} - \boldsymbol{q}^{\prime}),\\
\left<\tilde{\boldsymbol{\Lambda}}(\boldsymbol{q}, \omega)\cdot\tilde{\boldsymbol{\Lambda}}(\boldsymbol{q}^{\prime}, \omega^{\prime})^*\right> = (2\pi)^3 \frac{2 f^2 \tau}{1 + \omega^2 \tau^2} \, \delta(\omega - \omega^{\prime}) \, a^2 \delta(\boldsymbol{q} - \boldsymbol{q}^{\prime}),
\end{align}
\end{subequations}
where we have taken $\boldsymbol{q} = \boldsymbol{q}^{\prime}$ and $\omega = \omega^{\prime}$, outside of the Dirac delta functions $\delta(\boldsymbol{q} - \boldsymbol{q}^{\prime})$ and $\delta(\omega - \omega^{\prime})$, for simplicity.
We further assume by isotropy
\begin{equation}
\left<\tilde{\boldsymbol{\Lambda}}_{\parallel}(\boldsymbol{q}, \omega)\cdot\tilde{\boldsymbol{\Lambda}}_{\parallel}(\boldsymbol{q}^{\prime}, \omega^{\prime})^*\right> = \left<\tilde{\boldsymbol{\Lambda}}_{\perp}(\boldsymbol{q}, \omega)\cdot\tilde{\boldsymbol{\Lambda}}_{\perp}(\boldsymbol{q}^{\prime}, \omega^{\prime})^*\right> = \frac{1}{2} \left<\tilde{\boldsymbol{\Lambda}}(\boldsymbol{q}, \omega)\cdot\tilde{\boldsymbol{\Lambda}}(\boldsymbol{q}^{\prime}, \omega^{\prime})^*\right>.
\end{equation}
We thus obtain the following space and time Fourier fluctuations
\begin{subequations}
\begin{align}
\left<\tilde{\boldsymbol{U}}(\boldsymbol{q}, \omega) \cdot \tilde{\boldsymbol{U}}(\boldsymbol{q}, \omega)^*\right> &= (2\pi)^3 \frac{\sigma^2 \tau |\boldsymbol{q}|^2 \, \delta(\omega - \omega^{\prime}) \, a^2 \delta(\boldsymbol{q} - \boldsymbol{q}^{\prime})}{[\omega^2 \zeta^2 + |\boldsymbol{q}|^4 (B + \mu)^2][1 + \omega^2 \tau^2]} + (2\pi)^3 \frac{\sigma^2 \tau |\boldsymbol{q}|^2 \, \delta(\omega - \omega^{\prime}) \, a^2 \delta(\boldsymbol{q} - \boldsymbol{q}^{\prime})}{[\omega^2 \zeta^2 + |\boldsymbol{q}|^4 \mu^2][1 + \omega^2 \tau^2]},\\
\left<\tilde{\boldsymbol{U}}(\boldsymbol{q}, \omega) \cdot \tilde{\boldsymbol{U}}(\boldsymbol{q}, \omega)^*\right> &= (2\pi)^3 \frac{f^2 \tau \, \delta(\omega - \omega^{\prime}) \, a^2 \delta(\boldsymbol{q} - \boldsymbol{q}^{\prime})}{[\omega^2 \zeta^2 + |\boldsymbol{q}|^4 (B + \mu)^2][1 + \omega^2 \tau^2]} + (2\pi)^3 \frac{f^2 \tau \, \delta(\omega - \omega^{\prime}) \, a^2 \delta(\boldsymbol{q} - \boldsymbol{q}^{\prime})}{[\omega^2 \zeta^2 + |\boldsymbol{q}|^4 \mu^2][1 + \omega^2 \tau^2]}.
\end{align}
\end{subequations}
Using the following identities
\begin{subequations}
\begin{align}
&\int_{\mathbb{R}} \mathrm{d}\omega \frac{1}{a^2 + \omega^2 b^2} = \frac{\pi}{ab},\\
&\begin{aligned}
\int_{\mathbb{R}} \mathrm{d}\omega \int_{\mathbb{R}} \mathrm{d}\omega^{\prime} \, e^{\mathrm{i}(\omega - \omega^{\prime})t} \, \frac{\delta(\omega - \omega^{\prime})}{(a^2 + \omega^2 b^2)(c^2 + \omega^2 d^2)} &= \int_{\mathbb{R}} \mathrm{d}\omega \, \frac{1}{b^2 c^2 - a^2 d^2} \left[\frac{b^2}{a^2 + \omega^2 b^2} - \frac{d^2}{c^2 + \omega^2 d^2}\right]\\
&= \frac{1}{b^2 c^2 - a^2 d^2} \left[\frac{\pi b}{a} - \frac{\pi d}{c}\right]\\
&= \frac{\pi}{ac[bc + ad]},
\end{aligned}
\end{align}
\label{eq:idlor}%
\end{subequations}
we obtain the equal-time Fourier fluctuations
\begin{subequations}
\begin{align}
\left<\tilde{\boldsymbol{u}}(\boldsymbol{q}, t) \cdot \tilde{\boldsymbol{u}}(\boldsymbol{q}, t)^*\right> &= 2\pi \frac{\pi \sigma^2 \tau |\boldsymbol{q}|^2 \, a^2 \delta(\boldsymbol{q} - \boldsymbol{q}^{\prime})}{|\boldsymbol{q}|^2 (B + \mu)[\zeta + |\boldsymbol{q}|^2(B + \mu) \tau]} + 2\pi \frac{\pi \sigma^2 \tau |\boldsymbol{q}|^2 \, a^2 \delta(\boldsymbol{q} - \boldsymbol{q}^{\prime})}{|\boldsymbol{q}|^2 \mu[\zeta + |\boldsymbol{q}|^2\mu \tau]},\\
\left<\tilde{\boldsymbol{u}}(\boldsymbol{q}, t) \cdot \tilde{\boldsymbol{u}}(\boldsymbol{q}, t)^*\right> &= 2\pi \frac{\pi f^2 \tau \, a^2 \delta(\boldsymbol{q} - \boldsymbol{q}^{\prime})}{|\boldsymbol{q}|^2 (B + \mu)[\zeta + |\boldsymbol{q}|^2(B + \mu) \tau]} + 2\pi \frac{\pi f^2 \tau \, a^2 \delta(\boldsymbol{q} - \boldsymbol{q}^{\prime})}{|\boldsymbol{q}|^2 \mu[\zeta + |\boldsymbol{q}|^2\mu \tau]},
\end{align}
\end{subequations}
which correspond to \eqref{eq:uk2} given the longitudinal and transversal correlation length scales $\xi_{\parallel} = \sqrt{(B + \mu)\tau/\zeta}$ and $\xi_{\perp} = \sqrt{\mu\tau/\zeta}$.

\section{Displacement fluctuations in a two-dimensional triangular lattice}
\label{app:discreteuk2}

\begin{figure} 
\centering
\begin{tikzpicture}

\coordinate (A) at (0, 0);
\draw[black, fill=black] (A) circle (0.2) node[below, yshift=-5pt]{$i$};
\draw[black] (A) -- ++(3, 0) edge[dashed] ++(1, 0);
\draw[black] (A) -- ++(0, {sqrt(3)}) edge[dashed] ++(0, 1);
\draw [decorate,decoration={brace, amplitude=5pt, raise=5pt}] ($(A) + (0, {sqrt(3) + 1})$) -- ++(4, 0) node[midway, yshift=20pt]{$L_x \sim \sqrt{N}$};
\draw [decorate,decoration={brace, amplitude=5pt, mirror, raise=5pt}] ($(A) + (4, 0)$) -- ++(0, {sqrt(3) + 1}) node[midway, xshift=20pt, rotate=270, anchor=center]{$L_y \sim \sqrt{3 N/4}$};
\coordinate (B) at ($(A) + ({2*cos(0*60)}, {2*sin(0*60)})$);
\draw[black, fill=black, thick] (B) circle (0.2) node[above, yshift=5pt]{$j = i + 1$};
\draw[black, thick, -stealth, shorten >=0.2cm] (A) -- (B) node[midway, below]{$\boldsymbol{e}_1 = a \hat{\boldsymbol{e}}_x$};
\coordinate (C) at ($(A) + ({2*cos(1*60)}, {2*sin(1*60)})$);
\draw[black, fill=black, thick] (C) circle (0.2) node[above, yshift=5pt]{$j = i + 2$};
\draw[black, thick, -stealth, shorten >=0.2cm] (A) -- (C) node[midway, right]{$\boldsymbol{e}_2$};
\coordinate (D) at ($(A) + ({2*cos(2*60)}, {2*sin(2*60)})$);
\draw[black, fill=white, thick] (D) circle (0.2) node[above, yshift=5pt]{$j = i + 3$};
\draw[black, thick, -stealth, shorten >=0.2cm] (A) -- (D) node[midway, right]{$\boldsymbol{e}_3$};
\coordinate (E) at ($(A) + ({2*cos(3*60)}, {2*sin(3*60)})$);
\draw[black, fill=white, thick] (E) circle (0.2) node[above, yshift=5pt]{$j = i - 1$};
\draw[black, thick, dashed, -stealth, shorten >=0.2cm] (A) -- (E) node[midway, above]{$-\boldsymbol{e}_1$};
\coordinate (F) at ($(A) + ({2*cos(4*60)}, {2*sin(4*60)})$);
\draw[black, fill=white, thick] (F) circle (0.2) node[below, yshift=-5pt]{$j = i - 2$};
\draw[black, thick, dashed, -stealth, shorten >=0.2cm] (A) -- (F) node[midway, left]{$-\boldsymbol{e}_2$};
\coordinate (G) at ($(A) + ({2*cos(5*60)}, {2*sin(5*60)})$);
\draw[black, fill=white, thick] (G) circle (0.2) node[below, yshift=-5pt]{$j = i - 3$};
\draw[black, thick, dashed, -stealth, shorten >=0.2cm] (A) -- (G) node[midway, left]{$-\boldsymbol{e}_3$};
\coordinate (H) at ($(B) + ({2*cos(1*60)}, {2*sin(1*60)})$);
\draw[black, fill=black, thick] (H) circle (0.2);

\end{tikzpicture}
\caption{Notations within the triangular lattice.}
\label{fig:2dtri}
\end{figure}

We consider a periodic triangular lattice in two dimensions, see Fig.~\ref{fig:2dtri}, where $\boldsymbol{e}_{\alpha} = a (\cos((\alpha - 1) \pi/3),~\sin((\alpha - 1) \pi/3))$. We denote $N$ the total number of sites -- of the order of $\sqrt{N}$ in each dimension --, $\boldsymbol{r}_i^0$ the reference lattice position of particle $i \in \{1, \ldots, N\}$, and $\boldsymbol{u}_i = \boldsymbol{r}_i - \boldsymbol{r}_i^0$ its displacement from this lattice position.
We define in this lattice the discrete divergence operator \cite{cavagna2024noise} which to a quantity $\boldsymbol{v}_{i\to\alpha}$ defined on links $i\to\alpha$ associates
\begin{equation}
\nabla \cdot \underline{\boldsymbol{v}}_i = \sum_{\alpha=1}^3 [\boldsymbol{v}_{i\to\alpha} - \boldsymbol{v}_{(i-\alpha)\to\alpha}]
\end{equation}
defined on sites $i$, and the Laplacian operator $\nabla^2$ which to a quantity $\boldsymbol{w}_i$ defined on sites $i$ associates
\begin{equation}
\nabla^2 \boldsymbol{w}_i = \sum_{\alpha=1}^3 [\boldsymbol{w}_{i+\alpha} + \boldsymbol{w}_{i-\alpha} - 2 \boldsymbol{w}_i]
\end{equation}
also defined on sites $i$.

We introduce the following overdamped equation of motion
\begin{equation}
\zeta \dot{\boldsymbol{u}}_i(t) - \gamma \nabla^2 \dot{\boldsymbol{u}}_i = k \nabla^2 \boldsymbol{u}_i + \boldsymbol{\lambda}_i(t)
\label{eq:udiscrete}
\end{equation}
where $\zeta$ is a friction coefficient, $\gamma$ a pair friction coefficient, $k$ a spring constant, and $\boldsymbol{\lambda}_i(t)$ is the stochastic active term.

We define the discrete space Fourier transform
\begin{subequations}
\begin{align}
\tilde{\boldsymbol{v}}_{\boldsymbol{q}} &= \sum_{i=1}^N e^{-\mathrm{i}\boldsymbol{q}\cdot\boldsymbol{r}_i^0} \, \boldsymbol{v}_i,\\
\boldsymbol{v}_i &= \frac{1}{N} \sum_{\substack{m,n=0\\\boldsymbol{q}=\frac{2\pi}{\sqrt{N}a}(m, n)}}^{N - 1} e^{\mathrm{i}\boldsymbol{q}\cdot\boldsymbol{r}_i^0} \, \tilde{\boldsymbol{v}}_{\boldsymbol{q}},
\end{align}
\label{eq:dft}%
\end{subequations}
with the following orthogonality relation
\begin{equation}
\sum_{i=1}^N e^{\mathrm{i}(\boldsymbol{q} - \boldsymbol{q}^{\prime})\cdot\boldsymbol{r}_i^0} = N \, \delta_{\boldsymbol{q},\boldsymbol{q}^{\prime}}.
\label{eq:dftortho}
\end{equation}

We write \eqref{eq:udiscrete} in Fourier space
\begin{equation}
\mathrm{i}\omega\zeta\tilde{\boldsymbol{U}}_{\boldsymbol{q}}(\omega) + \mathrm{i} \omega Q(\boldsymbol{q})^2 \gamma \tilde{\boldsymbol{U}}_{\boldsymbol{q}}(\omega) = -Q(\boldsymbol{q})^2 k \tilde{\boldsymbol{U}}_{\boldsymbol{q}}(\omega) + \tilde{\boldsymbol{\Lambda}}_{\boldsymbol{q}}(\omega)
\end{equation}
where we have introduced the kernel
\begin{equation}
Q(\boldsymbol{q})^2 = 2 \sum_{\alpha=1}^3 (1 - \cos(\boldsymbol{q}\cdot\boldsymbol{e}_{\alpha})) \underset{|\boldsymbol{q}|\to0} = \frac{3}{2}|\boldsymbol{q}|^2 a^2.
\end{equation}
We thus obtain the following space and time Fourier fluctuations
\begin{equation}
\left<\tilde{\boldsymbol{U}}_{\boldsymbol{q}}(\omega)\cdot\tilde{\boldsymbol{U}}_{\boldsymbol{q}^{\prime}}(\omega^{\prime})^*\right> = \frac{\left<\tilde{\boldsymbol{\Lambda}}_{\boldsymbol{q}}(\omega) \cdot \tilde{\boldsymbol{\Lambda}}_{\boldsymbol{q}^{\prime}}(\omega^{\prime})\right>}{[i\mathrm{\omega}(\zeta + Q(\boldsymbol{q})^2 \gamma) + Q(\boldsymbol{q})^2 k][-i\mathrm{\omega}^{\prime}(\zeta + Q(\boldsymbol{q}^{\prime})^2 \gamma) + Q(\boldsymbol{q}^{\prime})^2 k]}.
\end{equation}

\subsection{Active stresses}

We first provide the derivation for a fluctuating stress with finite-time correlations, for simplicity in the absence of pair friction. We have
\begin{subequations}
\begin{align}
\boldsymbol{\lambda}_i(t) &= \nabla\cdot\underline{\boldsymbol{\sigma}}_i,\\
\left<\boldsymbol{\sigma}_{i\to\alpha}(t)\cdot\boldsymbol{\sigma}_{j\to\beta}(t^{\prime})\right> &= \sigma^2 a^{-2} e^{-|t - t^{\prime}|/\tau} \, \delta_{ij} \delta_{\alpha\beta} \delta(t - t^{\prime}),
\end{align}
\end{subequations}
which read in Fourier space
\begin{equation}
\left<\tilde{\boldsymbol{\Lambda}}_{\boldsymbol{q}}(\omega)\cdot\tilde{\boldsymbol{\Lambda}}_{\boldsymbol{q}^{\prime}}(\omega^{\prime})^*\right> = 2\pi \frac{2 \sigma^2 a^{-2} \tau Q(\boldsymbol{q})^2}{1 + \omega^2\tau^2} N\delta_{\boldsymbol{q},\boldsymbol{q}^{\prime}} \, \delta(\omega - \omega^{\prime}).
\end{equation}
such that the space and time Fourier fluctuations are
\begin{equation}
\left<\tilde{\boldsymbol{U}}_{\boldsymbol{q}}(\omega)\cdot\tilde{\boldsymbol{U}}_{\boldsymbol{q}^{\prime}}(\omega^{\prime})^*\right> = 2\pi \frac{2 \sigma^2 a^{-2} \tau Q(\boldsymbol{q})^2}{[\omega^2\zeta^2 + Q(\boldsymbol{q})^4 k^2][1 + \omega^2\tau^2]} \, N\delta_{\boldsymbol{q},\boldsymbol{q}^{\prime}} \, \delta(\omega - \omega^{\prime}).
\end{equation}
We use \eqref{eq:idlor} and write the equal-time Fourier fluctuations
\begin{equation}
\left<\tilde{\boldsymbol{u}}_{\boldsymbol{q}}(t)\cdot\tilde{\boldsymbol{u}}_{\boldsymbol{q}^{\prime}}(t)^*\right> = \frac{\sigma^2 a^{-2} \tau}{k(\zeta + Q(\boldsymbol{q})^2 k \tau)} \, N \delta_{\boldsymbol{q},\boldsymbol{q}^{\prime}} = \frac{\sigma^2\tau^2}{\zeta^2} \frac{\xi^{-2}}{1 + Q(\boldsymbol{q})^2 (\xi/a)^2} \, N \delta_{\boldsymbol{q},\boldsymbol{q}^{\prime}},
\end{equation}
where we have introduced the correlation length $\xi = a \sqrt{k \tau/\zeta}$.
We obtain the same large-wavelength scaling as \eqref{eq:uk2fdt}: the finite-time correlation of the stress only affects the small-wavelength fluctuations.

\subsection{Fluctuation-dissipation and the thermal limit}
\label{sec:thermallimituq2}

It is important to compare and contrast our results with those expected in thermal equilibrium. Given the dissipation term on the l.h.s. of \eqref{eq:udiscrete}, the second fluctuation-dissipation theorem \cite{kubo1966fluctuationdissipation,doi1988theory,tothova2022overdamped} dictates that the stochastic term $\boldsymbol{\lambda}_i(t)$ should have the following decomposition and correlations
\begin{subequations}
\begin{align}
\boldsymbol{\lambda}_i(t) &= \boldsymbol{\eta}_i(t) + \nabla\cdot\underline{\boldsymbol{\sigma}}_i,\\
\left<\boldsymbol{\eta}_i(t)\cdot\boldsymbol{\eta}_j(t^{\prime})\right> &= 2 k_{\mathrm{B}} T_1 \zeta \, \delta_{ij} \, \delta(t - t^{\prime}),\\
\left<\boldsymbol{\sigma}_{i\to\alpha}(t)\cdot\boldsymbol{\sigma}_{j\to\beta}(t^{\prime})\right> &= 2 k_{\mathrm{B}} T_2 \gamma \, \delta_{ij}\delta_{\alpha\beta} \, \delta(t - t^{\prime}),\\
\left<\boldsymbol{\eta}_j(t) \cdot \boldsymbol{\sigma}_{j\to\beta}(t^{\prime})\right> &= 0.
\end{align}
\label{eq:fdtnoise}%
\end{subequations}
with $T = T_1 = T_2$ the bath temperature.

The fluctuations of the driving noise deriving from \eqref{eq:fdtnoise} are
\begin{equation}
\left<\tilde{\boldsymbol{\Lambda}}_{\boldsymbol{q}}(\omega) \cdot \tilde{\boldsymbol{\Lambda}}_{\boldsymbol{q}^{\prime}}(\omega^{\prime})^*\right> = 4 \pi k_{\mathrm{B}} T_1 \zeta N \delta_{\boldsymbol{q},\boldsymbol{q}^{\prime}} \delta(\omega - \omega^{\prime}) + 4 \pi k_{\mathrm{B}} T_2 \gamma Q(\boldsymbol{q})^2 N \delta_{\boldsymbol{q},\boldsymbol{q}^{\prime}} \delta(\omega - \omega^{\prime}).
\end{equation}
We refine our expression of the displacement fluctuations
\begin{equation}
\left<\tilde{\boldsymbol{U}}_{\boldsymbol{q}}(\omega)\cdot\tilde{\boldsymbol{U}}_{\boldsymbol{q}^{\prime}}(\omega^{\prime})^*\right> = \frac{4 \pi k_{\mathrm{B}} T_1 \zeta N \delta_{\boldsymbol{q},\boldsymbol{q}^{\prime}} \delta(\omega - \omega^{\prime})}{\omega^2(\zeta + Q(\boldsymbol{q})^2 \gamma)^2 + Q(\boldsymbol{q})^4 k^2} + \frac{4 \pi k_{\mathrm{B}} T_2 \gamma Q(\boldsymbol{q})^2 N \delta_{\boldsymbol{q},\boldsymbol{q}^{\prime}} \delta(\omega - \omega^{\prime})}{\omega^2(\zeta + Q(\boldsymbol{q})^2 \gamma)^2 + Q(\boldsymbol{q})^4 k^2},
\end{equation}
and take the inverse time Fourier transform using \eqref{eq:idlor} to compute the equal-time Fourier fluctuations
\begin{equation}
\left<\tilde{\boldsymbol{u}}_{\boldsymbol{q}}(t) \cdot \tilde{\boldsymbol{u}}_{\boldsymbol{q}^{\prime}}(t)^*\right> = \frac{k_{\mathrm{B}} T_1 \zeta}{(\zeta + Q(\boldsymbol{q})^2 \gamma) Q(\boldsymbol{q})^2 k} N \delta_{\boldsymbol{q},\boldsymbol{q}^{\prime}} + \frac{k_{\mathrm{B}} T_2 \gamma}{(\zeta + Q(\boldsymbol{q})^2 \gamma) k} N \delta_{\boldsymbol{q},\boldsymbol{q}^{\prime}}.
\label{eq:uk2fdt}
\end{equation}
We stress here that $\left<|\tilde{\boldsymbol{u}}_{\boldsymbol{q}}(t)|^2\right>$ remains finite for $|\boldsymbol{q}| \to 0$ if and only if $T_1 = 0$ and $\zeta \neq 0$.
Therefore, it is the competition between a fluctuating stress ($T_2 \neq 0$) and a particle-wise drag force ($\zeta \neq 0$) in the absence of a particle-wise fluctuating force ($T_1 = 0$) which damps large-wavelength fluctuations.
Moreover, the effect of an additional viscosity-like term only affects the small-wavelength behaviour.

\section{Large-distance scaling of the translational order correlation function}
\label{app:ctr}

We consider the local translational order parameter \eqref{eq:psiq0}
\begin{equation}
\psi_{\boldsymbol{q}_0,i} = e^{\mathrm{i}\boldsymbol{q}_0\cdot(\boldsymbol{r}_i-\boldsymbol{r}_0)}
\end{equation}
where $\boldsymbol{q}_0$ is a reciprocal vector of the lattice, $\boldsymbol{r}_i = \boldsymbol{r}_i^0 + \boldsymbol{u}_i$ the position of particle $i$ which is displaced of $\boldsymbol{u}_i$ from its lattice position $\boldsymbol{r}_i^0$.
We write the fluctuations
\begin{equation}
\left<\psi_{\boldsymbol{q}_0,i}\psi_{\boldsymbol{q}_0,j}^*\right> = \left<e^{\mathrm{i}\boldsymbol{q}_0 (\boldsymbol{r}_i - \boldsymbol{r}_j)}\right> = \left<e^{\mathrm{i} \boldsymbol{q}_0 \cdot (\boldsymbol{r}_i^0 - \boldsymbol{r}_j^0)}e^{\mathrm{i} \boldsymbol{q}_0 \cdot (\boldsymbol{u}_i - \boldsymbol{u}_j)}\right>.
\label{eq:psiq0fluc}
\end{equation}
We note that for two lattice points $\boldsymbol{r}_i^0$ and $\boldsymbol{r}_j^0$ then $\boldsymbol{q}_0 \cdot (\boldsymbol{r}_i^0 - \boldsymbol{r}_j^0)$ is 0 modulo $2\pi$ therefore $e^{\mathrm{i} \boldsymbol{q}_0 \cdot (\boldsymbol{r}_i^0 - \boldsymbol{r}_j^0)} = 1$.
We also define for a random variable $X$ its characteristic function $t \mapsto \left<\exp(\mathrm{i} t X)\right>$.
For a normally distributed random variable $X$ with zero mean $\left<X\right> = 0$ then
\begin{equation}
\left<\exp(\mathrm{i} t X)\right> = \exp\left(-\frac{1}{2}t^2\left<X^2\right>\right).
\label{eq:gausschar}
\end{equation}
We assume for two distant particles $i$ and $j$ that $\boldsymbol{q}_0 \cdot (\boldsymbol{r}_i - \boldsymbol{r}_j)$ is normally distributed -- from the linearity of \eqref{eq:udiscrete} and the Gaussian nature of driving noises -- and assume from isotropy
\begin{equation}
\left<(\boldsymbol{q}_0 \cdot (\boldsymbol{u}_i - \boldsymbol{u}_j))^2\right> = \frac{1}{2} |\boldsymbol{q}_0|^2 \left<|\boldsymbol{u}_i - \boldsymbol{u}_j|^2\right> = |\boldsymbol{q}_0|^2 \left<|\boldsymbol{u}_i|^2\right> - |\boldsymbol{q}_0|^2 \left<\boldsymbol{u}_i\cdot\boldsymbol{u}_j\right>
\label{eq:hypiso}
\end{equation}
where we have used $\left<|\boldsymbol{u}_i|^2\right> = \left<|\boldsymbol{u}_j|^2\right>$ for all $i$ and $j$ in steady state.

As a matter of simplification we use the continuous-space spectrum of displacements \eqref{eq:uk2} to compute the correlations $\left<|\boldsymbol{u}_i|^2\right>$ and $\left<\boldsymbol{u}_i\cdot\boldsymbol{u}_j\right>$.
We then write
\begin{subequations}
\begin{align}
\left<|\boldsymbol{u}_i|^2\right> &= \frac{1}{(2 \pi)^4} \int_{\mathbb{R}^2} \mathrm{d}^2\boldsymbol{q} \int_{\mathbb{R}^2} \mathrm{d}^2\boldsymbol{q}^{\prime} \, \left<\tilde{\boldsymbol{u}}(\boldsymbol{q}, t)\cdot\tilde{\boldsymbol{u}}(\boldsymbol{q}^{\prime}, t)^*\right>,\\
\left<\boldsymbol{u}_i\cdot\boldsymbol{u}_j\right> = \left<\boldsymbol{u}_j\cdot\boldsymbol{u}_i\right> &= \frac{1}{(2 \pi)^4} \int_{\mathbb{R}^2} \mathrm{d}^2\boldsymbol{q} \int_{\mathbb{R}^2} \mathrm{d}^2\boldsymbol{q}^{\prime} \, \cos\left(\boldsymbol{q}\cdot\boldsymbol{r}_i^0 - \boldsymbol{q}^{\prime}\cdot\boldsymbol{r}_j^0\right) \left<\tilde{\boldsymbol{u}}(\boldsymbol{q}, t)\cdot\tilde{\boldsymbol{u}}(\boldsymbol{q}^{\prime}, t)^*\right>.
\end{align}
\label{eq:ucor}%
\end{subequations}
such that using \eqref{eq:hypiso}
\begin{equation}
\left<(\boldsymbol{q}_0 \cdot (\boldsymbol{u}_j - \boldsymbol{u}_i))^2\right> = \frac{|\boldsymbol{q}_0|^2}{(2\pi)^4} \int_{\mathbb{R}^2} \mathrm{d}^2 \boldsymbol{q} \int_{\mathbb{R}^2} \mathrm{d}^2\boldsymbol{q}^{\prime} \, \left[1 - \cos\left(\boldsymbol{q}\cdot\boldsymbol{r}_i^0 - \boldsymbol{q}\cdot\boldsymbol{r}_j^0\right)\right] \, \left<\tilde{\boldsymbol{u}}(\boldsymbol{q}, t)\cdot\tilde{\boldsymbol{u}}(\boldsymbol{q}^{\prime}, t)^*\right>.
\label{eq:disctocontft}
\end{equation}
We note that \eqref{eq:uk2} are symmetric by rotation of $\boldsymbol{q}$.
With
\begin{equation}
\boldsymbol{\Delta} = \boldsymbol{r}_i^0 - \boldsymbol{r}_j^0,~r = |\boldsymbol{\Delta}|,
\end{equation}
we define for all wave vector $\boldsymbol{q}$ the angle $\phi$ such that $\boldsymbol{q} \cdot \boldsymbol{\Delta} = q \, r \, \cos\phi$ where $q = |\boldsymbol{q}|$.
Therefore for any function $f(|\boldsymbol{q}|)$ we have the following identity
\begin{subequations}
\begin{align}
\int_{\mathbb{R}^2} \mathrm{d}^2\boldsymbol{q} \, \cos\left(\boldsymbol{q}\cdot\boldsymbol{\Delta}\right) \, f(|\boldsymbol{q}|) &= \int_{\mathbb{R}} \mathrm{d}q \int_0^{2\pi} \mathrm{d}\phi \, q \, \cos\left(q \, r \, \cos\phi\right) \, f(q) = \int_{\mathbb{R}} \mathrm{d}q \, 2 \pi q \, J_0(q\,r) \, f(q),\\
\int_{\mathbb{R}^2} \mathrm{d}^2\boldsymbol{q} \, f(|\boldsymbol{q}|) &= \int_{\mathbb{R}} \mathrm{d}q \, 2 \pi q \, f(q),
\end{align}
\label{eq:circavft}%
\end{subequations}
where $J_0$ is the 0-th Bessel function of the first kind.

\subsection{Fluctuating stress}

Using \eqref{eq:uk2stress}, \eqref{eq:ucor}, and \eqref{eq:circavft}, we write
\begin{subequations}
\begin{align}
\label{eq:u2stressint}
\left<|\boldsymbol{u}_i|^2\right> &= \frac{\sigma^2 \tau^2 a^2}{4 \pi \zeta^2} \left[\int_{2\pi/L}^{2\pi/a} \mathrm{d}q \, \frac{q}{\xi_{\parallel}^2 (1 + q^2 \xi_{\parallel}^2)} + \int_{2\pi/L}^{2\pi/a} \mathrm{d}q \, \frac{q}{\xi_{\perp}^2 (1 + q^2 \xi_{\perp}^2)}\right],\\
\label{eq:uustressint}
\left<\boldsymbol{u}_i \cdot \boldsymbol{u}_j\right> &= \frac{\sigma^2 \tau^2 a^2}{4 \pi \zeta^2} \left[\int_{2\pi/L}^{2\pi/a} \mathrm{d}q \, \frac{q \, J_0(q r)}{\xi_{\parallel}^2 (1 + q^2 \xi_{\parallel}^2)} + \int_{2\pi/L}^{2\pi/a} \mathrm{d}q \, \frac{q \, J_0(q r)}{\xi_{\perp}^2 (1 + q^2 \xi_{\perp}^2)}\right],
\end{align}
\label{eq:u2uustressint}%
\end{subequations}
where the upper cutoff in the integral translates the fact that the continuous description applies to scales greater than the coarse-graining scale $a$, and the lower cutoff corresponds to the system size $L$.
At fixed $r$ we first note that
\begin{equation}
J_0(q r) \underset{q \to 0}{=} 1 - \frac{1}{\Gamma(2)} \left(\frac{q r}{2}\right)^2
\label{eq:taylorj0}
\end{equation}
with $\Gamma$ the Euler gamma function.
It follows that all integrands in \eqref{eq:u2uustressint} converge to finite values for $q \to 0$ therefore all integrals converge for $L \to \infty$ and we thus take this limit for the derivation below.
To compute \eqref{eq:u2stressint} we use the following identity
\begin{equation}
\int_0^{2\pi/a} \mathrm{d}q \, \frac{q}{1 + q^2 \xi^2} = \frac{1}{2\xi^2} \log\left(1 + (2\pi)^2 \frac{\xi^2}{a^2}\right)
\end{equation}
and write
\begin{equation}
\left<|\boldsymbol{u}_i|^2\right> \underset{L \to \infty}{=} \frac{\sigma^2\tau^2 a^2}{8 \pi \zeta^2} \left[\frac{1}{\xi_{\parallel}^4}\log\Big(1 + (2\pi)^2 (\xi_{\parallel}/a)^2\Big) + \frac{1}{\xi_{\perp}^4}\log\Big(1 + (2\pi)^2 (\xi_{\perp}/a)^2\Big)\right] \equiv \left<u^2\right>_{\infty}^{\sigma}.
\label{eq:u2infstress}
\end{equation}
To compute \eqref{eq:uustressint} we use the following inequality
\begin{equation}
|J_0(q r)| \leq \frac{1}{\sqrt{q r}}
\label{eq:j0bound}
\end{equation}
to bound the following integral
\begin{equation}
\left|\int_0^{2\pi/a} \mathrm{d}q \, \frac{q \, J_0(q r)}{1 + q^2\xi^2}\right| \leq \frac{1}{\sqrt{r}} \int_0^{2\pi/a} \mathrm{d}q \, \frac{\sqrt{q}}{1 + q^2\xi^2} \xrightarrow[r \to \infty]{} 0
\end{equation}
where the latter limit derives from the fact that the integral is finite, which implies
\begin{equation}
\left<\boldsymbol{u}_i\cdot\boldsymbol{u}_j\right> \xrightarrow[\substack{L\to\infty\\r \to \infty}]{} 0.
\label{eq:uuinfstress}
\end{equation}
Therefore, using \eqref{eq:psiq0fluc}, \eqref{eq:gausschar}, \eqref{eq:hypiso}, \eqref{eq:u2infstress}, and \eqref{eq:uuinfstress}, we obtain
\begin{equation}
\left<\psi_{\boldsymbol{q}_0,i}\psi_{\boldsymbol{q}_0,j}^*\right> \underset{\substack{L\to\infty\\|\boldsymbol{r}_i^0 -\boldsymbol{r}_j^0| \to \infty}}{=} \exp\left(-\frac{1}{2}|\boldsymbol{q}_0|^2\left<u^2\right>_{\infty}^{\sigma}\right).
\end{equation}

\subsection{Fluctuating force}

Using \eqref{eq:uk2force}, \eqref{eq:ucor}, \eqref{eq:disctocontft}, and \eqref{eq:circavft}, we write
\begin{subequations}
\begin{align}
\label{eq:u2forceint}
\left<|\boldsymbol{u}_i|^2\right> &= \frac{f^2 \tau^2 a^2}{4 \pi \zeta^2} \left[\int_{2\pi/L}^{2\pi/a} \mathrm{d}q \, \frac{q}{q^2 \xi_{\parallel}^2(1 + q^2 \xi_{\parallel}^2)} + \int_{2\pi/L}^{2\pi/a} \mathrm{d}q \, \frac{q}{q^2 \xi_{\perp}^2(1 + q^2 \xi_{\perp}^2)}\right],\\
\label{eq:quforceint}
\left<(\boldsymbol{q}_0\cdot(\boldsymbol{u}_j - \boldsymbol{u}_i))^2\right> &= \frac{f^2 \tau^2 a^2 |\boldsymbol{q}_0|^2}{4 \pi \zeta^2} \left[\int_{2\pi/L}^{2\pi/a} \mathrm{d}q \, \frac{q(1 - J_0(q r))}{q^2 \xi_{\parallel}^2(1 + q^2 \xi_{\parallel}^2)} + \int_{2\pi/L}^{2\pi/a} \mathrm{d}q \, \frac{q(1 - J_0(q r))}{q^2 \xi_{\perp}^2(1 + q^2 \xi_{\perp}^2)}\right],
\end{align}
\end{subequations}
where the upper cutoff in the integral translates the fact that the continuous description applies to scales greater than the coarse-graining scale $a$, and the lower cutoff corresponds to the system size $L$.
To compute \eqref{eq:u2forceint} we note that $q[q^2(1 + q^2\xi^2)]^{-1} \sim q^{-1}$ for $q \to 0$ and $\int_{2\pi/L}^{2\pi/a} \mathrm{d}q \, q^{-1}$ diverges for $L \to \infty$, therefore we can use the equivalence of the integrals
\begin{equation}
\int_{2\pi/L}^{2\pi/a} \mathrm{d}q \, \frac{q}{q^2(1 + q^2\xi^2)} \underset{L \to \infty}{=} \int_{2\pi/L}^{2\pi/a} \mathrm{d}q \, \frac{1}{q} = \log(L/a),
\end{equation}
and write
\begin{equation}
\left<|\boldsymbol{u}_i|^2\right> \underset{L \to \infty}{=} \frac{f^2\tau^2a^2}{4 \pi \zeta^2} \left[\frac{1}{\xi_{\parallel}^2} + \frac{1}{\xi_{\perp}^2}\right] \log(L/a) \equiv C^f \log(L/a).
\label{eq:u2infforce}
\end{equation}
It follows from \eqref{eq:taylorj0} that both integrands in \eqref{eq:quforceint} converge in the limit $q \to 0$ therefore both integrals converge for $L \to \infty$ and we thus take this limit for the derivation below.
We focus on the following integral
\begin{equation}
\int_0^{2\pi/a} \mathrm{d}q \, \frac{q(1 - J_0(q r))}{q^2 (1 + q^2 \xi^2)} = \int_0^{2\pi/r} \mathrm{d}q \, \frac{q(1 - J_0(q r))}{q^2 (1 + q^2 \xi^2)} + \int_{2\pi/r}^{2\pi/a} \mathrm{d}q \, \frac{q}{q^2 (1 + q^2 \xi^2)} - \int_{2\pi/r}^{2\pi/a} \mathrm{d}q \, \frac{q \, J_0(q r)}{q^2 (1 + q^2 \xi^2)}.
\label{eq:decompint}
\end{equation}
This decomposition is such that the first term on the r.h.s. cancels in the limit $r \to \infty$.
For the second term we note once again that $q [q^2 (1 + q^2 \xi^2)]^{-1} \sim q^{-1}$ for $q \to 0$ and $\int_{2\pi/r}^{2\pi/a} \mathrm{d}q \, q^{-1}$ diverges for $r \to \infty$, therefore we can use the equivalence on the integrals
\begin{equation}
\int_{2\pi/r}^{2\pi/a} \mathrm{d}q \, \frac{q}{q^2 (1 + q^2\xi^2)} \underset{r \to \infty}{=} \int_{2\pi/r}^{2\pi/a} \mathrm{d}q \, \frac{1}{q} = \log(r/a).
\end{equation}
Finally for the third term, using \eqref{eq:j0bound} we bound the integral in the following way
\begin{equation}
\left|\int_{2\pi/r}^{2\pi/a} \mathrm{d}q \, \frac{q J_0(q r)}{q^2 (1 + q^2 \xi^2)} \right| \leq \frac{1}{\sqrt{r}} \int_{2\pi/r}^{2\pi/a} \mathrm{d}q \, \frac{1}{q^{3/2} (1 + q^2 \xi^2)},
\end{equation}
where we note once again that $[q^{3/2} (1 + q^2\xi^2)]^{-1} \sim q^{-3/2}$ for $q \to 0$ and $\int_{2\pi/r}^{2\pi/a} \mathrm{d}q \, q^{-3/2}$ diverges for $r \to \infty$, therefore we can use the equivalence on the integrals
\begin{equation}
\frac{1}{\sqrt{r}} \int_{2\pi/r}^{2\pi/a} \mathrm{d}q \, \frac{1}{q^{3/2} (1 + q^2 \xi^2)} \underset{r \to \infty}{=} \frac{1}{\sqrt{r}} \int_{2\pi/r}^{2\pi/a} \mathrm{d}q \, \frac{1}{q^{3/2}} = \sqrt{\frac{2}{\pi}}\left[1 - \sqrt{\frac{a}{r}}\right],
\end{equation}
which is finite in the $r \to \infty$ limit.
Therefore only the second term in the r.h.s. of \eqref{eq:decompint} diverges, and we have the following equivalence
\begin{equation}
\int_0^{2\pi/a} \mathrm{d}q \, \frac{q (1 - J_0(qr))}{q^2 (1 + q^2 \xi^2)} \underset{r \to \infty}{=} \log(r/a).
\end{equation}
which eventually leads to
\begin{equation}
\left<(\boldsymbol{q}_0\cdot(\boldsymbol{u}_j-\boldsymbol{u}_i))^2\right> \underset{\substack{L\to\infty\\|\boldsymbol{r}_i^0-\boldsymbol{r}_j^0|\to\infty}}{=} \frac{f^2 \tau^2 a^2 |\boldsymbol{q}_0|^2}{4 \pi \zeta^2} \left[\frac{1}{\xi_{\parallel}^2} + \frac{1}{\xi_{\perp}^2}\right] \log(r/a) = |\boldsymbol{q}_0|^2 \, C^f \, \log(r/a).
\label{eq:quinfforce}
\end{equation}
Therefore, using \eqref{eq:psiq0fluc}, \eqref{eq:gausschar}, \eqref{eq:u2infforce}, and \eqref{eq:quinfforce}, we obtain
\begin{equation}
\left<\psi_{\boldsymbol{q}_0,i}\psi_{\boldsymbol{q}_0,j}^*\right> \underset{\substack{L\to\infty\\|\boldsymbol{r}_i^0 -\boldsymbol{r}_j^0| \to \infty}}{=} (r/a)^{-\frac{1}{2}|\boldsymbol{q}_0|^2 C^f}.
\end{equation}

\section{Large-wavelength scaling of the structure factor}
\label{app:S}

We consider the structure factor \eqref{eq:S}
\begin{equation}
S(\boldsymbol{q}) = \frac{1}{N} \sum_{i,j=1}^N \left<e^{\mathrm{i}\boldsymbol{q} \cdot (\boldsymbol{r}_j - \boldsymbol{r}_i)}\right> = \frac{1}{N} \sum_{i,j=1}^N e^{\mathrm{i}\boldsymbol{q} \cdot (\boldsymbol{r}_j^0 - \boldsymbol{r}_i^0)} \left<e^{\mathrm{i}\boldsymbol{q} \cdot (\boldsymbol{u}_j - \boldsymbol{u}_i)}\right>.
\end{equation}
For normally distributed driving noise then, due to the linearity of \eqref{eq:udiscrete}, displacements are also normally distributed.
Using \eqref{eq:gausschar} we write
\begin{equation}
\left<\exp\left(\mathrm{i}\boldsymbol{q}\cdot(\boldsymbol{u}_j - \boldsymbol{u}_i)\right)\right> = \exp\left(-\frac{1}{2}\left<(\boldsymbol{q}\cdot(\boldsymbol{u}_j - \boldsymbol{u}_i))^2\right>\right) = \exp\left(-\frac{1}{4}|\boldsymbol{q}|^2\left<|\boldsymbol{u}_j - \boldsymbol{u}_i|^2\right>\right)
\end{equation}
where the second equality derives from isotropy.
It is possible to Taylor expand the latter exponential if
\begin{equation}
|\boldsymbol{q}|^2\left<|\boldsymbol{u}_j - \boldsymbol{u}_i|^2\right> \ll 1.
\label{eq:qtaylorcond}
\end{equation}
In the limit $L \to \infty$ we consider wavevectors which scale as
\begin{equation}
|\boldsymbol{q}| \sim \frac{2\pi}{L}.
\end{equation}
Therefore, considering $\left<|\boldsymbol{u}_i - \boldsymbol{u}_j|^2\right> \leq 2 \left<|\boldsymbol{u}_i|^2\right>$ in steady state, and the following scalings
\begin{subequations}
\begin{align}
\left<|\boldsymbol{u}_i|^2\right> \underset{L\to\infty}&{=} \left<u^2\right>_{\infty}^{\sigma},\\
\left<|\boldsymbol{u}_i|^2\right> \underset{L\to\infty}&{=} C^f \log(L/a),
\end{align}
\end{subequations}
for fluctuating stress \eqref{eq:u2infstress} and fluctuating force \eqref{eq:u2infforce} respectively, we have that condition \eqref{eq:qtaylorcond} is satisfied for large enough $L \gg a$.
We then write
\begin{equation}
\begin{aligned}
S(\boldsymbol{q}) \underset{\substack{L \to \infty\\|\boldsymbol{q}|\sim2\pi/L}}&{=} \frac{1}{N} \sum_{i,j=1}^N e^{\mathrm{i}\boldsymbol{q}\cdot(\boldsymbol{r}_j^0 - \boldsymbol{r}_i^0)} \left(1 - \frac{1}{4} |\boldsymbol{q}|^2 \left<|\boldsymbol{u}_j - \boldsymbol{u}_i|^2\right>\right)\\
&= \frac{1}{N} \sum_{i,j=1}^N e^{\mathrm{i}\boldsymbol{q}\cdot(\boldsymbol{r}_j^0 - \boldsymbol{r}_i^0)} \left(1 - \frac{1}{2} |\boldsymbol{q}|^2 \left<|\boldsymbol{u}_i|^2\right> + \frac{1}{2} |\boldsymbol{q}|^2 \left<\boldsymbol{u}_i \cdot \boldsymbol{u}_j\right>\right),
\end{aligned}
\end{equation}
where the first two terms cancel due to the orthogonality relation \eqref{eq:dftortho}.
Using \eqref{eq:dft} we thus obtain
\begin{equation}
S(\boldsymbol{q}) \underset{\substack{L \to \infty\\|\boldsymbol{q}|\sim2\pi/L}}{=} \frac{1}{2N} |\boldsymbol{q}|^2 \left<|\tilde{\boldsymbol{u}}_{\boldsymbol{q}}|^2\right>.
\end{equation}

\clearpage
\section{Dynamical transition in active Brownian particles (ABP) and pair active Brownian particles (pABP)}
\label{app:abpdyn}

\begin{figure}[H]
\centering
\begin{tikzpicture}
\matrix (abpdyn)[row sep=0mm, column sep=0mm, inner sep=0mm,  matrix of nodes] at (0,0) {
\includegraphics[width=0.3\textwidth]{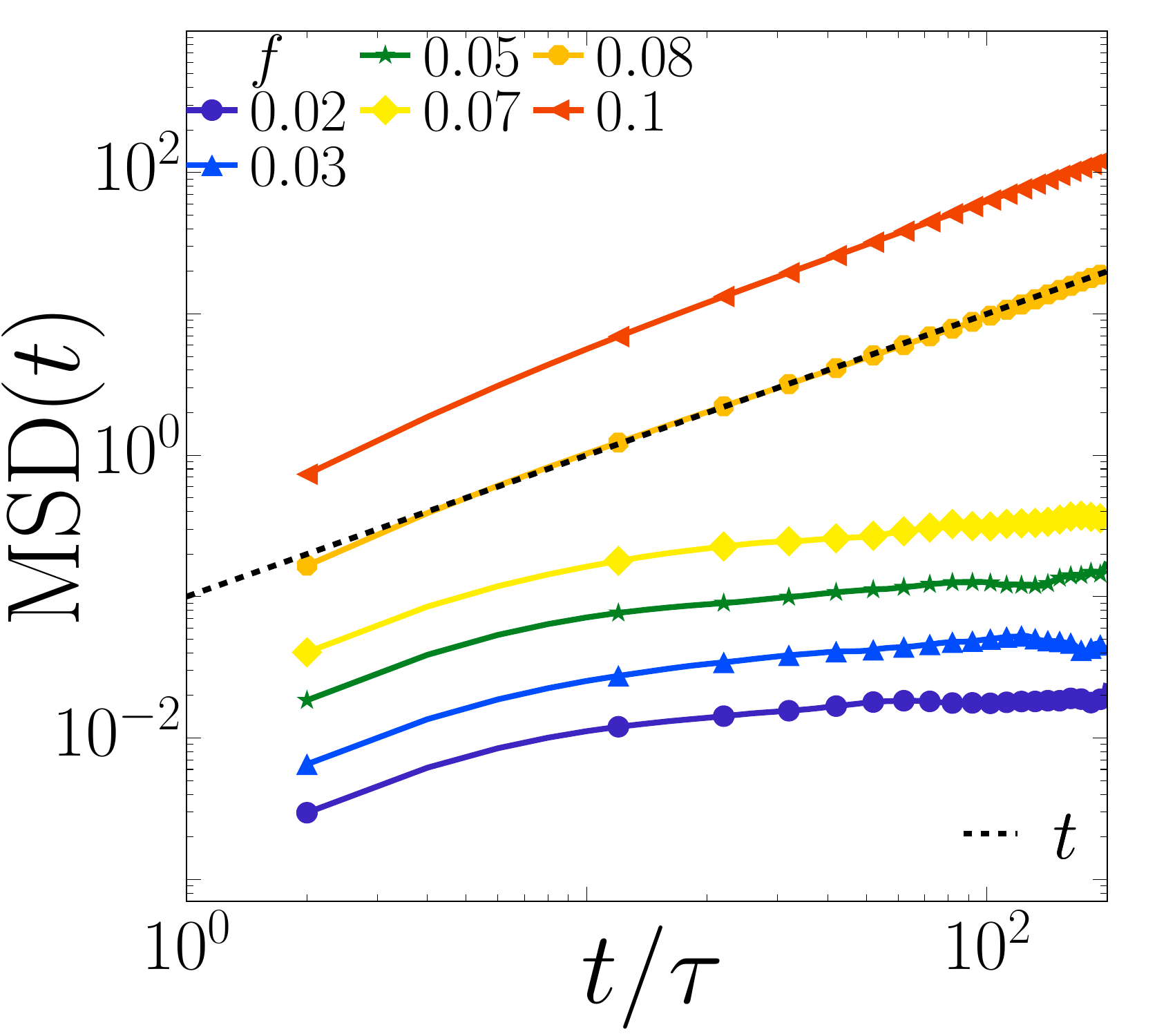}&
\includegraphics[width=0.3\textwidth]{ABP_MSD_pair.pdf}\\
\includegraphics[width=0.3\textwidth]{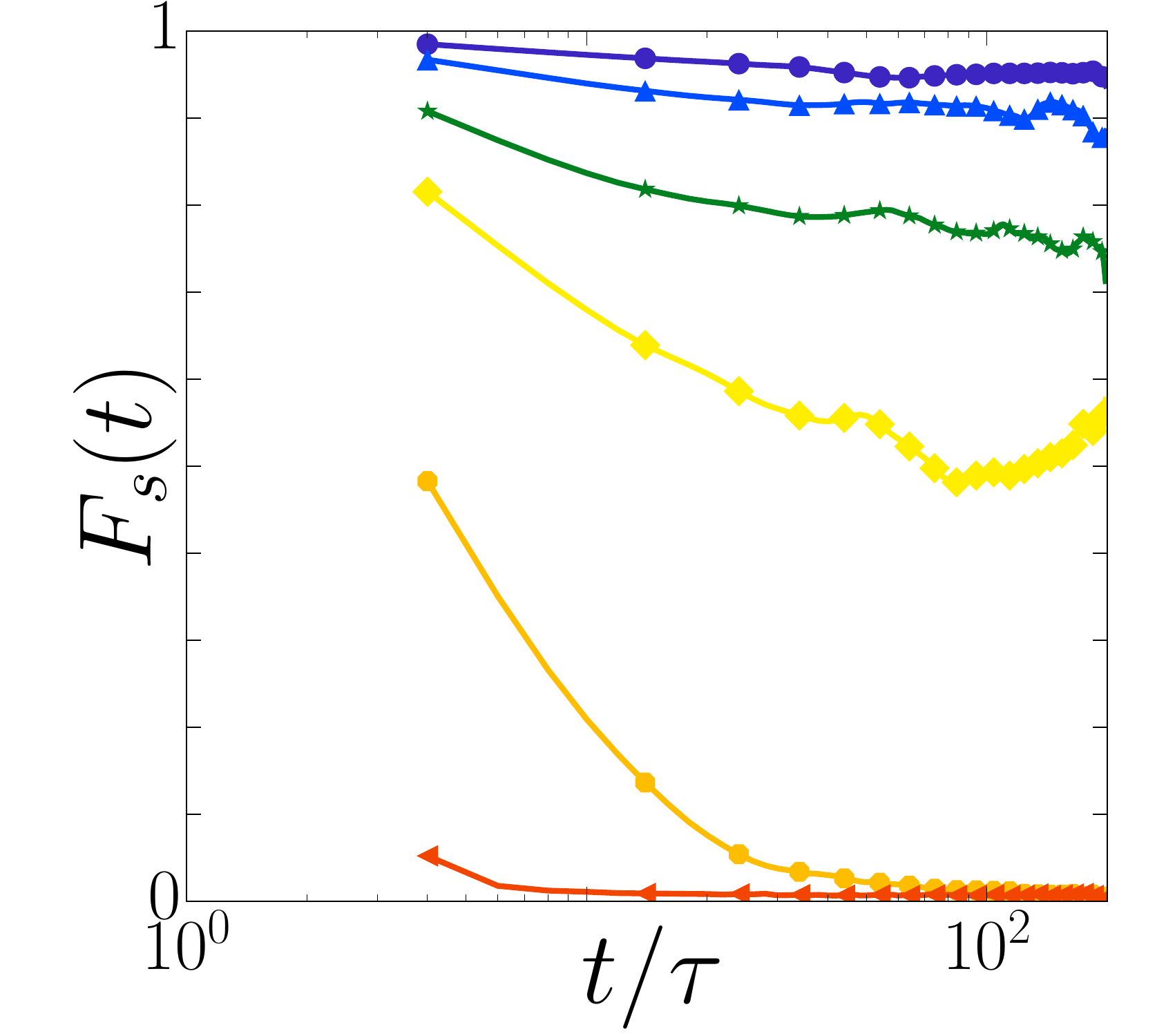}&
\includegraphics[width=0.3\textwidth]{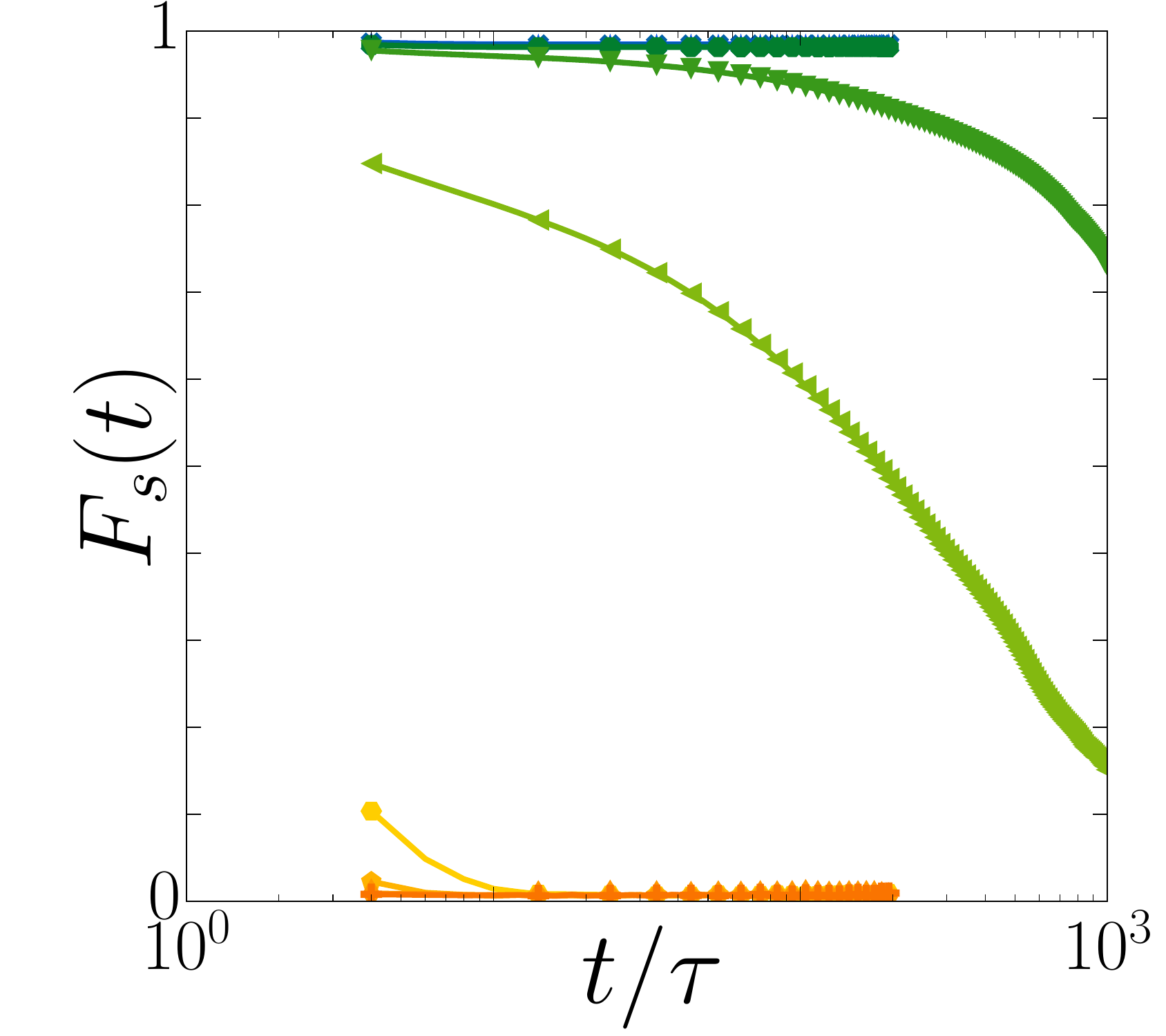}\\
};
\draw (abpdyn-1-1.north) node[above, xshift=5pt, yshift=-5pt]{\strut{\bf ABP}};
\draw (abpdyn-1-2.north) node[above, xshift=5pt, yshift=-5pt]{\strut{\bf pABP}};
\draw (abpdyn-1-1.south west) node[fill=white, inner sep=0.5pt, xshift=32.25pt, yshift=57pt]{(a)};
\draw (abpdyn-1-2.south west) node[fill=white, inner sep=0.5pt, xshift=32.25pt, yshift=57pt]{(b)};
\draw (abpdyn-2-1.south west) node[fill=white, inner sep=0.5pt, xshift=32.25pt, yshift=57pt]{(c)};
\draw (abpdyn-2-2.south west) node[fill=white, inner sep=0.5pt, xshift=32.25pt, yshift=57pt]{(d)};
\end{tikzpicture}
\caption{(a, b) Mean sqaured displacements as functions of time $\mathrm{MSD}(t)$ \eqref{eq:msd} in steady state for different stochastic force amplitudes $f$.
Dashed lines are linear functions of time as guides to the eye.
(c, d) Self-intermediate scattering function $F_s(t)$ \eqref{eq:fs} in steady state for different stochastic force amplitudes $f$.
(a, c) Active Brownian particles (ABP) and (b, d) pair active Brownian particles (\hyperref[sec:abp]{pABP}), with $N = 16384$ particles and persistence time $\tau_p=25$.
Identical markers between (a) and (c) and between (b) and (d) correspond to identical data sets.}
\label{fig:dynabp}
\end{figure}

We analyse the melting transition in both our particle models -- active Brownian particles (ABP) and pair active Brownian particles (pABP) -- from a dynamical point of view, using two-time correlation functions borrowed from glass physics.
First, the mean square displacement $\mathrm{MSD}(t)$ \eqref{eq:msd} \cite{kob1995testing} which corresponds to the variance of displacements over particles as a function of time, and the self-intermediate scattering function \cite{kob1995testinga}
\begin{equation}
F_s(\boldsymbol{k}, t) = \left<\left|\frac{1}{N} \sum_{i=1}^N \exp\Big(\mathrm{i}\boldsymbol{k} \cdot ([\boldsymbol{r}_i(t) - \overline{\boldsymbol{r}}(t)] - [\boldsymbol{r}_i(0) - \overline{\boldsymbol{r}}(0)])\Big)\right|\right>,
\label{eq:fs}
\end{equation}
with $\overline{\boldsymbol{r}}(t) = (1/N) \sum_{i=1}^N \boldsymbol{r}_i(t)$ the position of the centre of mass at time $t$ and $\boldsymbol{k} = 2\pi (1/D, 0)$, which may be interpreted as the proportion of particles whose displacements at time $t$ are less than $2\pi/|\boldsymbol{k}| = D$ in norm, assuming isotropy.
We show both functions in Fig.~\ref{fig:dynabp}.

In the ABP model, the melting transition of the soft crystal lies between $f = 0.07$ and $0.08$, with a bounded MSD (Fig.~\ref{fig:dynabp}(a)) and no or few defects below that value, and a defect-rich liquid above that value.
Melting is also evident in the more rapid decay of the self-intermediate scattering function above the transition (Fig.~\ref{fig:dynabp}(c)).

In the pABP model, there is an abrupt transition from solid to liquid between $f = 0.112$ and  $0.115$, evident in the change between a state with very low plateau values of the MSD (Fig.~\ref{fig:dynabp}(b)) and a correspondingly high plateau value of the self-intermediate scattering function (Fig.~\ref{fig:dynabp}(d)) on the one hand, and a rapidly diffusing liquid state where correlations are lost on the other hand.

\begin{figure}[H]
\centering
\includegraphics[width=0.3\textwidth]{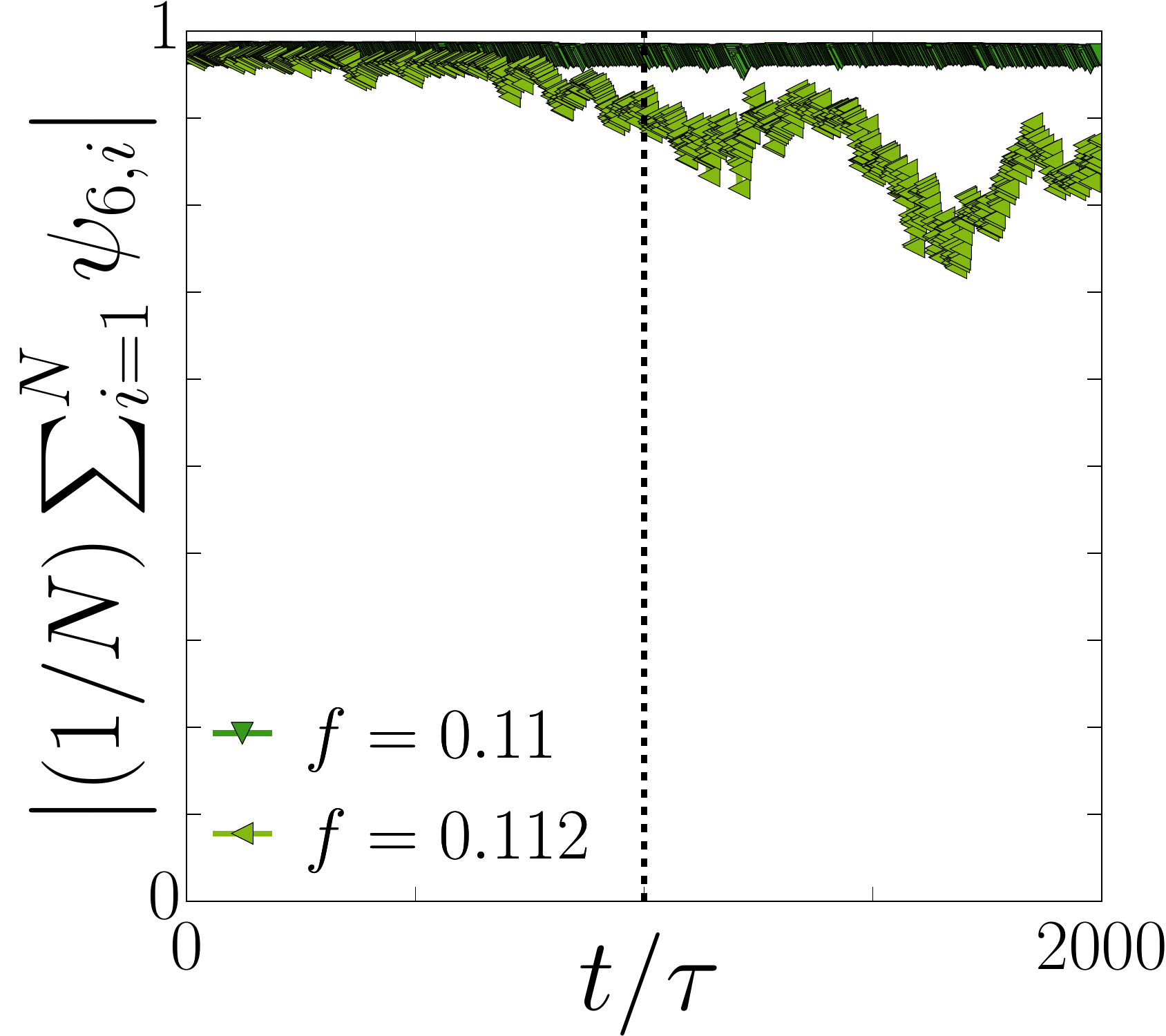}
\caption{
Time lapse of the ensemble-averaged hexatic order parameter $\left|(1/N)\sum_{i=1}^N \psi_{6,i}\right|$ for pair active Brownian particles systems from an initial ordered configuration.
Vertical dashed line represents half of the time lapse.
}
\label{fig:psi6long}
\end{figure}

We confirm that pair active Brownian particles systems at $f=0.11$ (solid phase) and $f=0.112$ (phase-separated) have reached steady by reproducing the time evolution of the hexatic order parameter in these systems, starting from an initial ordered configuration, in Fig.~\ref{fig:psi6long}.
We have used the data starting roughly from the half of this time lapse (highlighted with a vertical dashed line) to compute the mean squared displacement and self-intermediate scattering function of Fig.~\ref{fig:dynabp}.
We also provide videos highlighting the argument of the local hexatic order parameter (see \textit{e.g.} Fig.~\ref{fig:defects}(a-c)) at $f=0.11$ (\texttt{supp\_video\_1.mp4}) and $f=0.112$ (\texttt{supp\_video\_2.mp4}).

\clearpage
\section{Defects in pair active Brownian particles (pABP)}
\label{app:abpdefects}

\begin{figure}[H]
\centering
\begin{tikzpicture}
\matrix (abpdefects)[row sep=0mm, column sep=0mm, inner sep=0mm,  matrix of nodes] at (0,0) {
\includegraphics[width=0.2625\textwidth]{ABP_snap_psi6_pair_v0_0.1.pdf}&
\includegraphics[width=0.2625\textwidth]{ABP_snap_psi6_pair_v0_0.112.pdf}&
\includegraphics[width=0.2625\textwidth]{ABP_snap_psi6_pair_v0_0.15.pdf}\\
\includegraphics[width=0.2625\textwidth]{ABP_snap_psiq0_pair_v0_0.1.pdf}&
\includegraphics[width=0.2625\textwidth]{ABP_snap_psiq0_pair_v0_0.112.pdf}&
\includegraphics[width=0.2625\textwidth]{ABP_snap_psiq0_pair_v0_0.15.pdf}\\
\includegraphics[width=0.2625\textwidth]{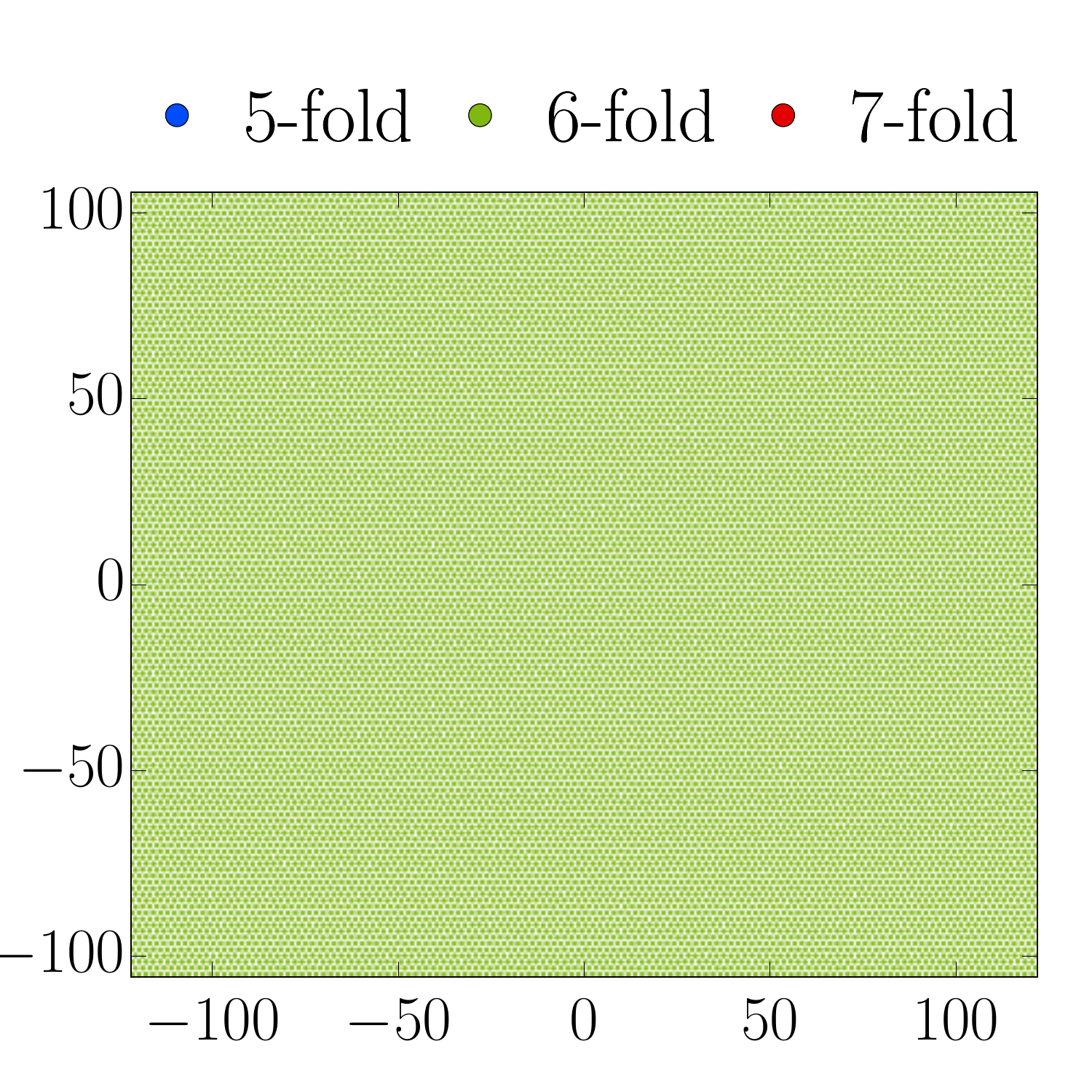}&
\includegraphics[width=0.2625\textwidth]{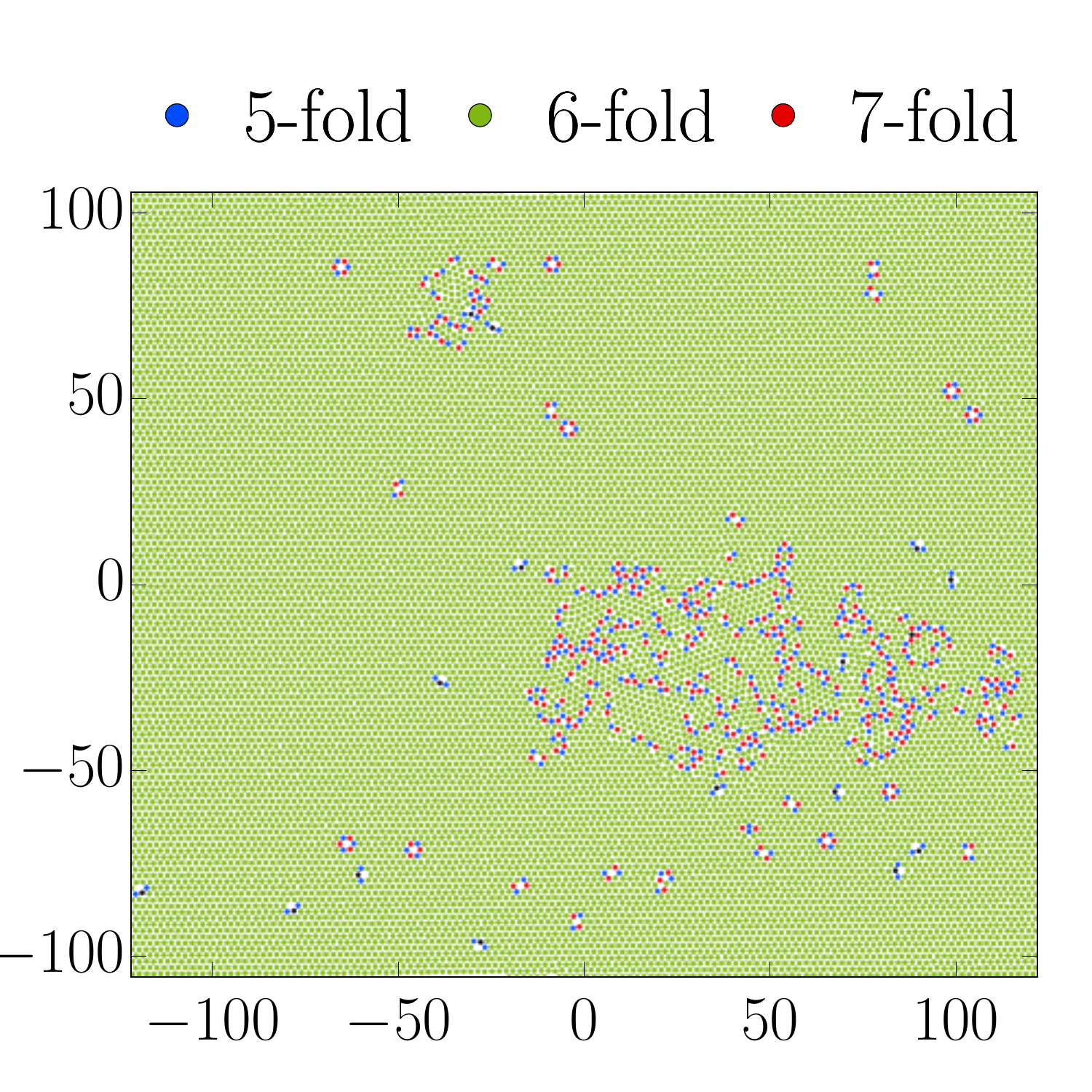}&
\includegraphics[width=0.2625\textwidth]{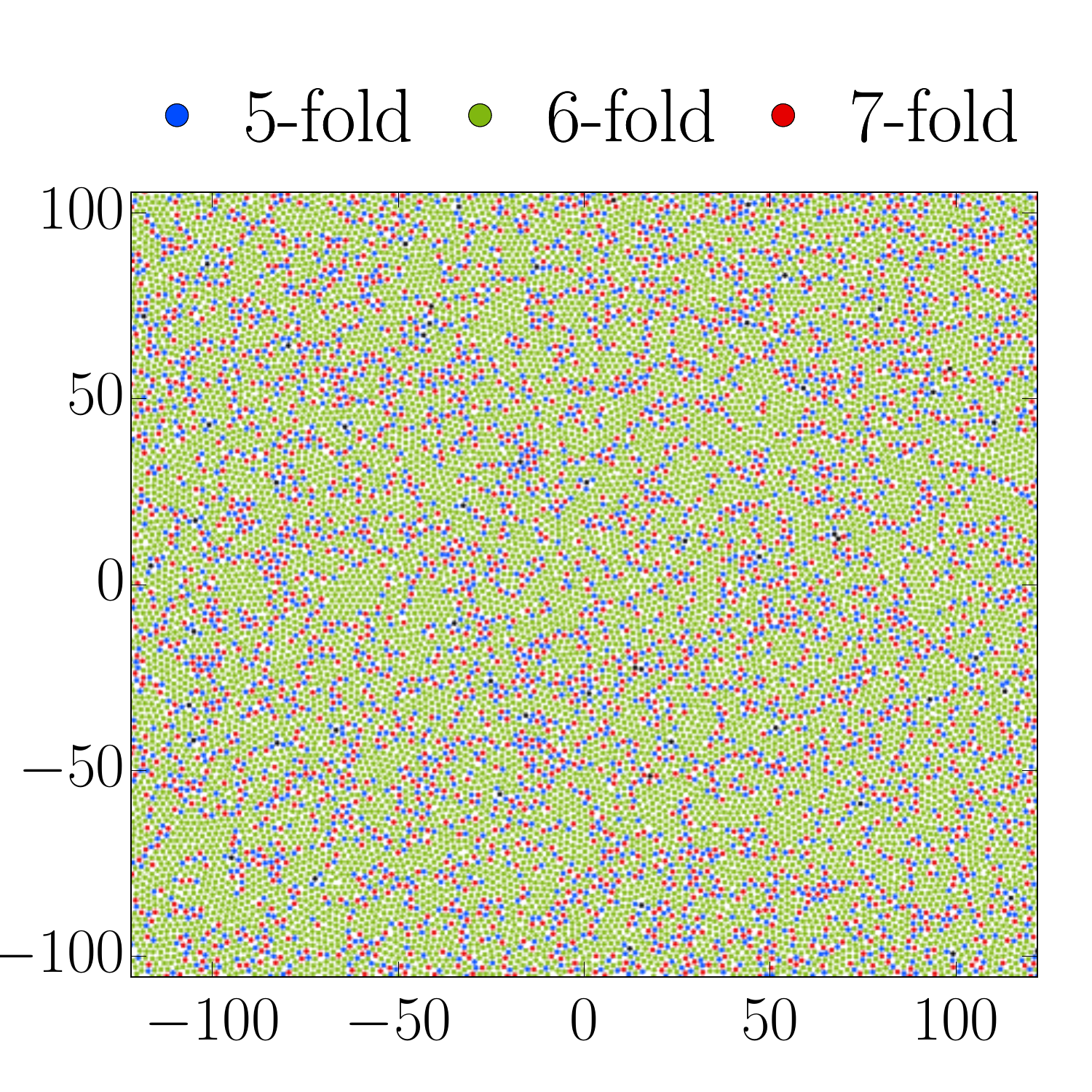}\\
};
\draw (abpdefects-1-1.north) node[above, xshift=5pt, yshift=-5pt]{\large\strut{$f=0.1$}};
\draw (abpdefects-1-2.north) node[above, xshift=5pt, yshift=-5pt]{\large\strut{$f=0.112$}};
\draw (abpdefects-1-3.north) node[above, xshift=5pt, yshift=-5pt]{\large\strut{$f=0.15$}};
\draw (abpdefects-1-1.west) node[rotate=90, xshift=-5pt, yshift=5pt]{\large\strut{$\psi_{6,i}$}};
\draw (abpdefects-2-1.west) node[rotate=90, xshift=-5pt, yshift=5pt]{\large\strut{$\psi_{\boldsymbol{q}_0,i}$}};
\draw (abpdefects-3-1.west) node[rotate=90, xshift=-5pt, yshift=5pt]{\large\strut{$z_i$}};
\draw (abpdefects-1-1.north west) node[fill=white, inner sep=0.5pt, xshift=25pt, yshift=-31pt]{(a)};
\draw (abpdefects-1-2.north west) node[fill=white, inner sep=0.5pt, xshift=25pt, yshift=-31pt]{(b)};
\draw (abpdefects-1-3.north west) node[fill=white, inner sep=0.5pt, xshift=25pt, yshift=-31pt]{(c)};
\draw (abpdefects-2-1.north west) node[fill=white, inner sep=0.5pt, xshift=25pt, yshift=-31pt]{(d)};
\draw (abpdefects-2-2.north west) node[fill=white, inner sep=0.5pt, xshift=25pt, yshift=-31pt]{(e)};
\draw (abpdefects-2-3.north west) node[fill=white, inner sep=0.5pt, xshift=25pt, yshift=-31pt]{(f)};
\draw (abpdefects-3-1.north west) node[fill=white, inner sep=0.5pt, xshift=25pt, yshift=-31pt]{(g)};
\draw (abpdefects-3-2.north west) node[fill=white, inner sep=0.5pt, xshift=25pt, yshift=-31pt]{(h)};
\draw (abpdefects-3-3.north west) node[fill=white, inner sep=0.5pt, xshift=25pt, yshift=-31pt]{(i)};
\end{tikzpicture}
\caption{Order and disorder for pair active Brownian particles (\hyperref[sec:abp]{pABP}).
Visualisation of the argument of the local hexatic order parameter $\mathrm{arg}(\psi_{6,i})$ \eqref{eq:psi6} (a-c), the argument of the local translational order parameter $\mathrm{arg}(\psi_{\boldsymbol{q}_{0,i}})$ \eqref{eq:psiq0} (d-f), and the coordination number $z_i$ of particles (g-i).
We used $N=16384$ particles and persistence time $\tau = 25$.
Stochastic force amplitude is (a, d, g) $f=0.1$, (b, e, h) $f=0.112$, (c, f, i) $f=0.15$.}
\label{fig:defects}
\end{figure}

We reproduce in Fig.~\ref{fig:defects}(a-d) the snapshots of Fig.~\ref{fig:snap} and add the corresponding visualisation of the coordination number $z_i$ of particles (Fig.~\ref{fig:defects}(e-f).

We highlight structural order with three descriptors: the local orientational order parameter $\psi_{6,i}$ \eqref{eq:psi6} (Fig.~\ref{fig:defects}(a-b)) which characterises the local orientation of the lattice, the local translational order parameter $\psi_{\boldsymbol{q}_0,i}$ (Fig.~\ref{fig:defects}(c-d)) which characterises the translational regularity of the lattice, and the coordination $z_i$ of particles (Fig.~\ref{fig:defects}(e-f)), \textit{i.e.} the number of neighbours of each particle.
We compute these neighbours with a Delaunay triangulation from the particles' positions -- this is done with library \texttt{CGAL} \cite{cgal}.
We stress that in a regular triangular lattice, $z_i = 6$.
Coordination numbers $z_i \neq 6$ (typically $z_i = 5$ or $7$) characterise topological defects known as \textit{disclinations} which locally break translational order.
However orientational order is not broken if these defects are bound.
In this latter case they correspond to a \textit{dislocation}.

\clearpage
\section{Unscaled structure factor}
\label{app:unscaledS}

\begin{figure}[H]
\centering
\begin{tikzpicture}
\matrix (uS)[row sep=0mm, column sep=0mm, inner sep=0mm,  matrix of nodes] at (0,0) {
\includegraphics[width=0.3\textwidth]{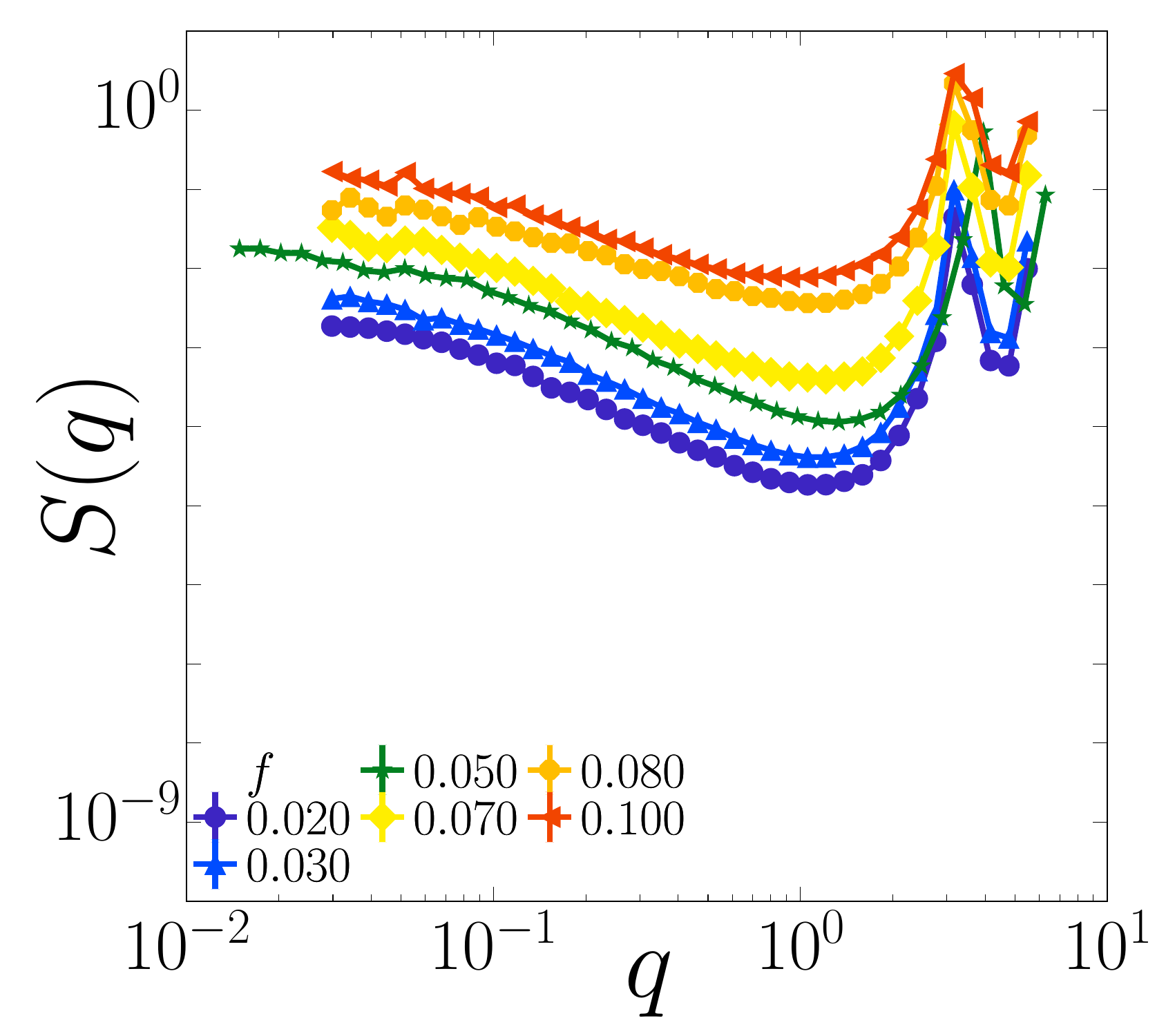}&
\includegraphics[width=0.3\textwidth]{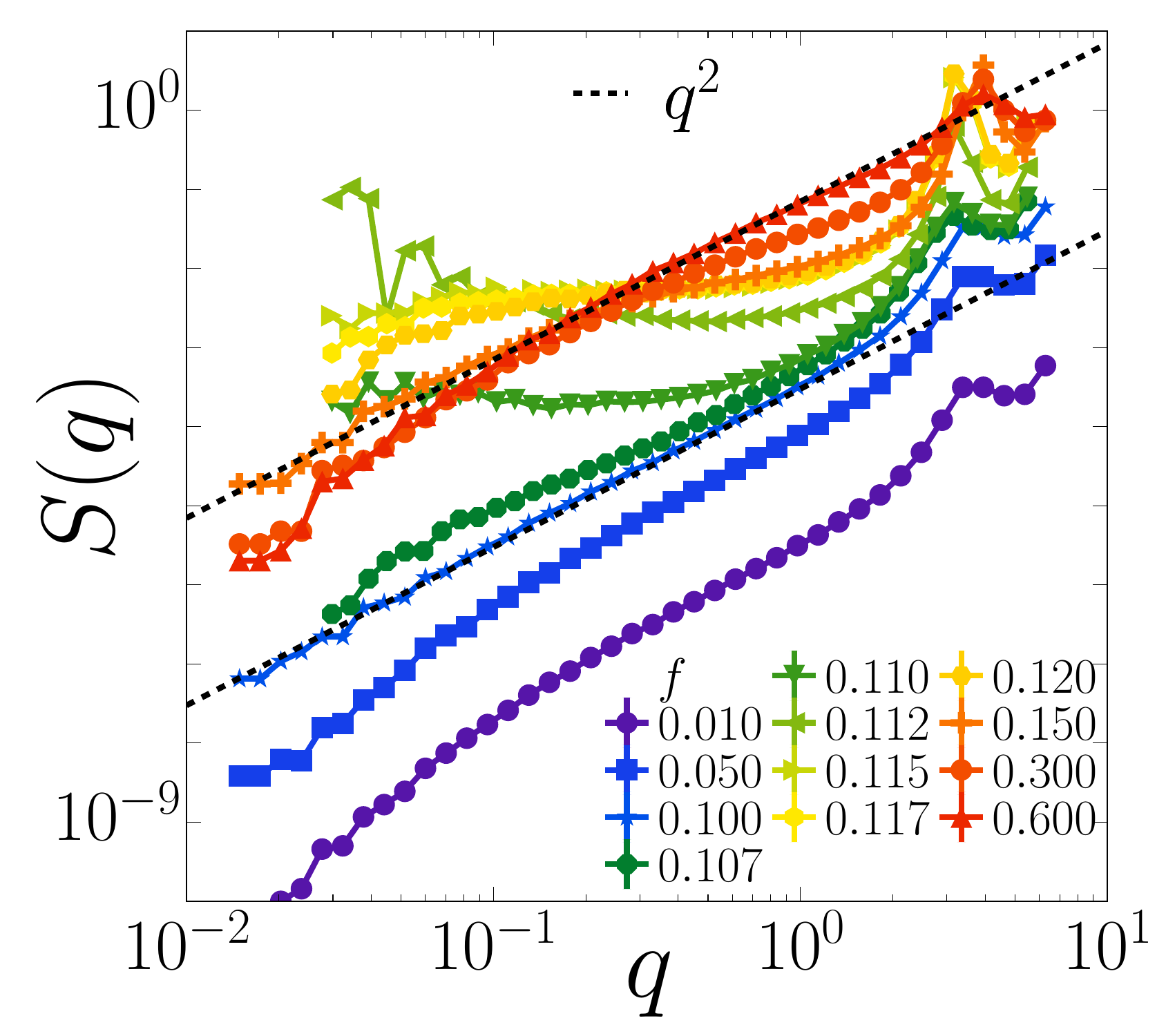}\\
};
\draw (uS-1-1.north) node[above, xshift=5pt, yshift=-5pt]{\strut{\bf ABP}};
\draw (uS-1-2.north) node[above, xshift=5pt, yshift=-5pt]{\strut{\bf pABP}};
\draw (uS-1-1.north east) node[fill=white, inner sep=0.5pt, xshift=-15pt, yshift=-50pt]{(a)};
\draw (uS-1-2.north east) node[fill=white, inner sep=0.5pt, xshift=-15pt, yshift=-50pt]{(b)};
\end{tikzpicture}
\caption{Structure factor $S(q)$ \eqref{eq:S} defined as cylindrical average of $S(\boldsymbol{q})$ over wave-vectors $\boldsymbol{q} = (2\pi m/L_x, 2\pi n/L_y)$ which satisfy $|\boldsymbol{q}| \in [q -\delta q/2, q + \delta q/2]$ with $\delta q=10^{-2}$.
We plot the structure factor for different stochastic force amplitudes $f$ in (a) for the particle model with particle-wise stochastic force (ABP), and in (b) for the particle model with pair-wise stochastic forces (\hyperref[sec:abp]{pABP}).
We used $N=16384$ or $65536$ particles, and persistence $\tau = 25$.}
\label{fig:uS}
\end{figure}

We reproduce in Fig.~\ref{fig:uS} the unscaled data of Fig.~\ref{fig:S}.

\end{document}